\documentclass[aps,prc,10pt,showpacs,showkeys,onecolumn,superscriptaddress,groupedaddress]{revtex4-1}
\usepackage{graphicx}
\usepackage{graphics}
\usepackage{xspace}
\usepackage{amssymb}
\usepackage{amsmath}
\usepackage[dvips]{color}
\usepackage{latexsym}
\usepackage{mathtools}
\usepackage{mathrsfs}
\usepackage{url}
\usepackage[english]{babel}
\usepackage{color}
\usepackage{bm}
\newcommand{\inieq}{\begin{eqnarray}}            
\newcommand{\fineq}{\end{eqnarray}}            
\newcommand{\diff}{{\rm\,d}}                    
\newcommand{\be}{\begin{equation}}
\newcommand{\ee}{\end{equation}}
\newcommand{\ba}{\begin{eqnarray}}
\newcommand{\ea}{\end{eqnarray}}

\def\r{\mbox{{\bf  r}}}

\def\q{\mbox{\boldmath $q$}}

\def\ee{\mbox{$\left(e,e^{\prime}\right)$\ }}
\def\eep{\mbox{$\left(e,e^{\prime}p\right)$\ }}

\DeclarePairedDelimiter{\abs}{\lvert}{\rvert}

\begin{document}
\noindent
\title{Elastic and quasi-elastic electron scattering on the 
$N = 14,  20$, and $28$ isotonic chains}
\author{Andrea Meucci}
\affiliation{Dipartimento di Fisica,  
Universit\`a degli Studi di Pavia and \\
INFN, Sezione di Pavia,  Via A. Bassi 6, I-27100 Pavia, Italy}
\author{Matteo Vorabbi}
\affiliation{Dipartimento di Fisica,  
Universit\`a degli Studi di Pavia and \\
INFN, Sezione di Pavia,  Via A. Bassi 6, I-27100 Pavia, Italy}
\author{Carlotta Giusti}
\affiliation{Dipartimento di Fisica,  
Universit\`a degli Studi di Pavia and \\
INFN, Sezione di Pavia,  Via A. Bassi 6, I-27100 Pavia, Italy}
\author{Franco Davide Pacati}
\affiliation{Dipartimento di Fisica,  
Universit\`a degli Studi di Pavia and \\
INFN, Sezione di Pavia,  Via A. Bassi 6, I-27100 Pavia, Italy}
\author{Paolo Finelli}
\affiliation{Dipartimento di Fisica e Astronomia, 
Universit\`{a} degli Studi di Bologna and \\
INFN,
Sezione di Bologna, Via Irnerio 46, I-40126 Bologna, Italy}
\date{\today}

\begin{abstract} 
We present theoretical predictions for  
electron scattering  on  the 
$N = 14,  20$, and $28$ isotonic
chains from proton-deficient to proton-rich nuclei. 
The calculations are performed within the framework of the distorted-wave
Born approximation and the proton  and neutron density distributions are 
evaluated adopting a Relativistic Hartree-Bogoliubov (RHB) approach with a density dependent 
meson-exchange interaction. 
We present results 
for the elastic and quasi-elastic cross sections and for the parity-violating
asymmetry parameter.
Owing to the correlations between
the evolution of the electric charge form factors along each  chain  with the
underlying proton shell structure of the isotones, 
elastic electron scattering experiments on
isotones can provide useful informations about the 
 occupation and filling of the single-particle levels of protons.
\end{abstract}

\pacs{25.30.Bf; 21.10.-k; 25.30.Fj;  24.10.Jv}
\keywords{elastic electron scattering; 
Relativistic mean field models, Ground-state properties}

\maketitle

\section{Introduction}
\label{intro}
The study of nuclear properties has gained large and detailed information 
from reactions investigating the nuclear response to 
external probes 
\cite{Hofstadter:1956qs,Donnelly:1975ze,Donnelly:1984rg,Boffi:1993gs,RevModPhys.80.189}. 
Electron induced reactions are particularly well suited to explore such 
properties, as the predominant electromagnetic interaction of electrons with
nuclei is well known and relatively weak. Electrons 
are able to penetrate nuclear matter deeply  and without a large perturbation of 
the structure, being their mean free path in nuclei much larger than the nuclear 
dimensions. Moreover, as the energy $\omega$ and the momentum $\q$ 
transferred to the nucleus 
can be varied independently, the obtained spectral function can be mapped with a 
resolution that is adjusted to the scale of the process that is considered. 
A lot of experimental and theoretical work on elastic and inelastic electron scattering 
at different energies has provided detailed information on the charge density 
distribution of the nuclear ground state and on the energy, strength, and 
quantum numbers of the excited states produced by single particle (s.p.) or 
collective excitation mechanisms \cite{DeJ:1987qc,Fricke:1995zz,Donnelly:1975ze,Heisenberg:1984hb}. 

At a transferred energy of about $\omega = q^2/(2m_N)$, where $m_{N}$ is the
nucleon mass, the quasi-elastic (QE) peak appears. In this kinematic region 
the probe interacts essentially with one single nucleon which is then ejected 
and all the other nucleons behave as spectators. 
If only the scattered electron is detected, we have the inclusive reaction 
where all final nuclear states are included. A considerable number of data 
have been collected in the QE region \cite{book,RevModPhys.80.189,web-benhar}, 
concerning not only the differential cross sections but also the separation 
of longitudinal and transverse response functions. 

When also an emitted proton is detected in coincidence with the outgoing 
electron and the residual nucleus is left in a discrete eigenstate we have an 
exclusive reaction. A large number of coincidence $(e,e^{\prime}p)$  
experiments have been performed, which have confirmed the assumption of a 
direct knockout mechanism and have provided  detailed information on the s.p. 
properties of light-to-heavy nuclei \cite{Boffi:1993gs,book,Frullani:1984nn,
Bernheim:1981si,Lapikas:1003zz,deWittHuberts:1990zy,Udias:1993xy,gao00,Meucci:2001qc}.
In particular, the s.p. energies and the quantum numbers of the emitted 
nucleon inside the nucleus have been determined, directly relating to shell model 
properties. Moreover, the comparison between experimental data and theoretical 
calculations made it possible to extract spectroscopic factors, which 
revealed a partial occupation of the different shells and therefore the 
importance of nuclear correlations and the limits of a mean-field description 
of nuclear structure.

The use of electron scattering can be extended to the study of exotic nuclei 
far from stability line. The evolution of nuclear properties with the 
increasing asymmetry between the number of protons and neutrons is one of the 
most interesting topics of nuclear physics. Of particular interest is the 
behavior of the s.p. properties, with a consequent modification of the shell model 
magic numbers. The exclusive $(e,e^{\prime}p)$ reaction would be the best 
suited tool for this study \cite{Giusti:2011it,esotici1}. However, the 
measurement of $(e,e^{\prime}p)$ cross sections requires a double 
coincidence detection which is very difficult. Experiments for elastic 
scattering, and possibly inclusive QE scattering, appear easier to perform and 
are therefore to be considered as a first step.

In the next years radioactive ion beam (RIB) facilities \cite{tan95,gei95,mue01}  
will produce a large amount of data on unstable nuclei. In particular, 
electron-RIB colliders using storage rings are under construction at RIKEN 
(Japan) \cite{sud01,sud09,kat03} and GSI (Germany) \cite{gsi06}. Proposals 
have been presented for the ELISe experiment at FAIR in Germany 
\cite{elise,Simon:2007zz,Antonov:2011zza} and the SCRIPT project in Japan 
\cite{sud10,Suda01012012}.

From the theoretical point of view, several studies of electron 
scattering on unstable nuclei have already been published 
\cite{Garrido:1999zy,Garrido:2000ht,Ershov:2005kq,PhysRevC.70.034303,Antonov04,
PhysRevC.72.044307,PhysRevC.75.024606,Khan:2007ji,PhysRevC.79.034318,
PhysRevC.79.044313,RocaMaza:2008cg,RocaMaza:2012hv,PhysRevC.79.014317,
PhysRevC.77.064302,Liu:2012zj,Dong:2012ed,Giusti:2011it,esotici1}.  
In a recent paper \cite{esotici2} we have presented and discussed 
numerical predictions for elastic and inclusive QE electron scattering on 
oxygen and calcium isotopic chains, with the aim of investigating the evolution
of some nuclear properties with increasing asymmetry between the number of neutrons 
and protons.
The elastic electron scattering gives information on the global properties of a nucleus 
and, in particular, on the charge density distributions and on the properties of
proton wave functions. The inclusive QE scattering is the integral of the 
spectral density function over all the available final states. As such, it is 
affected by the dynamical properties and preferably exploits the nuclear s.p. 
aspects. 
 
It is much more difficult to measure neutron density distributions. 
Direct access to the neutron distribution can be obtained from the  
parity-violating asymmetry parameter $A_{pv}$, which 
is defined as the difference between the cross sections for the scattering of 
right- and left-handed longitudinally polarized electrons 
\cite{Donnelly1989589,Donnelly19791,PhysRevC.57.3430}. 
This quantity is related
to the radius of the neutron distribution $R_n$, because $Z^0$-boson exchange, 
which mediates the weak neutral interaction, couples mainly to neutrons and 
gives a model-independent measurement of $R_n$.

The first measurement of  $A_{pv}$ was performed by the PREX 
experiment \cite{Abrahamyan:2012gp,PhysRevC.85.032501} on $^{208}$Pb  
with a poor resolution. 
More stringent results are expected from an improved experiment 
(PREX-II \cite{prex2}) which has been recently approved. A recent result for 
the neutron skin on $^{208}$Pb has been extracted from coherent pion 
photoproduction cross sections measured at MAMI Mainz electron 
microtron \cite{mainz-neutronskin}.
These data, combined with the results of the CREX experiment on $^{48}$Ca 
\cite{crex}, which has also been conditionally approved, will provide an 
important test on the validity  of microscopic models concerning the 
dependence of $R_n$ on the mass number $A$ and the properties of the neutron 
skin. 

In \cite{esotici2} we have compared our calculations 
for the asymmetry parameter $A_{pv}$ with the result of the first PREX 
experiment on $^{208}$Pb and have obtained a good agreement with the 
empirical value. Moreover, we have provided numerical predictions for the 
future experiment CREX on $^{48}$Ca and we have studied the behavior of 
$A_{pv}$ along oxygen and calcium isotonic chains.  

In this work we extend our study to isotonic chains. We present
and discuss numerical predictions for the cross sections of elastic and 
inclusive QE electron scattering and for the parity-violating asymmetry 
parameter $A_{pv}$ on the
$N =$ 14, 20, and 28  isotonic chains. This is complementary to our previous
study on isotopic chains of \cite{esotici2}. Electron scattering on
an isotopic chain gives information on the dependence of the charge density
distribution and of the proton wave functions on the neutron number. 
In an isotonic chain we can investigate the 
behavior of  the charge distribution and of the proton s.p. states 
when a new proton is added to the nucleus. Nuclei with both neutron and 
proton excess are considered in our study.
Our choice of the isotonic chains is motivated by the fact that data for light 
nuclei, such as those  with 
$N = 14$, are likely to be obtained in future electron scattering 
facilities such as SCRIT and ELISe;  the  $N =$ 20, and 28 
isotonic chains correspond to nuclei with an intermediate mass and a magic
number of neutrons. 
Medium-size nuclei 
are very interesting to study the details of nuclear forces.  
In particular, semi-magic 
$N =  20$ and $28$ isotones are composed by a moderately large number
of nucleons with orbits clearly separated from the neighboring ones,
so that smooth but continuous modifications in nuclear shapes 
can be observed when protons are added or removed. 
Therefore, the peculiar features of
nuclear forces between nucleons moving in orbits with specific quantum numbers
are much easier to disentangle in medium than in heavy nuclei, where 
the modification of shell structures occurs only when many
nucleons are involved.  A recent review of
the evolution of the $N = 28$ shell closure far from stability can be found in
\cite{sorlin}.

The basic ingredients of the calculations for both elastic and QE
scattering are the ground state wave functions of proton and neutron s.p. 
states. Models based on the relativistic mean-field (RMF) approximation
have been successfully applied in analyses of nuclear structure  
from light  to superheavy nuclei.
RMF models are phenomenological because the 
parameters of their effective Lagrangian are adjusted to reproduce
the nuclear matter equation of state and a set of global properties of 
spherical closed-shell nuclei \cite{Serot:1984ey,Rein:1989,Rign:1996,Serot:1997}. 
In recent years effective hadron field 
theories with additional non-linear terms and density-dependent coupling 
constants adjusted to Dirac-Brueckner self-energies in nuclear matter 
have been able to obtain a satisfactory nuclear matter equation of state  
and to reproduce the empirical bulk properties of 
finite nuclei \cite{Vretenar:2005zz}. 
To obtain a proper description of open-shell nuclei a unified and self-consistent treatment
of mean-field and pairing correlations is necessary. The relativistic
Hartree-Bogoliubov (RHB) model provides a unified treatment of the nuclear
mean-field and pairing correlations, which is crucial for an accurate description 
of the properties of the ground and excited states in weakly bound nuclei, 
and has been successfully employed in the analyses of exotic nuclei far from the valley of 
stability. For most nuclei here considered, however, pairing 
effects are negligible or small and the RHB model gives wave functions  
practically equivalent to the ones that can be obtained with the relativistic 
Dirac-Hartree model used in \cite{esotici2}.

The cross sections for elastic electron scattering are 
obtained solving the partial wave Dirac equation and include Coulomb 
distortion effects. For the inclusive QE electron scattering calculations are
performed with the relativistic Green's function (RGF) model, 
which  has already been widely and successfully applied to the analysis of
QE electron and neutrino-nucleus scattering data on different 
nuclei \cite{Meucci:2003uy,Meucci:2003cv,Meucci:2005pk,Meucci:2009nm,
Meucci:2011pi,Meucci:ant,Meucci:2011nc,Meucci:2011vd}.

The paper is organized as follows. In Sec. \ref{sec.rmf}  we give the main ingredients 
of  the relativistic model for the calculation of the ground state observables. 
In Sec. \ref{sec.scattering} we outline the main features of the elastic electron scattering, 
including the definition of the parity-violating asymmetry. The results for 
elastic electron scattering on the $N =$ 28, 20, and 14 isotonic chains are 
presented and discussed in Sec. \ref{results}.  In Sec. \ref{sec.qe} 
we present the inclusive QE 
scattering and show and discuss the corresponding numerical results. 
Some conclusions are drawn in Sec. \ref{conc}.

\section{Relativistic model for ground state observables }
\label{sec.rmf}

The RHB model \cite{Vretenar:2005zz} 
represents the most sophisticated
method to include particle-hole ({\it ph}) and particle-particle ({\it pp}) interactions
in a mean-field approximation. In this framework, the ground state 
of a nucleus $|\Phi_0\rangle$ 
is represented by the product of independent single-quasiparticle states that are 
eigenvectors of a Hamiltonian containing two average potentials: 
a self-consistent mean-field $\hat{h}$, which encloses
all the long range {\it ph} correlations, and a pairing field $\hat{\Delta}$, which includes
{\it pp} correlations. For the {\it ph} part we use a finite range model in which the nuclear interaction
is mediated by meson exchange \cite{Serot:1984ey}. 
This effective interaction is characterized by
meson masses and density dependent coupling constants.
Even if recently there have been
some improvements \cite{Finelli:2005ni, Bogner:2009bt} towards an ab-initio derivation,
the most successful parametrizations are purely phenomenological with parameters
adjusted to reproduce the nuclear matter equation of state and a set of bulk properties
of closed-shell nuclei (we employed a DDME1 parameterization \cite{PhysRevC.66.024306}).
On the other hand, pairing correlations are described by the 
corresponding {\it pp}-part of the finite 
range Gogny interaction \cite{Decharge:1979fa, Berger:1984zz}. The pairing interaction
is treated at a non relativistic level, as discussed in  \cite{Serra:2001pv}.

The relativistic Hartree-Bogoliubov equations can be written as follows
\begin{equation}
\left(\begin{array}{cc}
\hat{h} - m - \lambda & \hat{\Delta} \\
-\hat{\Delta}^* & - \hat{h} + m + \lambda
\end{array}
\right)
\left(
\begin{array}{c}
U (r) \\
V (r)
\end{array}
\right) =
E 
\left(
\begin{array}{c}
U (r) \\
V (r)
\end{array}
\right) \; ,
\end{equation}
where $m$ is the nucleon mass and $\lambda$ is the chemical
potential determined by the particle number condition. 
The column vectors denote the quasiparticle wave functions and $E$ are the
quasiparticle energies. 
The RHB equations are solved self-consistently, together with the Klein-Gordon 
equations for the meson fields and the Poisson equation for the photon field. 
The relevant density for electron scattering calculations is defined as follows
\begin{equation}
\rho_p (r) = \sum_{E_k>0} V^\dagger_k \frac{1-\tau_3}{2} V_k \; .
\end{equation}
We refer the reader to \cite{Vretenar:2005zz} for more details.


\section{Elastic electron scattering}
\label{sec.scattering}
The differential cross section for the elastic scattering of an
electron, with momentum transfer $q$, off a spherical spin-zero nucleus  is given 
in the plane-wave Born approximation (PWBA) by
\inieq
\left(\frac{\diff \sigma}{ \diff \Omega^{\prime}}\right)_{{EL}} =
\sigma_{{M}} 
\abs{F_p(q)}^2
\ , \label{eq.csel}
\fineq
where  $\Omega^{\prime}$ is the scattered electron solid angle, 
$\sigma_{{M}}$ is the Mott cross section \cite{Boffi:1993gs,book}
and 
\inieq
F_p(q) = 
\int \diff \r ~ \jmath_0 (qr) \rho_p (r) \     \label{eq.csfp}
\fineq
is the charge form factor for a spherical nuclear charge density 
$\rho_p (r)$  and $\jmath_0$ is the
zeroth order spherical Bessel function.

In the case of medium and heavy nuclei the
distortion produced on the electron wave functions by the nuclear Coulomb
potential  from $\rho_p (r)$ cannot be neglected and the elastic 
cross sections are obtained in 
the distorted-wave Born approximation (DWBA)  
from the numerical solution of the partial wave Dirac equation.

\begin{figure}
	\centering \vskip 1mm
		\includegraphics[scale=0.35]{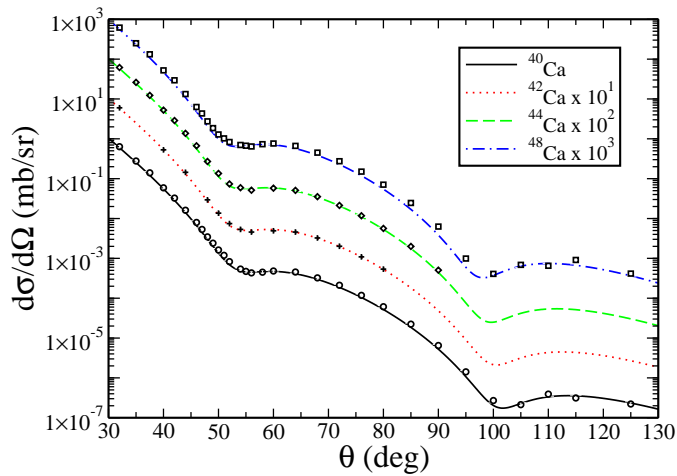}
\caption{ (Color online)
Differential cross sections for elastic electron
scattering on calcium isotopes at an electron energy $\varepsilon = 250$  MeV 
as  functions of the scattering angle $\theta$.
Experimental data from \cite{PhysRev.174.1380}.}	\vspace {5mm}\label{fig:calcium}
\end{figure}

\begin{figure}
	\centering \vskip 1mm
		\includegraphics[scale=0.35]{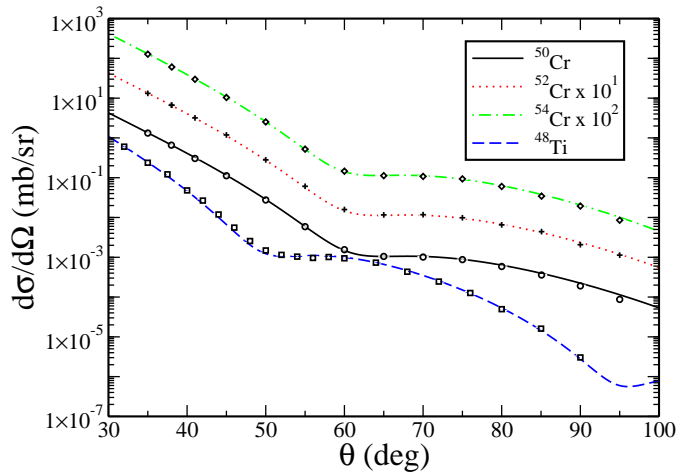}
\caption{ (Color online)
Differential cross sections for elastic electron
scattering on  $^{50,52,54}$Cr at $\varepsilon = 200$  
MeV and $^{48}$Ti at $\varepsilon = 250$  MeV
as  functions of $\theta$.
Experimental data from \cite{PhysRev.174.1380}
  ($^{48}$Ti) and
\cite{PhysRevC.27.113} ($^{50,52,54}$Cr).}	\vspace {5mm}\label{fig:cromo}
\end{figure}

Our DWBA results are compared with the experimental differential cross sections 
for elastic electron scattering on four calcium isotopes ($^{40,42,44,48}$Ca) 
at an electron energy $\varepsilon = 250$ MeV in Fig.~\ref{fig:calcium} and on  
$^{50,52,54}$Cr at $\varepsilon = 200$  MeV and $^{48}$Ti at 
$\varepsilon = 250$  MeV in Fig.~\ref{fig:cromo}. 
The DWBA calculations are able to reproduce the general trend of the data 
and give a good description of the experimental cross sections considered, 
except for small discrepancies 
at large scattering angles. 

 


Another interesting quantity that can be measured in elastic electron
scattering is the  parity-violating asymmetry, which is defined as
the difference between the cross sections for the elastic scattering of
electrons longitudinally polarized parallel  and antiparallel  
to their momentum. This difference  arises from the 
interference between photon and $Z^0$ exchange and represents an almost 
direct measurement of the Fourier transform of the neutron density 
\cite{Donnelly1989589,PhysRevC.61.064307}.
In fact, in Born approximation,  neglecting strangeness contributions and the 
electric neutron form factor, the parity-violating asymmetry can be expressed
as \cite{PhysRevC.63.025501,1742-6596-312-9-092044} 
\begin{eqnarray}\label{AP2}
A_{pv} = \dfrac{G_F\ q^2}{4 \sqrt{2}\ \pi \alpha} \left[
4 \sin^2 \Theta_W - 1 + 
\dfrac{F_n(q)}{F_p(q)} \right] \ ,
\end{eqnarray}
where $G_F\simeq 1.16639 \times 10 ^{-11}$ MeV$^{-2}$ is the Fermi constant and
$\sin^2\Theta_W\simeq 0.23$ is the Weinberg angle.
Since $4 \sin^2 \Theta_W - 1$ is small and the proton form factor 
$F_p(q)$ is known, we see that  
$A_{pv}$ provides a practical method to measure
the neutron form factor $F_n(q)$ and hence the neutron radius.
For these reasons parity-violating electron
scattering (PVES) has been suggested 
as a clean and powerful tool for
measuring the spatial distribution of neutrons in nuclei.

The first measurement of $A_{pv}$ in the elastic scattering of polarized 
electrons from $^{208}$Pb has been performed in Hall A at the Jefferson Lab 
(experiment PREX) \cite{Abrahamyan:2012gp,PhysRevC.85.032501}. Another 
experiment with improved electronics (PREX-II) \cite{prex2} has been recently
approved. In addition, the experiment CREX  \cite{crex}, with the goal of 
measuring the neutron skin of $^{48}$Ca, has also been conditionally approved.

Our DWBA results are in good agreement with the parameter $A_{pv}$  
measured by the first experiment PREX. The comparison can be found in 
\cite{esotici2}, where our numerical predictions for the experiment CREX 
are also given.

\section{Results for elastic electron scattering on the $N = 28, 20$ and $14$ isotonic chains }
\label{results}
In this Section we present our results for the evolution of 
some well-established observables in elastic electron 
scattering on the $N = 28, 20$ and $14$ isotonic chains. Many of
these nuclei lie in the region of the nuclear chart that is likely
to be explored in future electron-scattering experiments. 
We first consider $N= 28$, then we extend our analysis to $N= 20$ and $14$. 

\begin{figure}
	\centering \vskip 1mm
		\includegraphics[bb=0 20 717 539,scale=0.36]{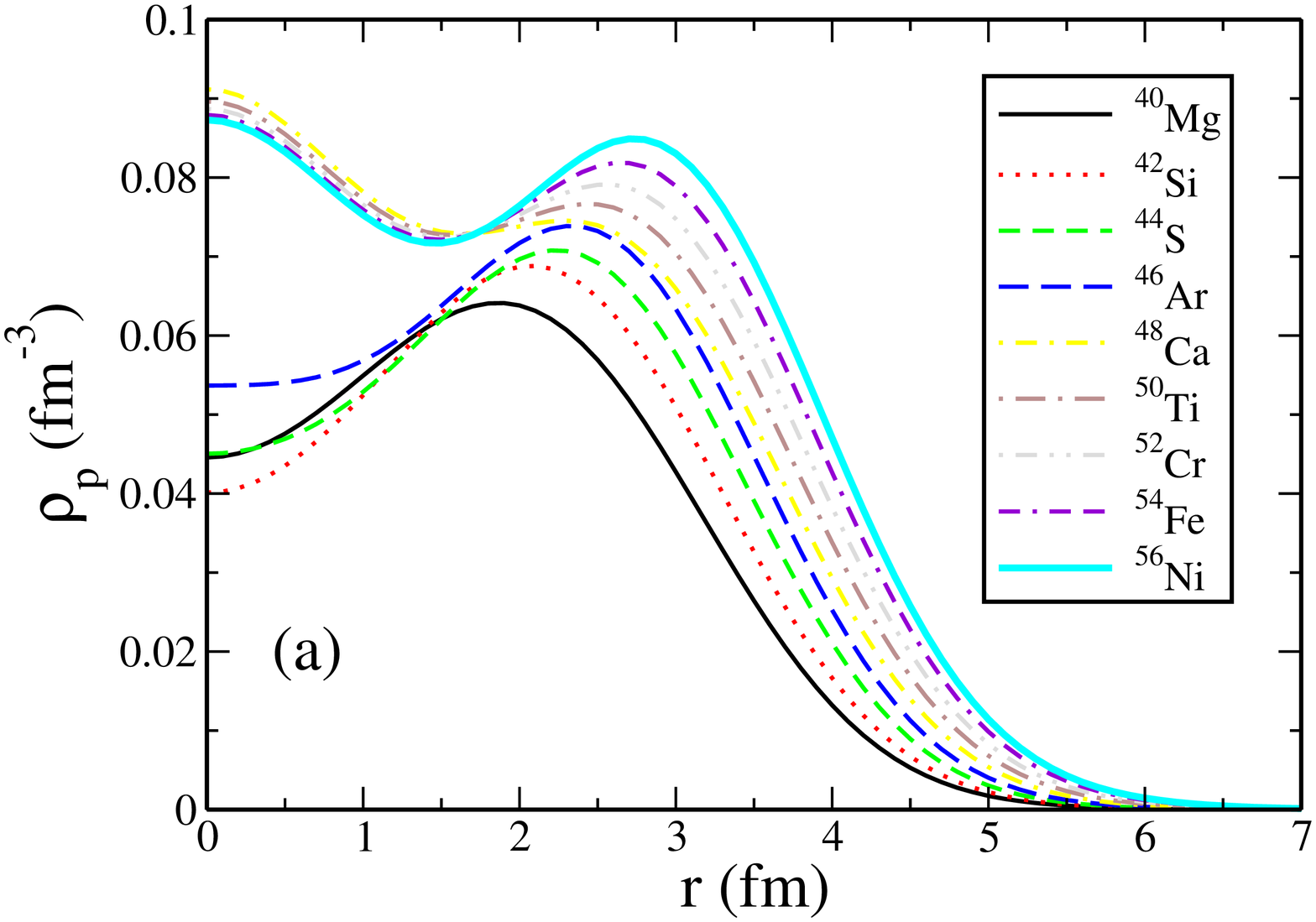}\vskip 7mm
		\includegraphics[bb=0 20 757 539,scale=0.34]{N28_sigma_850MeV.eps}\vskip 8mm
		\includegraphics[bb=0 20 757 539,scale=0.37]{N28_minima.eps}
\caption{ (Color online)
Panel (a): proton distributions  along the 
$N = 28$ isotonic chain. Panel (b): differential cross sections for elastic electron
scattering  at $\varepsilon = 850$  MeV as  functions of $\theta$.
Panel (c): evolution of the first minimum of the
differential cross section with the scattering angle $\theta$.
}	\vspace {5mm}\label{fig:n28-a}
\end{figure}
\begin{figure}
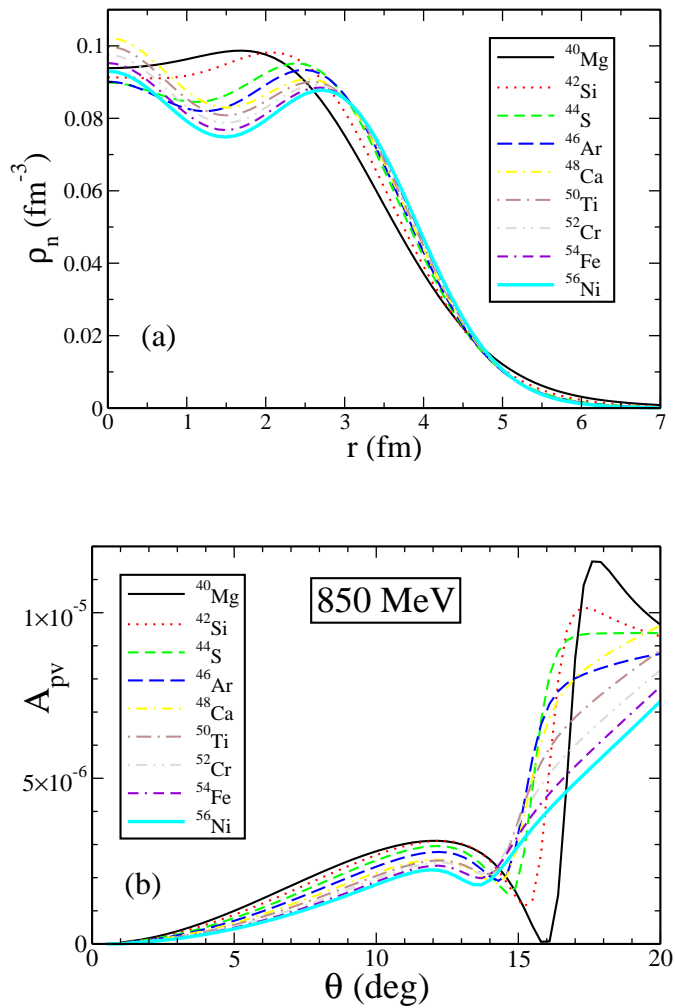

	\centering
		\vskip 2mm
		\includegraphics[bb=0 20 757 539,scale=0.35]{N28_rho_neu.eps}\vskip 7mm
		\includegraphics[bb=0 20 757 539,scale=0.35]{N28_apv_850MeV.eps}
\caption{ (Color online)
Panel (a): neutron distributions  along the 
$N = 28$ isotonic chain. Panel (b): parity-violating asymmetry parameters
$A_{pv}$ for elastic electron scattering at $\varepsilon = 850$ MeV as  functions of
the scattering angle $\theta$.
}	
\vspace {5mm}\label{fig:n28-b}
\end{figure}
\begin{figure}
	\centering
		\includegraphics[bb=0 20 717 539,scale=0.35]{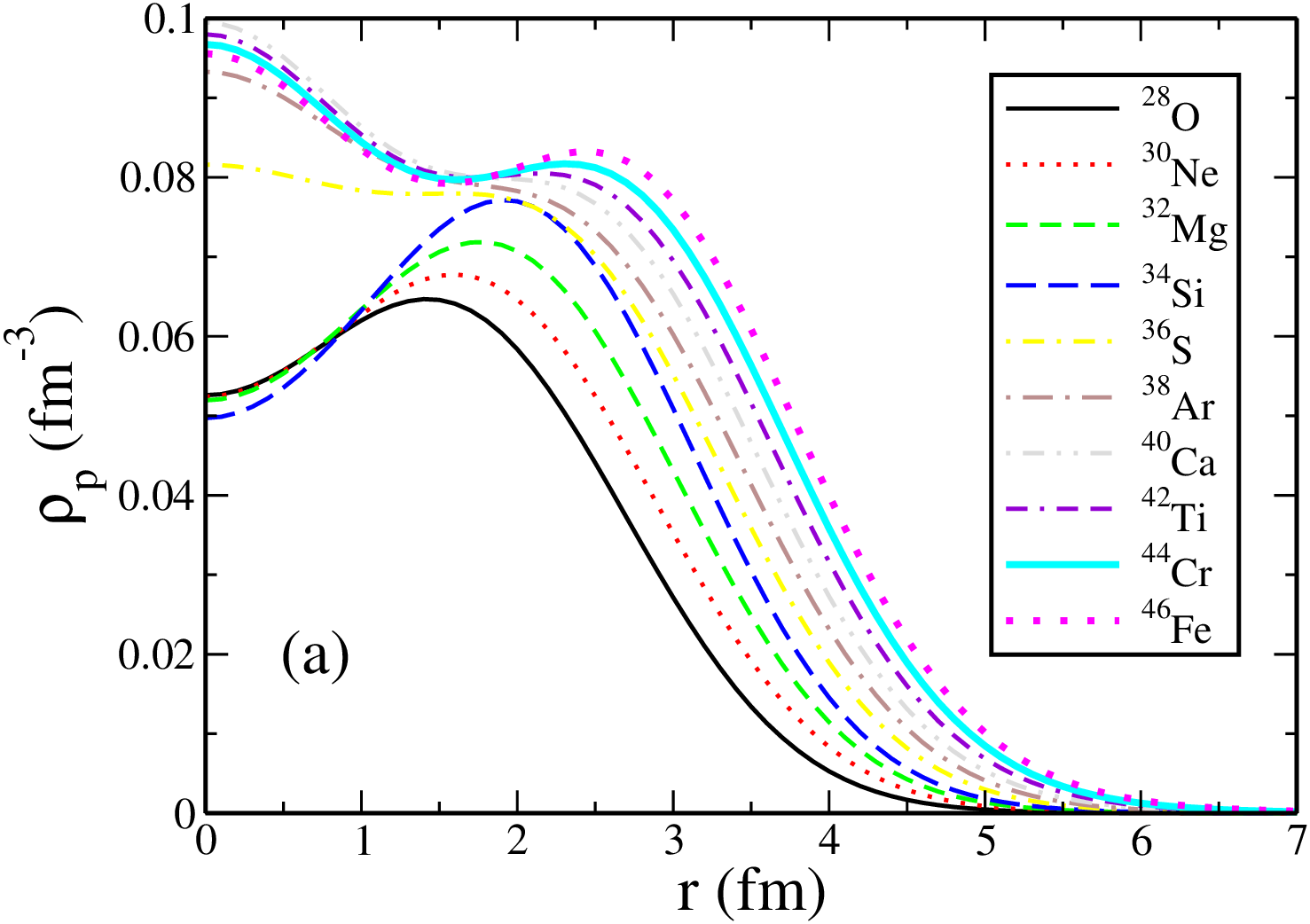}\vskip 7mm
		\includegraphics[bb=0 20 757 539,scale=0.34]{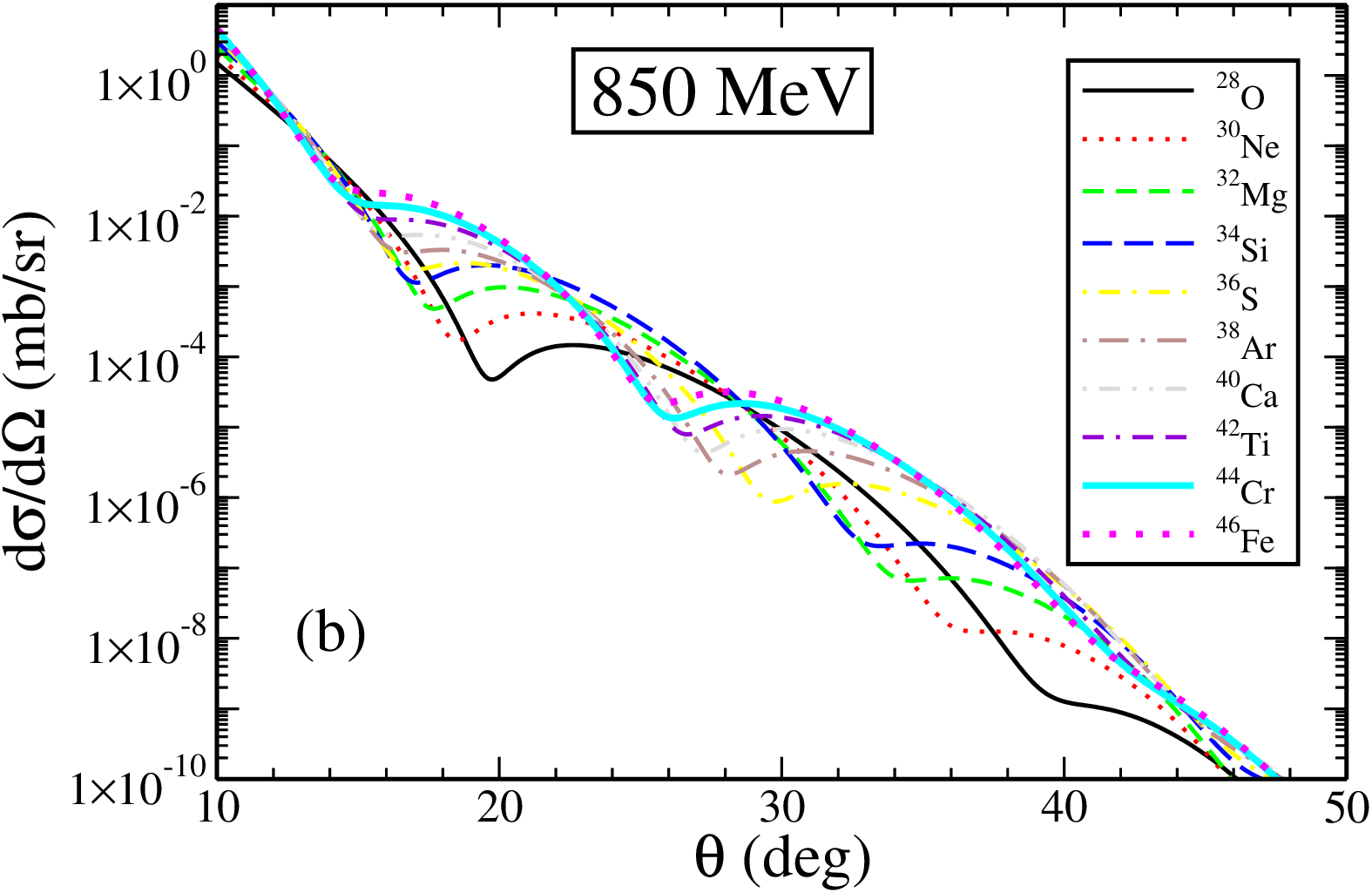}\vskip 8mm
		\includegraphics[bb=0 20 757 539,scale=0.35]{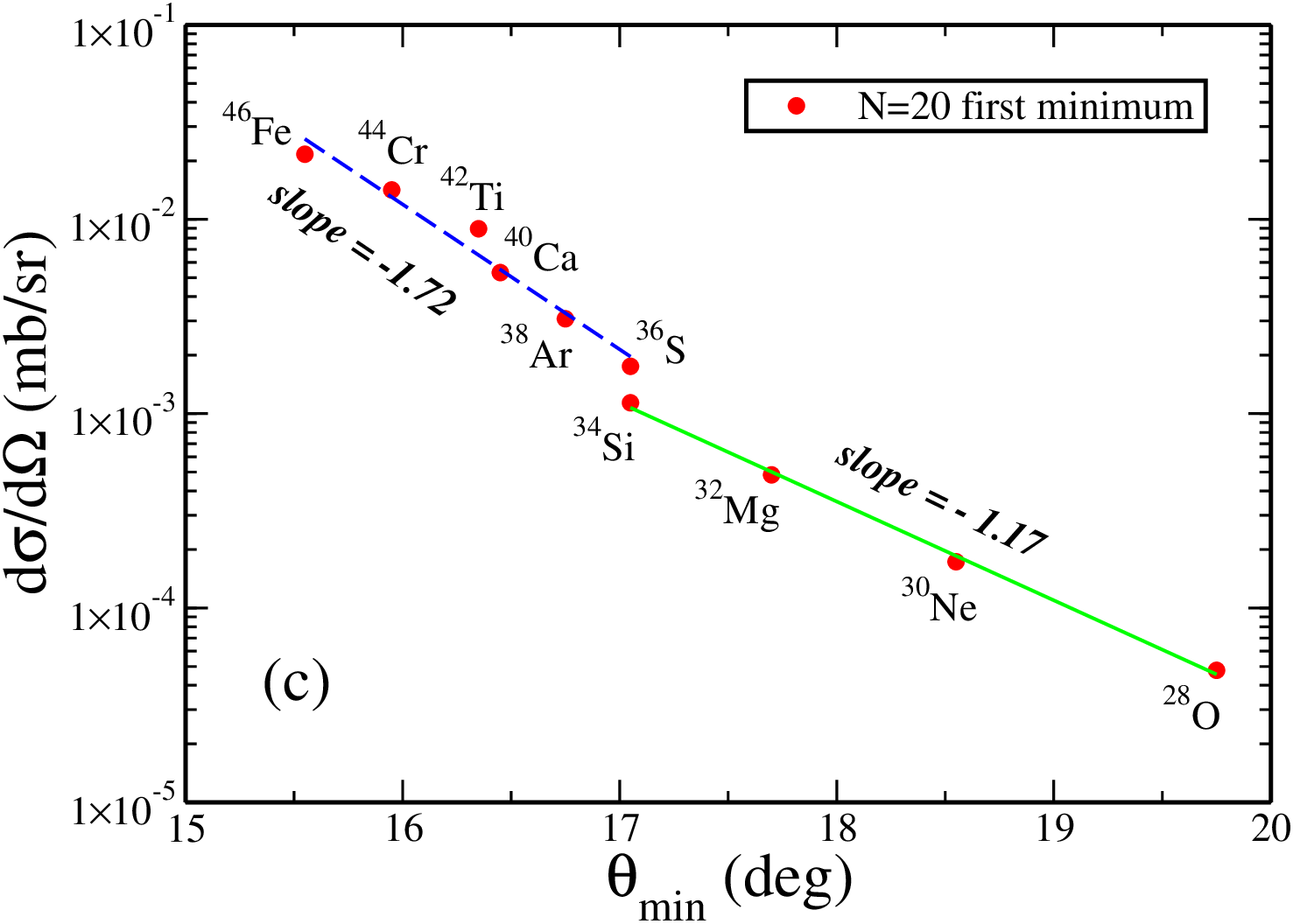}
\caption{  (Color online) The same as in Fig. \ref{fig:n28-a} but for the
$N = 20$ isotonic chain.
}	\vspace {5mm}\label{fig:n20-a}
\end{figure}
\begin{figure}
	\centering
		\includegraphics[bb=0 20 757 539,scale=0.35]{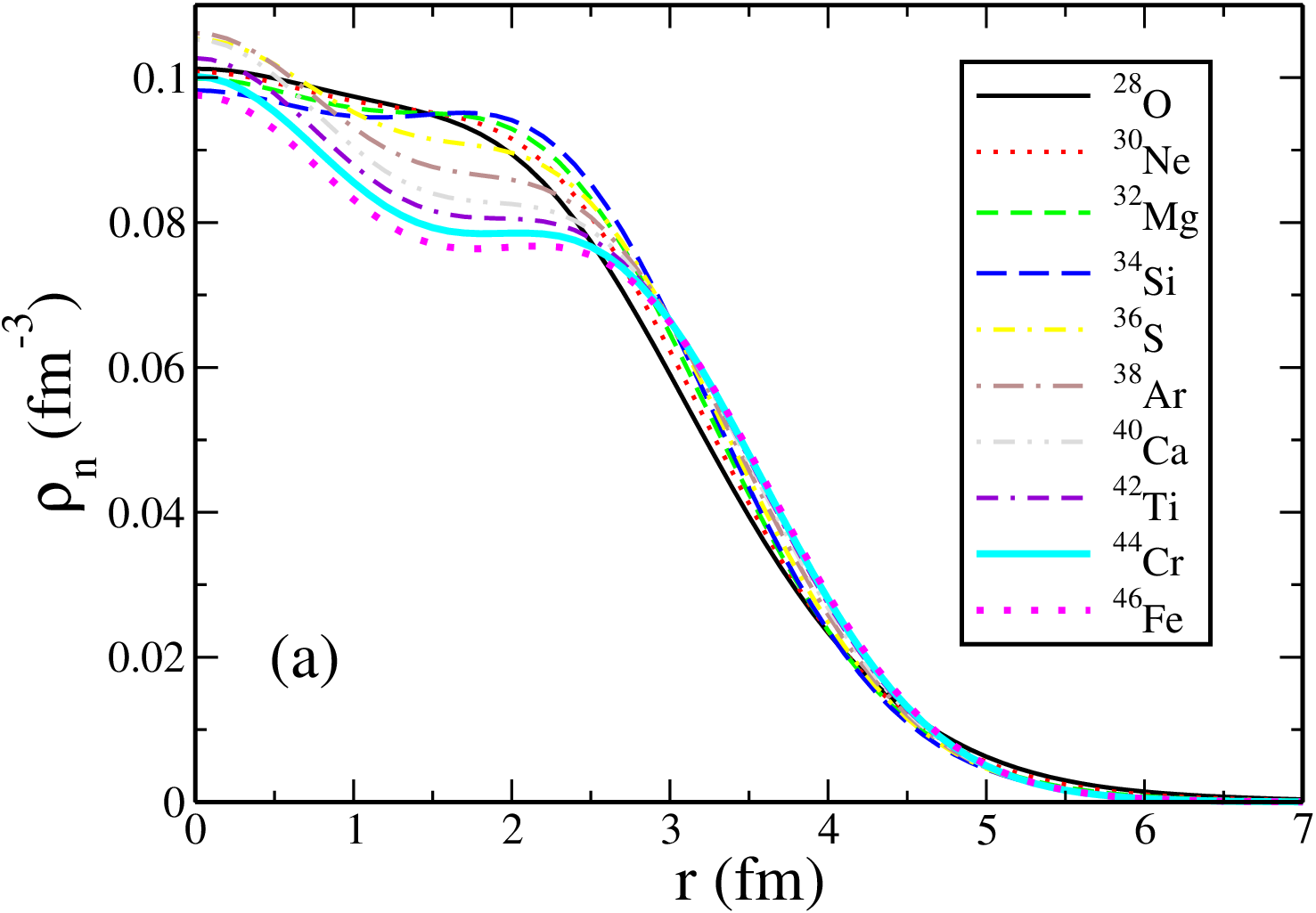}\vskip 7mm
		\includegraphics[bb=0 20 757 539,scale=0.35]{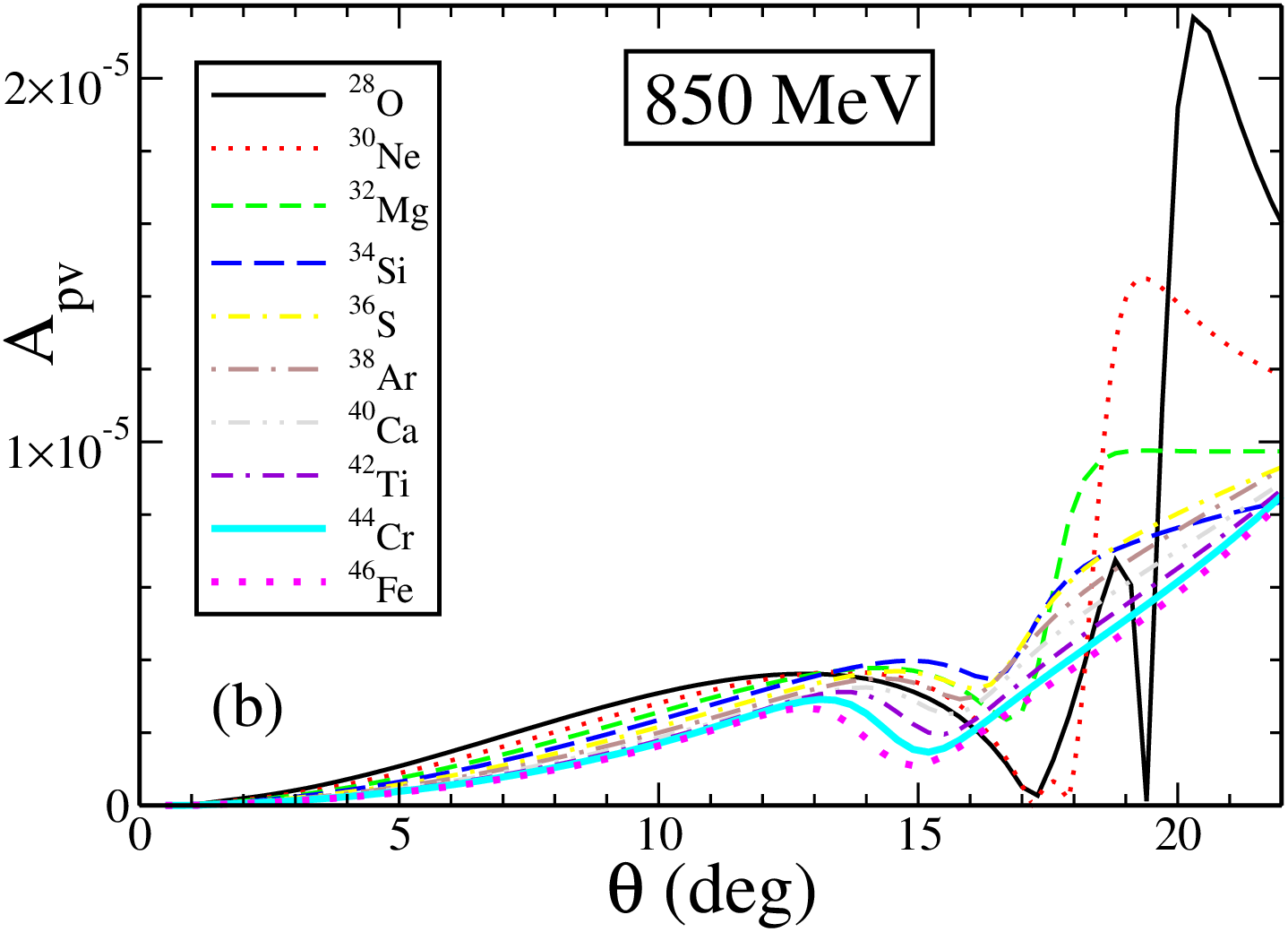}
\caption{ (Color online)
The same as in Fig. \ref{fig:n28-b} but for the
$N = 20$ isotonic chain.}	\vspace {5mm}\label{fig:n20-b}
\end{figure}
\begin{figure}
	\centering
		\includegraphics[bb=0 20 717 539,scale=0.35]{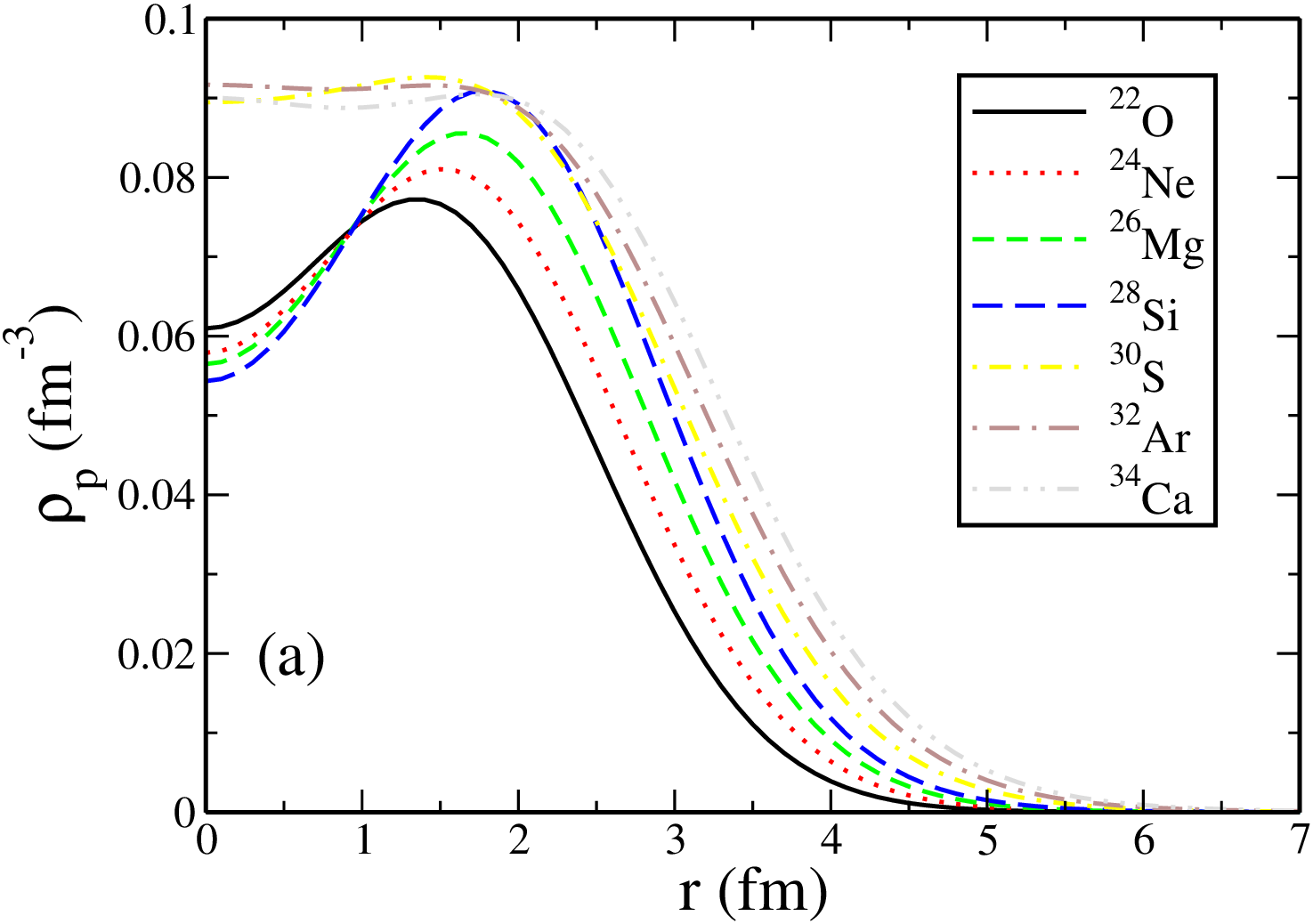}\vskip 7mm
		\includegraphics[bb=0 20 757 539,scale=0.34]{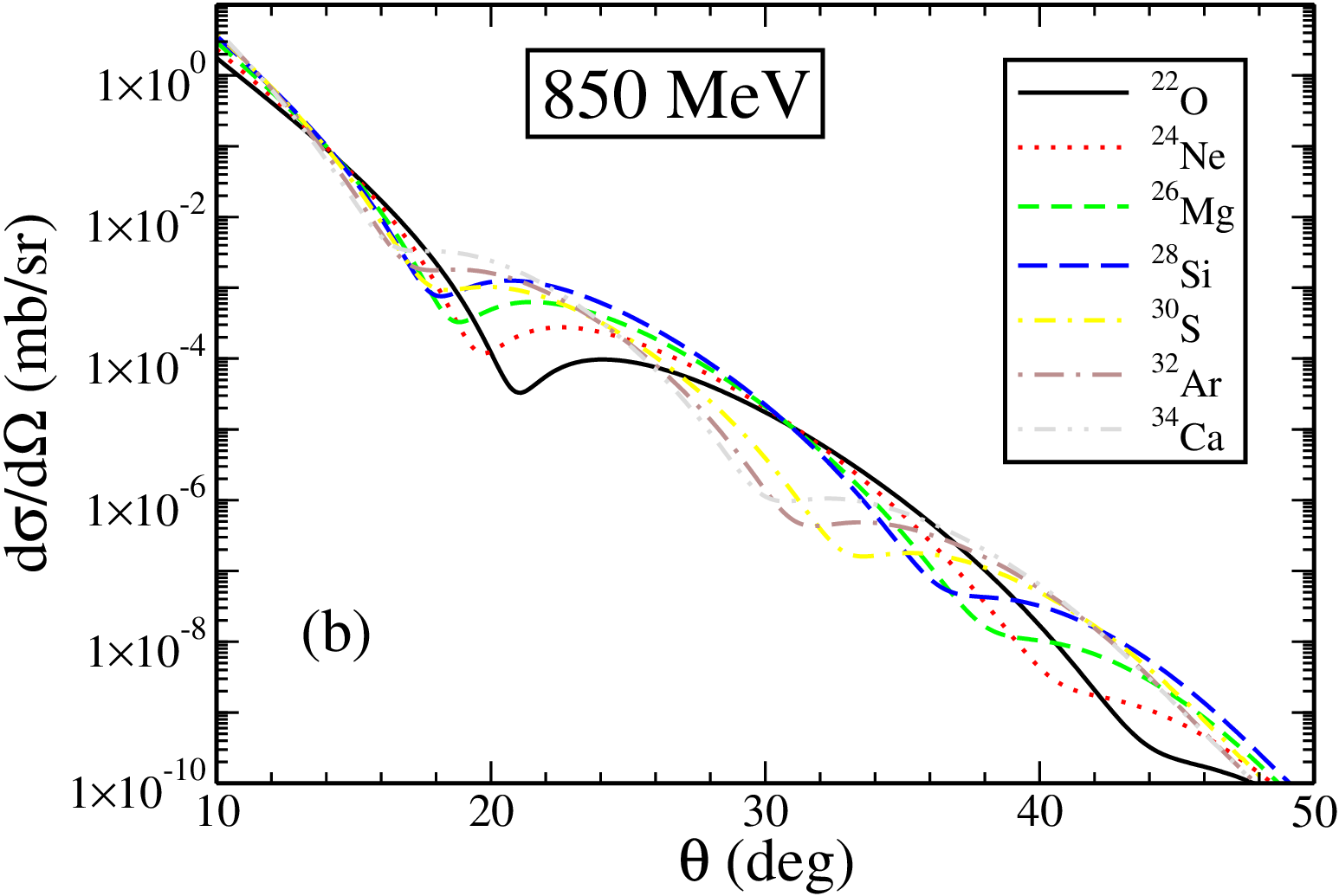}\vskip 8mm
		\includegraphics[bb=0 20 757 539,scale=0.35]{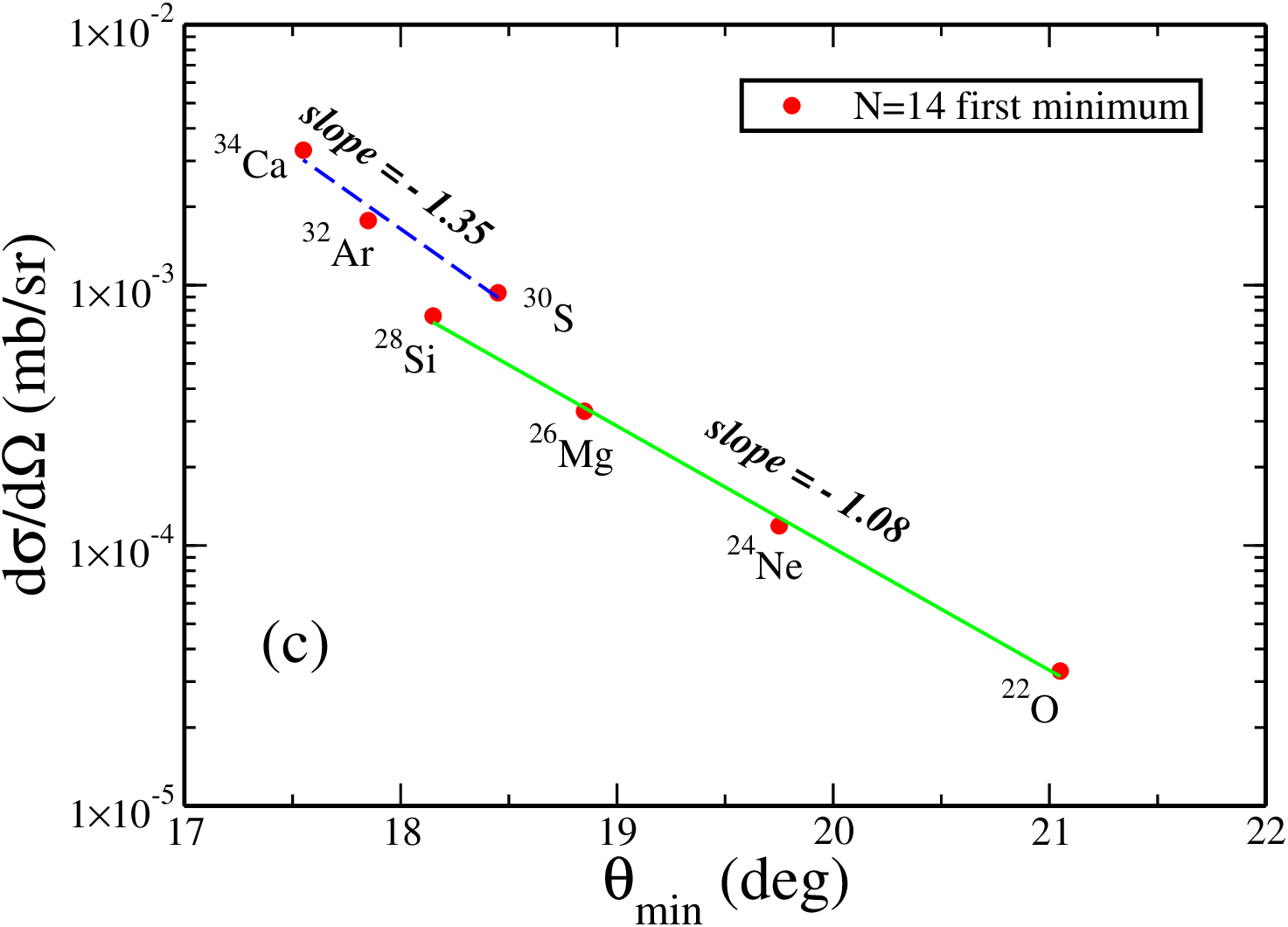}
\caption{ (Color online) The same as in Fig. \ref{fig:n28-a} but for the
$N = 14$ isotonic chain.
}	\vspace {5mm}\label{fig:n14-a}
\end{figure}
\begin{figure}
	\centering
		\includegraphics[bb=0 20 757 539,scale=0.35]{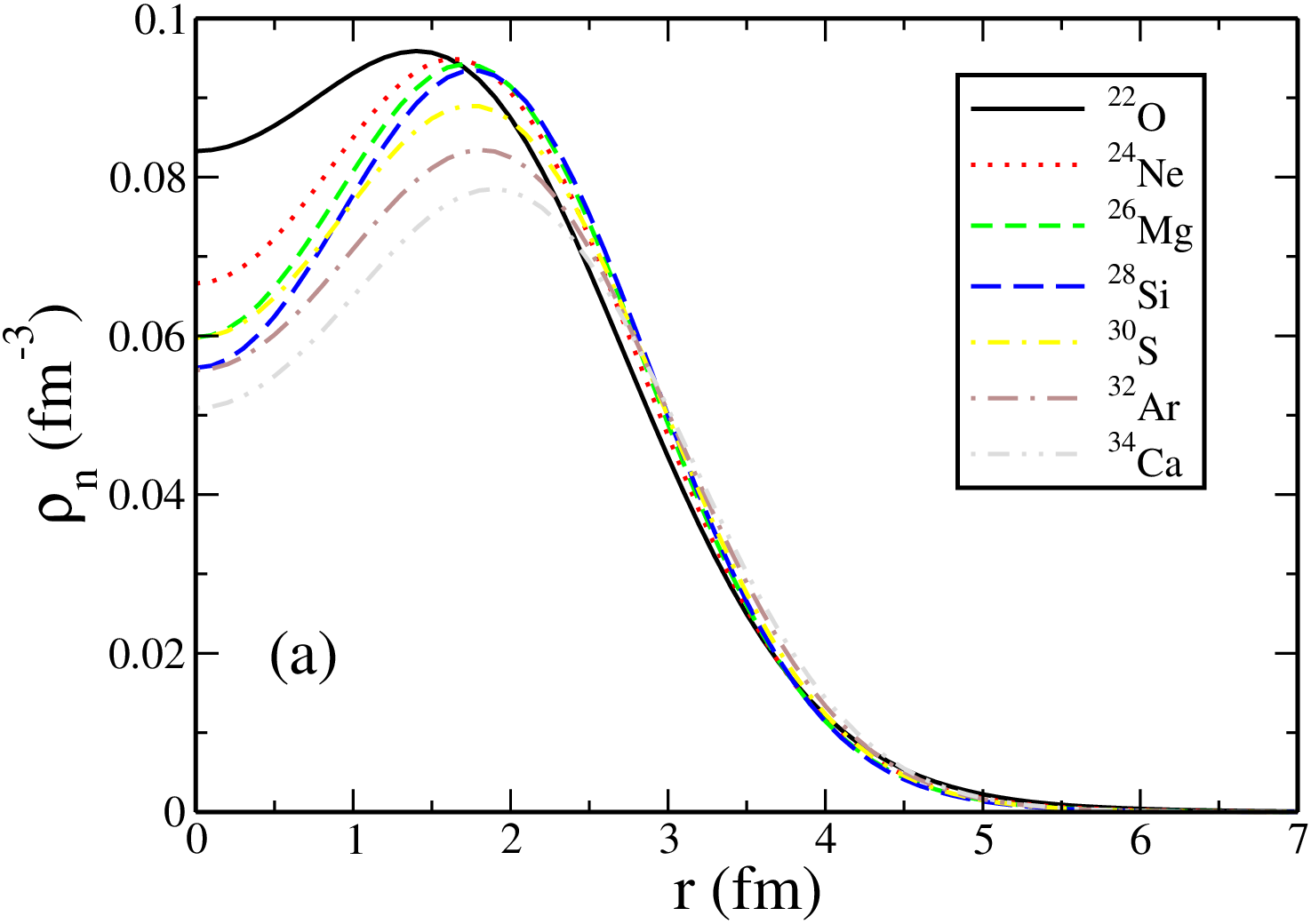}\vskip 7mm
		\includegraphics[bb=0 20 757 539,scale=0.35]{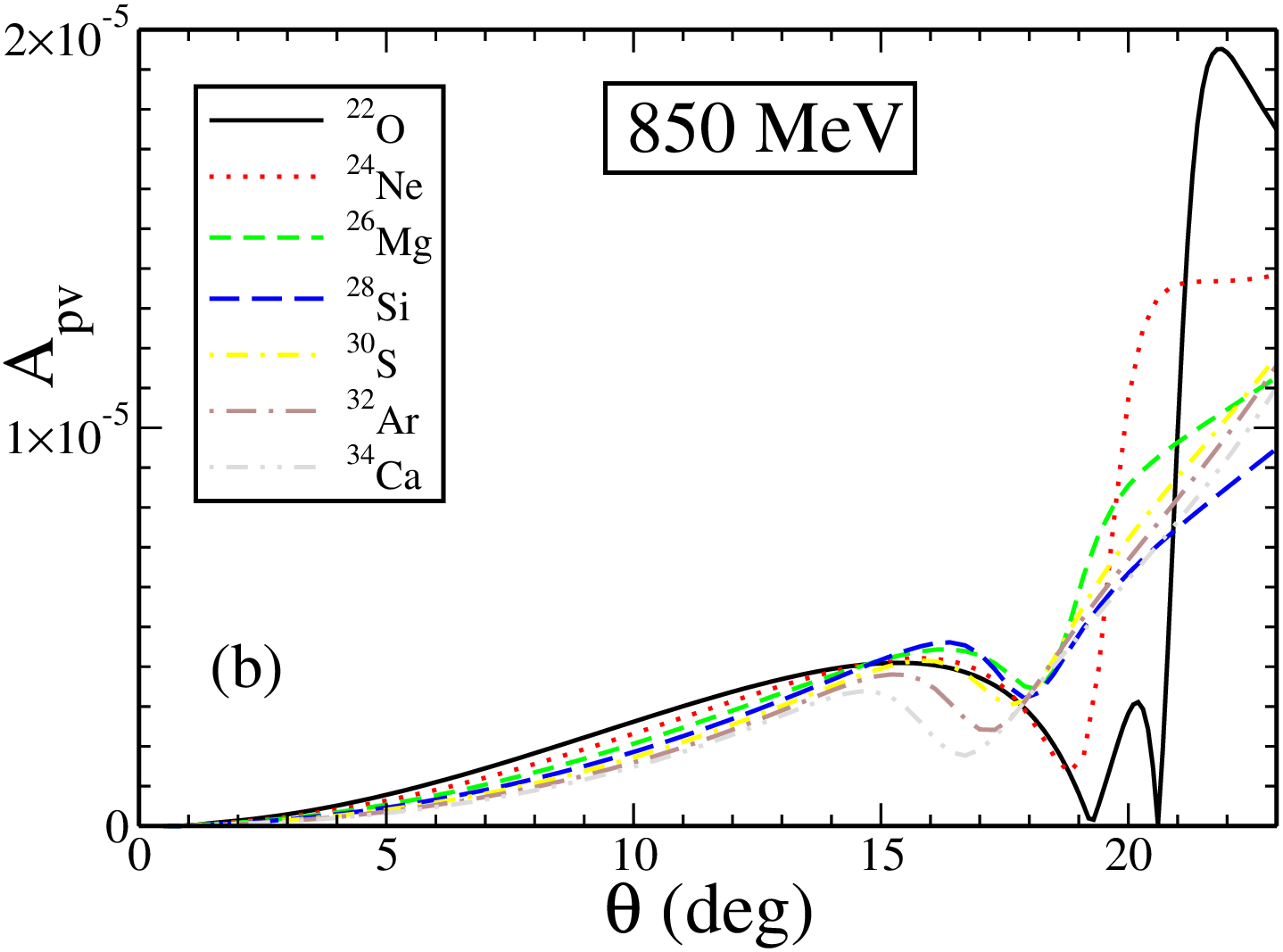}
\caption{ (Color online)
The same as in Fig. \ref{fig:n28-b} but for the
$N = 14$ isotonic chain.}	\vspace {5mm}\label{fig:n14-b}
\end{figure}

\subsection{Results for $N = 28$  }

In panel (a) of Fig. \ref{fig:n28-a} we plot the proton density
distributions $\rho_p$ as functions of the radial coordinate $r$ along the 
$N= 28$ isotonic chain. These density distributions are obtained summing the squared 
moduli of the s.p. wave-functions described in Section \ref{sec.rmf}. 
All the nuclei that we consider result to be bound but, experimentally, 
there are  proton-deficient
nuclei (from $^{40}$Mg to $^{46}$Ar), stable nuclei
(from $^{48}$Ca to $^{54}$Fe) and one proton-rich nucleus ($^{56}$Ni).
The most significant effect of adding  protons is to 
populate and extend the proton densities. 

The differences of the proton density profiles in the nuclear
interior display pronounced shell effects; the most relevant proton s.p. levels in our
analysis of  the $N = 28$ chain are the $2$s$_{1/2}$ and the $1$f$_{7/2}$.  
In the case of the heavier nuclei, starting from $^{48}$Ca up to $^{56}$Ni, 
we obtain similar results for $\rho_p$ in the central region whereas the differences 
at large $r$ can be ascribed to  the filling of the $1$f$_{7/2}$ shell which starts
from $^{50}$Ti. 

The proton density of the proton-deficient  $^{46}$Ar nucleus  also gets
a non negligible contribution from the $2$s$_{1/2}$ shell, which has an occupation
number of $0.330$,  and, as a consequence, in the central region 
it is approximately $20\%$ larger than those of lighter isotones. 
In our model the  $^{42}$Si nucleus, with $14$ protons, behaves like a magical nucleus and
its proton density does not get any contribution from neither the $2$s$_{1/2}$ nor
the $1$d$_{3/2}$ shells which, on the contrary, contribute to $^{40}$Mg and to 
$^{44}$S densities which are, therefore,  larger in the central region. 
{However, the experimental evidence of a $2^+_1$ state at $ 770 \pm 19$ keV 
\cite{PhysRevLett.99.022503}, much smaller than for other nuclei in this chain, has been
interpreted as a signal of the disappearance of the $N = 28$
shell closure around $^{42}$Si and has suggested that
proton-core excitations  and  the tensor interactions cannot be neglected \cite{sorlin,Sorlin2008602}.}

In panel (b) of Fig. \ref{fig:n28-a} we present the differential cross sections for
elastic electron scattering  at $\varepsilon = 850$ MeV
as functions of the scattering angle $\theta$. 
These cross sections have been calculated in the DWBA 
and with the self-consistent relativistic ground-state charge densities. 
Although Coulomb distortion is included in the calculations, the elastic cross 
sections are still related to the behavior of the corresponding proton charge 
density as in Eq. \ref{eq.csel}.

With increasing proton number along the chain the positions of the diffraction minima
usually shift toward smaller scattering angles, i.e., smaller
values of the momentum transfer $q$. The shift of the minima
is also accompanied by a simultaneous
increase in the height of the corresponding maxima of the cross sections.
The differential cross section 
provides  information on the proton density distribution and we can 
look for possible correlations 
between the cross sections that can be directly related to the behavior 
of the proton distributions.
In panel (c) of Fig. \ref{fig:n28-a} we plot the evolution of the 
position of the first minimum of the  elastic cross sections for each 
isotone in the chain. 
There is an evident transition between proton-poor and proton-rich
isotones and it is not possible to fit the positions of the first minima 
with a straight line. In addition, the minimum for $^{48}$Ca occurs at a larger
scattering angle than for $^{46}$Ar. However, it is still possible to draw a straight line
that connects the minima for the lighter isotones up to $^{46}$Ar and another line
for the heavier isotones. The slope of the two lines are different.
The isotones from $^{48}$Ca up to $^{56}$Ni have
similar proton densities  at small $r$ and present only small  differences 
at large $r$, which are ascribed to  the filling of the $1$f$_{7/2}$ shell.
In the case of the lighter isotones, we see that the $^{46}$Ar and $^{44}$S minima 
do not lie on the fitting lines. This is  related to the
non-negligible contribution of the  proton 
in the $2$s$_{1/2}$ shell, which has occupation number  $0.330$ for $^{46}$Ar 
and $0.153$ for $^{44}$S.
The densitiy of $^{40}$Mg gets some small contribution from the $2$s$_{1/2}$ and
the $1$d$_{3/2}$ shells but the  minimum of its cross section is aligned with that of
 $^{42}$Si.

The evolution of the neutron density distribution along an isotonic chain is less significant
than that of the proton density distribution; generally, there is a decrease of the density in the
nuclear interior and an extension toward large $r$ to preserve the 
normalization to the constant
number of neutrons, as it can be seen in panel (a) of  Fig. \ref{fig:n28-b}.
The cross sections for positive and negative helicity electron states at 
$\varepsilon = 850$ MeV
have also been calculated and the resulting parity-violating 
asymmetry parameter $A_{pv}$ is presented
in panel (b) of Fig. \ref{fig:n28-b} as a function of $\theta$. 
In the case of oxygen and calcium isotopic chains we 
found that the evolution of the positions of the
first minima as  functions of the neutron excess is approximated by 
a linear fit \cite{esotici2}. In the case of the $N = 28$ isotones the position of the
first minima of $A_{pv}$  usually evolves toward smaller $\theta$ as the number of 
protons increases. 
However, $^{46}$Ar and $^{48}$Ca, as well as  $^{50}$Ti and $^{52}$Cr, seem to
have almost coincident minima. 
The value of $A_{pv}$ at the minimum increases from $^{40}$Mg  to $^{46}$Ar, then
it decreases monotonically starting from $^{52}$Cr.

\subsection{Results for $N = 20$  }

In panel (a) of Fig. \ref{fig:n20-a} we present our results for the proton density
distributions  as functions of $r$ along
the $N = 20$ isotonic chain. 
In our model all these nuclei result to be bound but, along this chain, 
there are  proton-deficient
nuclei (from $^{28}$O to $^{34}$Si), stable nuclei
($^{36}$S, $^{38}$Ar, and $^{40}$Ca) and proton-rich nuclei ($^{42}$Ti, $^{44}$Cr,
and $^{46}$Fe). The densities of the proton-rich isotones are significantly extended toward
larger $r$ with respect to those of the proton-deficient ones.
Also in this chain pronounced shell effects are visible in the nuclear interior. 
In particular, we find 
large differences in the radial profiles of  the proton density of  $^{34}$Si and 
$^{36}$S; in our model the proton occupation number
of the $2$s$_{1/2}$ shell of $^{36}$S is not negligible and, owing to the observation 
that the squared wave function of the $2$s$_{1/2}$ state has a main peak at the center,
the proton central density is enlarged. We obtain that also 
the proton occupation number
of the $1$d$_{3/2}$ state for $^{36}$S  is significant;  
the corresponding peak of the squared wave function is away from the center 
and, therefore, the filling of the $1$d$_{3/2}$ state by
protons  increases the  density away from the center.
In the case of the heavier nuclei, starting from $^{38}$Ar up to $^{46}$Fe, we obtain
similar results for $\rho_p$ in the central region, whereas the differences 
at large $r$ can be ascribed to  the addition of  protons that
populate and extend the  densities.

In panel (b) of Fig. \ref{fig:n20-a} we present the DWBA differential cross 
sections for elastic electron scattering  at $\varepsilon = 850$ MeV
as functions of the scattering angle $\theta$. 
With increasing proton number along the chain the positions of the diffraction minima
shift toward smaller scattering angles and, correspondingly, there is an
increase in the height of the corresponding maxima of the cross sections.
In panel (c) of Fig. \ref{fig:n20-a} we plot the evolution of the 
position of the first minimum of the differential elastic cross section for each 
isotone in the chain.  
Owing to the large differences in the radial profiles of the densities of  
$^{34}$Si and $^{36}$S, it is not possible to fit the position of the first minima 
with a straight line. In addition, the minimum for $^{34}$Si occurs at the same
scattering angle as for $^{36}$S. Also in this chain 
we can draw a straight line that connects the minima for the lighter isotones up to $^{34}$Si and another line
for the heavier isotones. The slope of the two lines is different.

Recently, there has been a significant interest in looking for \lq\lq bubble\rq\rq\ nuclei,
i.e., nuclear systems where the central density is significantly reduced or
vanishing  with respect  to the saturation density.
Recent theoretical calculations, including RMF 
parametrizations \cite{todd,chu,Wang:2013rwg} and 
Skyrme energy density functional \cite{Khan:2007ji,PhysRevC.79.034318,wang}, 
predict a central depletion of the charge density distributions for $^{34}$Si and for
some Ar isotopes but, usually,  $^{34}$Si is  the only candidate for which 
the studies agree. However, a nuclear \lq\lq bubble\rq\rq\ has never been experimentally observed. 
In \cite{yao} 
the existence of a  proton \lq\lq bubble\rq\rq\ structure in the low-lying
excited $0^+_2$ and $2^+_1$ states of $^{34}$Si is found to be very unlikely, 
owing to the results of a new RMF method which includes mixing of both 
particle-number and angular-momentum projected quadrupole deformed states.
In panel (a) of Fig. \ref{fig:n20-a}  
large differences can be seen between  the proton density 
of  $^{34}$Si and $^{36}$S. This result agrees with similar findings of other 
RMF models \cite{todd,chu}. In particular, as it is shown in panel (c) of 
Fig. \ref{fig:n20-a}, the positions of the first minima of the elastic cross 
sections of $^{34}$Si and $^{36}$S do not fall on the same straight line 
and this should provide a useful test that could be investigated at future 
electron scattering facilities. 

In panel (a) of  Fig. \ref{fig:n20-b} we present the 
evolution of the neutron density distribution along the $N= 20$ isotonic chain.
For the heavier isotones we observe a decrease of the density in the
nuclear interior and a corresponding enhancement away from the center
to preserve the normalization to the constant
number of neutrons.   
In panel (b) of Fig. \ref{fig:n20-b} we present the parity-violating 
asymmetry parameter $A_{pv}$  as a function of $\theta$ evaluated 
at $\varepsilon = 850$ MeV. 
The neutron densities of $^{28}$O and $^{30}$Ne are very similar
and their corresponding $A_{pv}$ are almost coincident up to the first minimum.
Starting from $^{32}$Mg the minimum of $A_{pv}$ shifts toward smaller angles.

\subsection{Results for $N = 14$  }

In panel (a) of Fig. \ref{fig:n14-a} we present our results for the proton 
density distributions  as functions of $r$ along
the $N = 14$ isotonic chain. 
The densities in the nuclear interior of proton-deficient isotones up to the 
stable nucleus $^{28}$Si are very similar. For all these nuclei only the two protons in 
the $1$s$_{1/2}$ state contribute to
 the density at the center. Starting from $^{30}$S also the protons in the $2$s$_{1/2}$ state
 contribute and there is an evident transition in the density profile.
 As usual, the densities of the proton-rich isotones are significantly extended
 towards
larger $r$ with respect to those of the proton-deficient ones.

The DWBA differential cross sections for
elastic electron scattering  at $\varepsilon = 850$ MeV
as functions of $\theta$ are displayed in panel (b) of Fig. \ref{fig:n14-a}. 
The positions of the diffraction minima
shift toward smaller scattering angles as the proton number increases 
and, in addition, there is an
enhancement in the height of the corresponding maxima of the cross sections.
The evolution of the 
position of the first minimum of the differential cross section for each 
isotone in the chain, which is shown in panel (c) of Fig. \ref{fig:n14-a}, can 
adequately be described by the two different straight lines drawn in the figure,
one connecting the minima of the lighter isotones up to $^{28}$Si and the other
one connecting the minima of the heavier isotones.

In panel (a) of  Fig. \ref{fig:n14-b} we present the 
evolution of the neutron density distribution along the $N= 14$ isotonic chain.
The profiles of proton-deficient and proton-rich isotones are very different, 
in particular in the nuclear interior, but the normalization to the constant
number of neutrons is always preserved.   
In panel (b) of Fig. \ref{fig:n20-b} the parity-violating 
asymmetry parameter $A_{pv}$ is presented as a function of $\theta$.
The neutron densities of $^{22}$O and $^{24}$Ne are very different from those 
of the other isotones in the chain and, as a consequence,
their corresponding $A_{pv}$ are very different. The position of the first
 minimum shifts toward smaller angles and the 
 value of $A_{pv}$ at the minimum increases from $^{22}$O  to $^{26}$Mg, then
it decreases starting from $^{28}$Si.

\section{Inclusive quasi-elastic electron scattering}
\label{sec.qe}

The inclusive differential cross 
section for the quasi-elastic (QE) $(e,e^{\prime})$ scattering on a nucleus 
is obtained from the contraction between the lepton and hadron tensors as  \cite{book} 
\inieq
\left(\frac{\diff \sigma}{\diff \varepsilon^{\prime} \diff
\Omega^{\prime}}\right)_{{QE}} =
\sigma_{M} 
\left[ v_L R_L + v_{T} R_{T}\right] 
\ , \label{eq.csqe}
\fineq
where  $\varepsilon^{\prime} $ is the energy of the scattered electron and 
the coefficients $v$ come from the components of the lepton tensor that, under
the assumption of the plane-wave approximation for the 
electron wave functions, depend only on the lepton kinematics  \cite{book}.
All relevant nuclear structure information is contained in the longitudinal and 
transverse response functions $R_L$ and $R_T$. The response functions can be 
expressed in terms of suitable linear combinations of the components of the 
hadron tensor, which are given by products of the matrix elements of the 
nuclear current between initial and final nuclear states. 

In the  QE region the nuclear response is dominated by one-nucleon 
processes where the scattering occurs with only one nucleon, which is
subsequently emitted  from the nucleus by a direct knockout mechanism, and 
the remaining nucleons behave as spectators. Therefore, QE electron scattering can
adequately be described in the relativistic impulse approximation (RIA) by the
sum of incoherent processes involving only one nucleon scattering and the
components of the hadron tensor are obtained from the sum, over all the s.p. 
shell-model states, of the squared absolute value of the transition matrix 
elements of the single-nucleon current  \cite{book}. 

A reliable description of final-state interactions (FSI) between the emitted
nucleon and the residual nucleus is a crucial ingredient for the 
description of $(e,e^{\prime})$ data. 
The relevance of final state interactions (FSI) has been clearly stated for the exclusive 
\eep reaction, where the use of a
complex optical potential (OP) in the distorted-wave impulse approximation (DWIA) 
is required \cite{Boffi:1993gs,book,Udias:1993xy,Meucci:2001qc,
Meucci:2001ja,Meucci:2001ty,Radici:2003zz,Giusti:2011it}. The imaginary part 
of the OP produces an absorption that reduces
the cross section and accounts for the loss of part of the incident flux 
in the elastically scattered beam to 
the inelastic channels which are open.
In the inclusive scattering, where only the final lepton is detected and the 
final nuclear state is not determined, all elastic and inelastic channels 
contribute, all final-state channels should be retained and the flux, although 
redistributed among all possible channels, must be conserved.

\begin{figure}
	\centering
\includegraphics[scale=0.38]{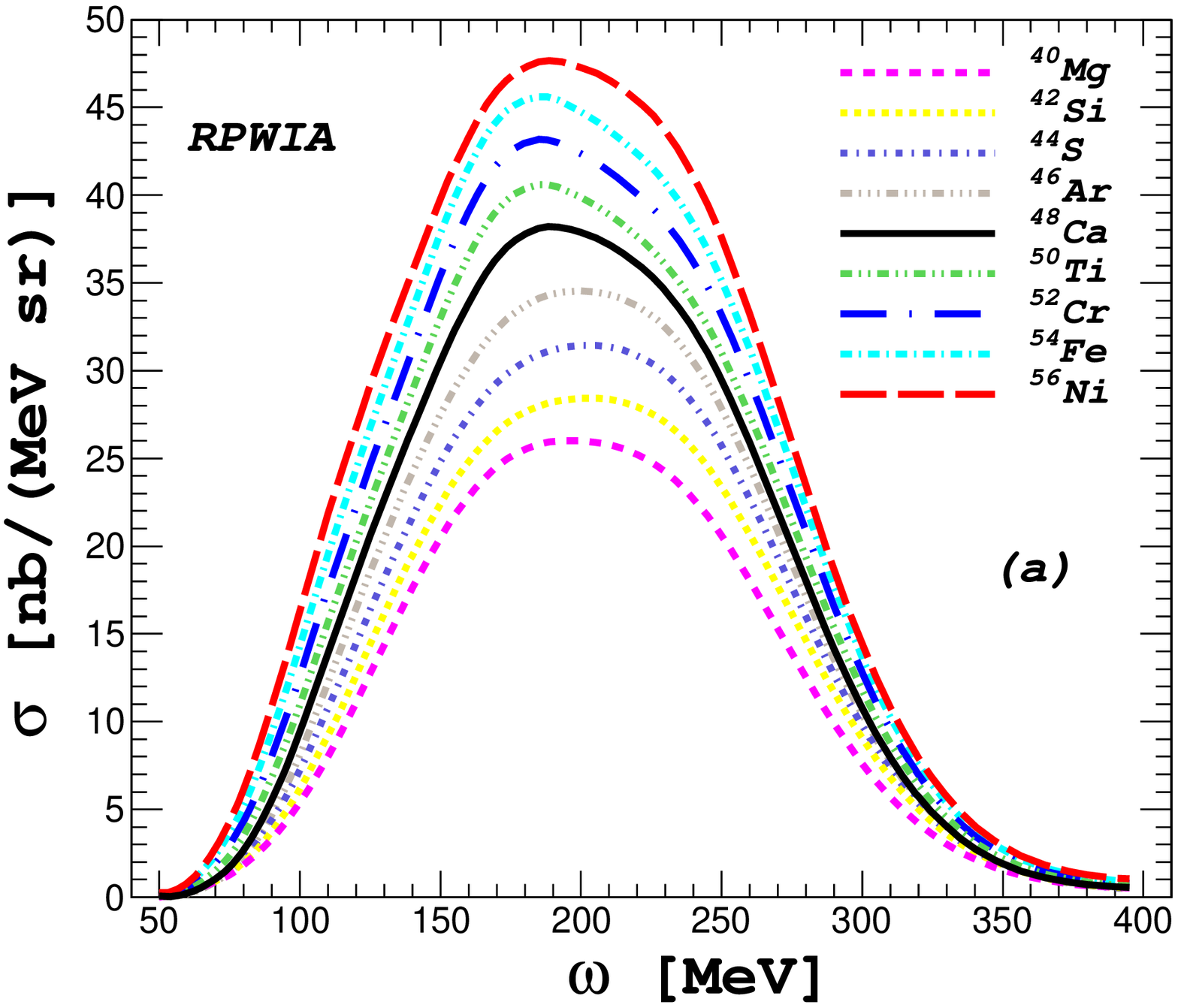} 
\includegraphics[scale=0.38]{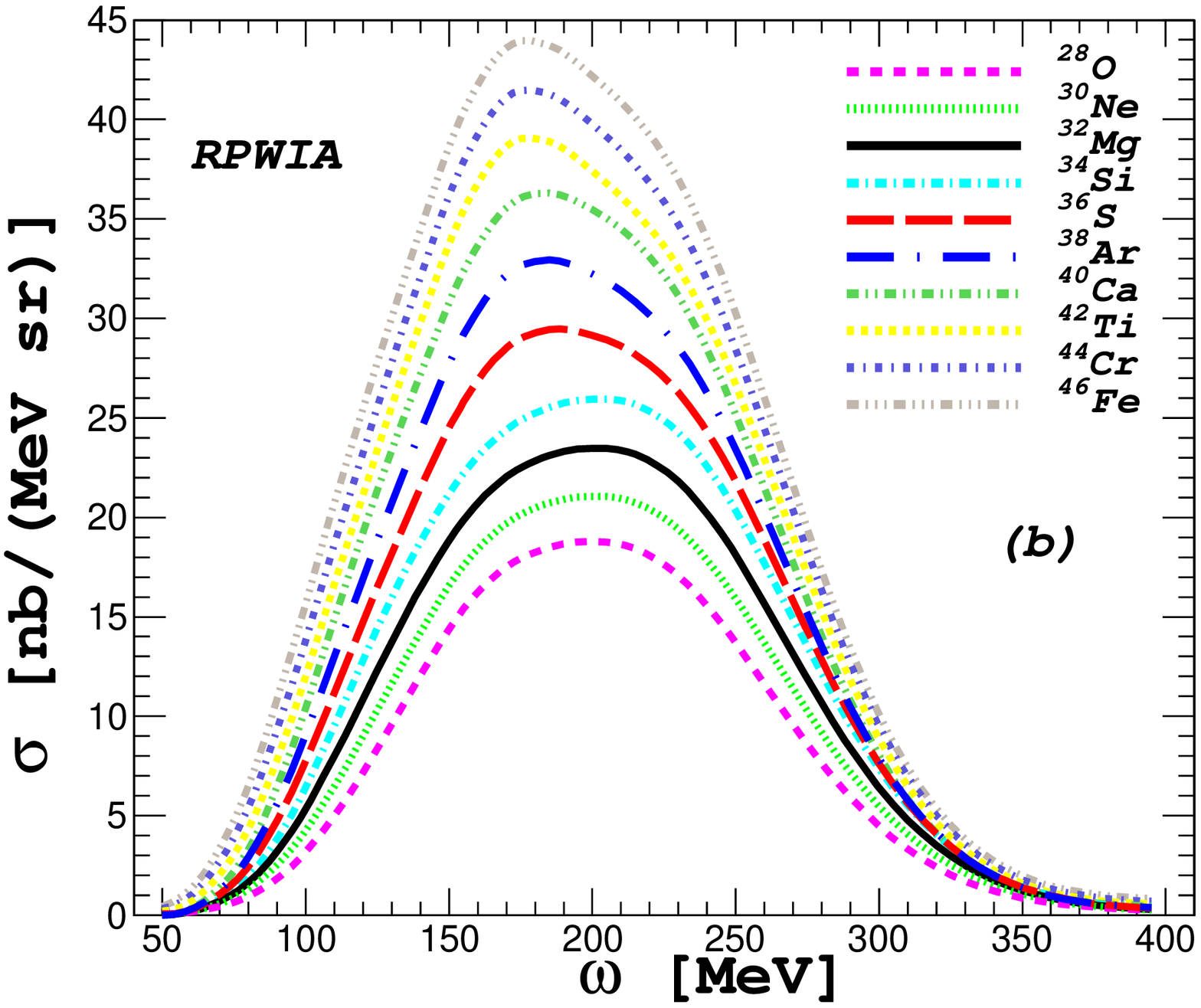} 
\includegraphics[scale=0.38]{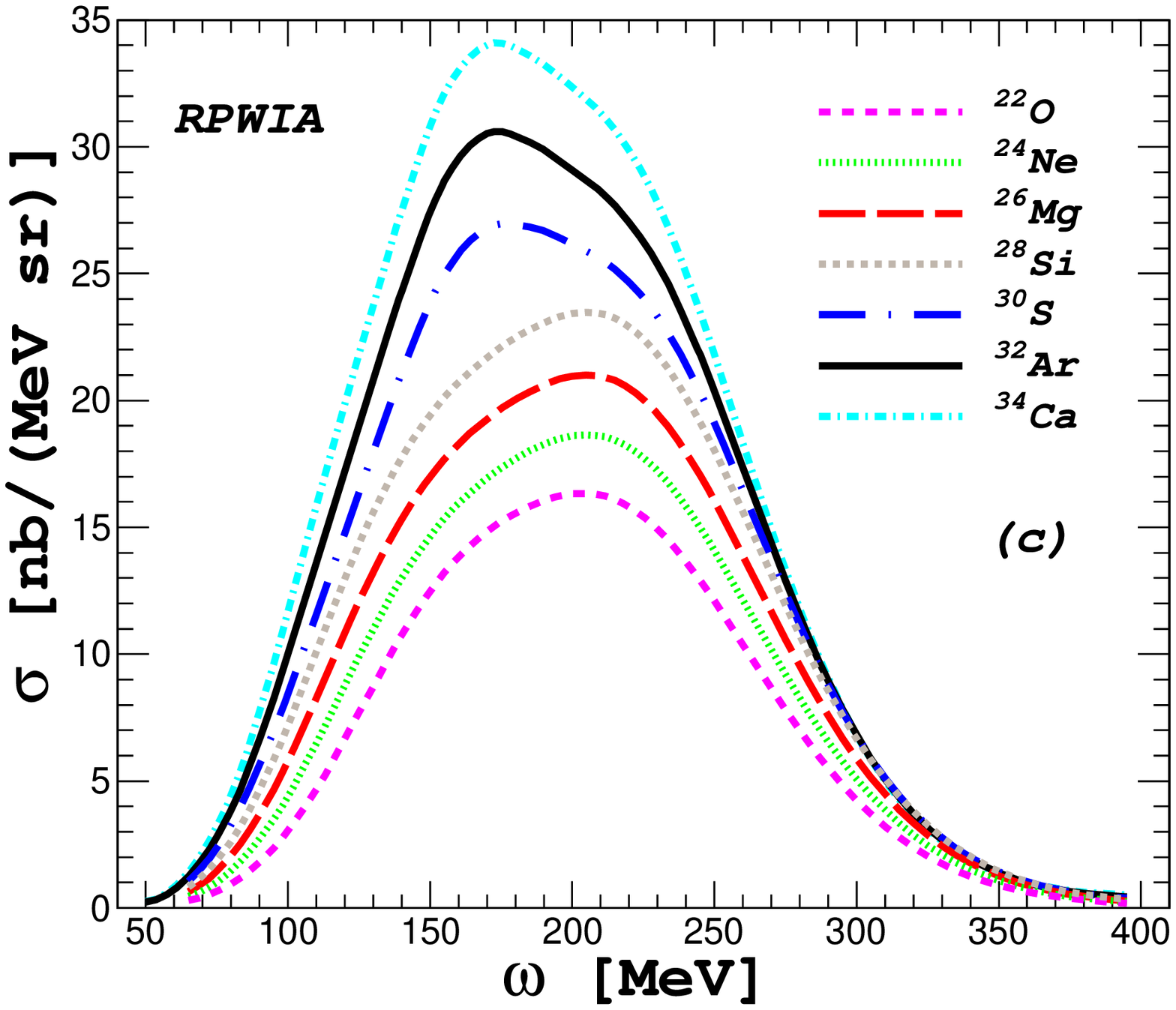} 
\caption{ (Color online)
Differential RPWIA cross sections for the inclusive QE $(e,e^{\prime})$ reaction
on isotones with $N = 28, 20$, and $14$ at $\varepsilon = 1080$ MeV 
and 
$\theta = 32^{\mathrm{o}}$ as functions of the energy transfer $\omega$.}	
\vspace {5mm}\label{fig:rpwia}
\end{figure}
\begin{figure}
	\centering
\includegraphics[scale=0.38]{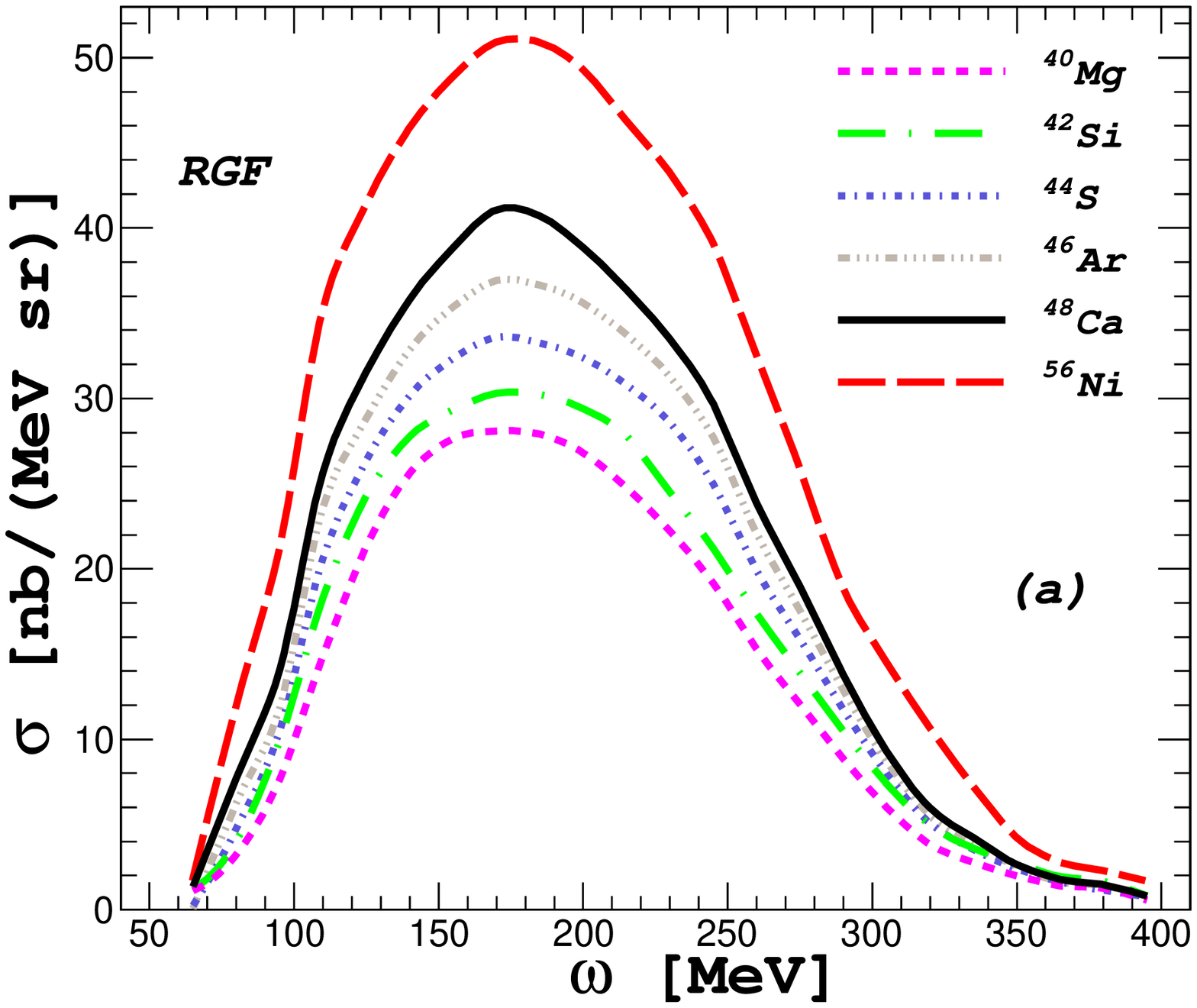} 
\includegraphics[scale=0.38]{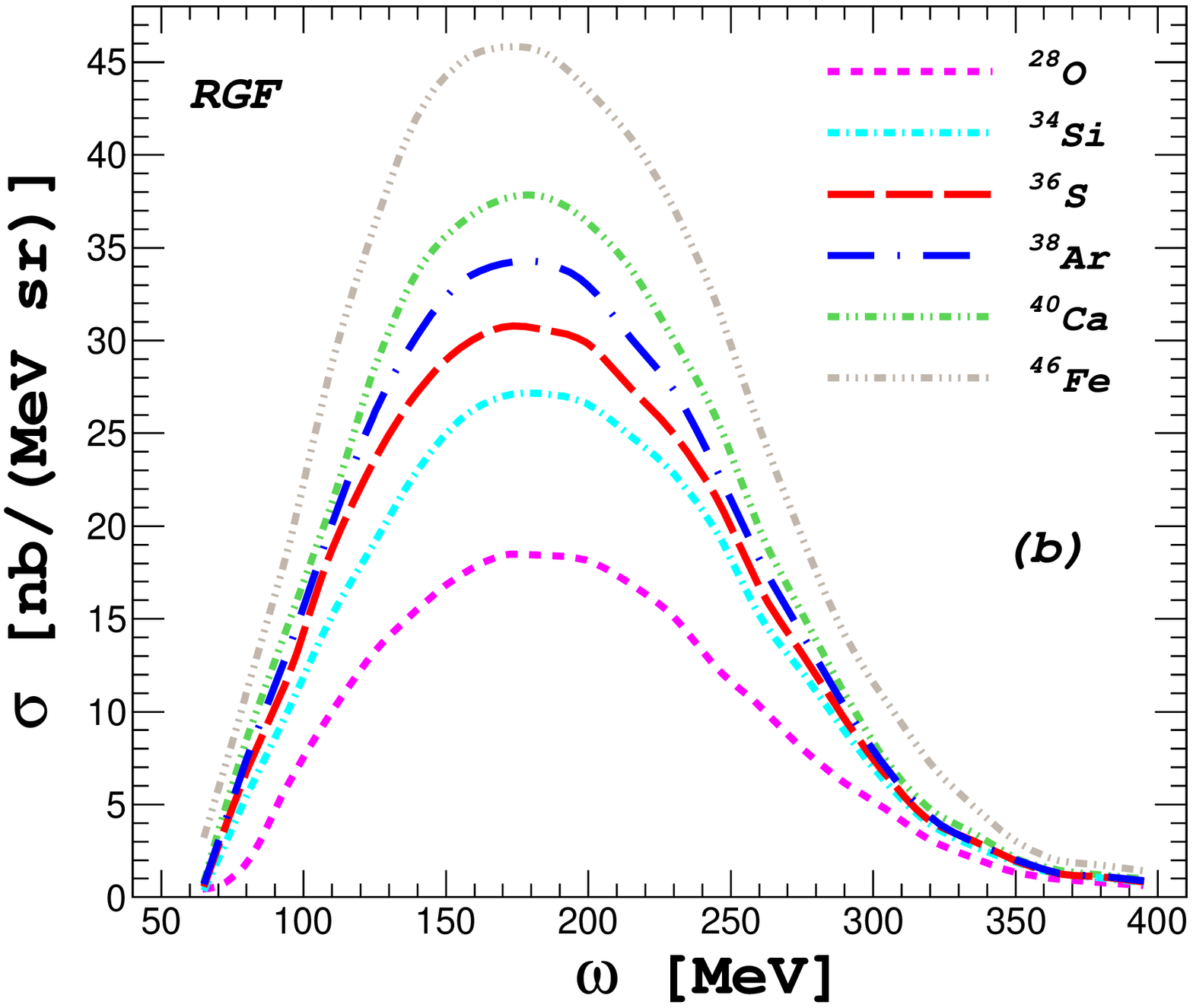} 
\includegraphics[scale=0.38]{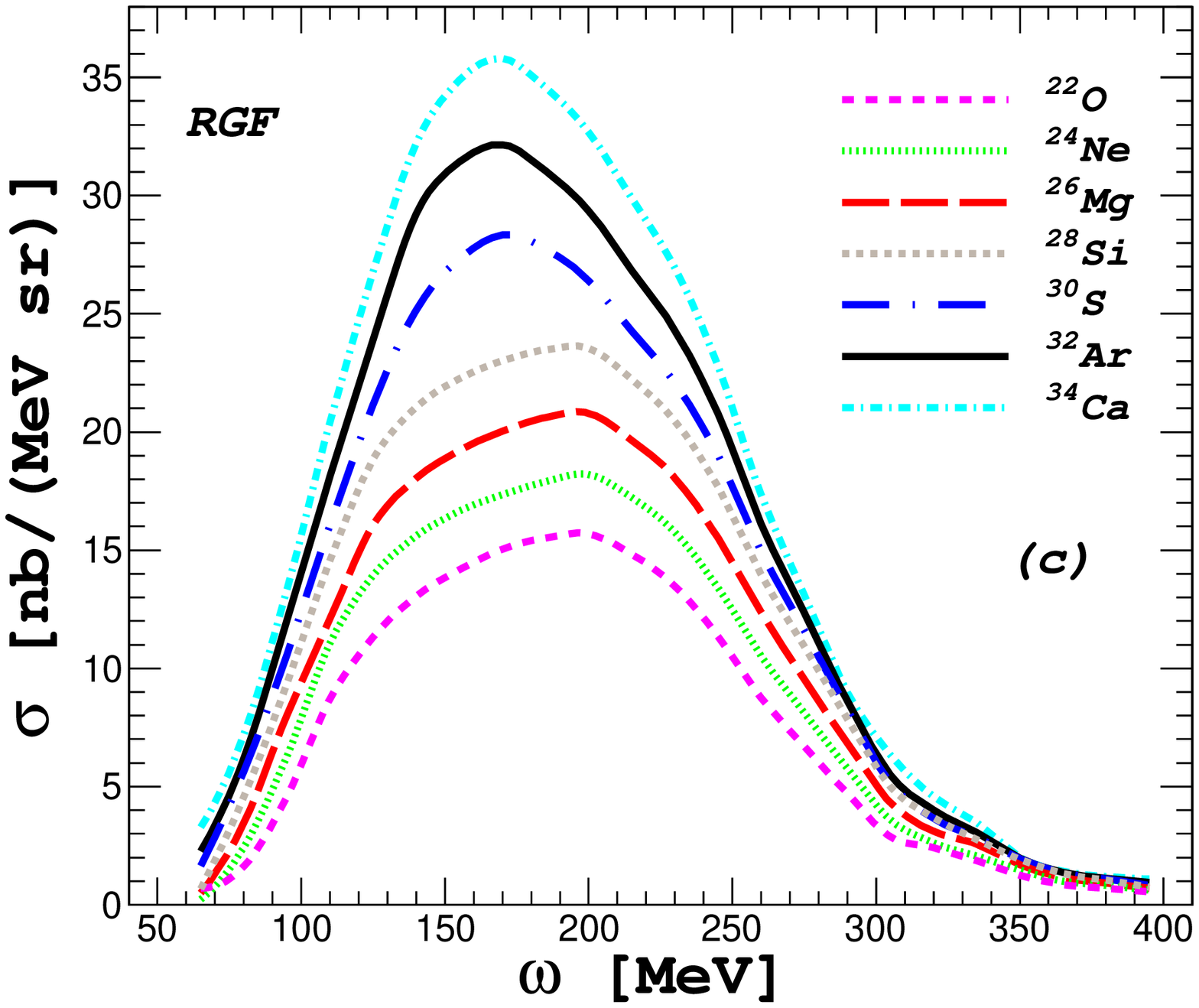} 
\caption{ (Color online)
Differential RGF cross sections for the inclusive QE $(e,e^{\prime})$ reaction 
on some selected isotones
with $N = 28, 20$, and $14$ at $\varepsilon = 1080$ MeV and 
$\theta = 32^{\mathrm{o}}$ as functions of $\omega$.}	\vspace {5mm}\label{fig:rgf}
\end{figure}

In the relativistic plane-wave impulse approximation (RPWIA) FSI 
are simply neglected.  In other approaches FSI are included in relativistic 
DWIA (RDWIA) calculations where the final nucleon state is evaluated with real 
potentials. In a different description of FSI relativistic Green's function 
(RGF) techniques \cite{Capuzzi:1991qd,Meucci:2003cv,Meucci:2003uy,
Capuzzi:2004au,Meucci:2005pk,
Meucci:2009nm,Meucci:2011pi,Giusti:cortona11,esotici2,Meucci:2013gja,PhysRevC.88.025502}
are used. 

In the RGF model, under suitable approximations, which are basically related 
to the IA, the components of the nuclear response are written 
in terms of the single particle optical model Green's function;
its spectral representation, that is
based on a biorthogonal expansion in terms of a non-Hermitian
optical potential $\cal H$ and of its Hermitian conjugate $\cal H^{\dagger}$,
is then exploited to obtain the 
components of the hadron tensor \cite{Meucci:2003uy,Meucci:2003cv} in terms of 
matrix elements of the same type as the DWIA ones of the exclusive \eep process, 
but involve eigenfunctions of both $\cal H$ and 
$\cal H^{\dagger}$, where the imaginary part has an opposite sign and gives 
in one case a loss and in the other case a gain of strength. The RGF formalism 
allows us to reconstruct the flux lost into nonelastic channels in the case of 
the inclusive response starting from the complex OP which 
describes elastic nucleon-nucleus scattering data and to include contributions 
which are not included in other models based on the IA. Moreover, with the use
of the same complex OP, it provides a consistent treatment of 
FSI in the exclusive and in the inclusive scattering. Because of the 
analyticity properties of the OP, the RGF model fulfills the Coulomb sum 
rule \cite{hori,Capuzzi:1991qd,Meucci:2003uy}.
More details about the RGF model can be found in \cite{Capuzzi:1991qd,
Meucci:2003cv,Meucci:2003uy,Capuzzi:2004au,Meucci:2005pk,
Meucci:2009nm,Meucci:2011pi,Giusti:cortona11,esotici2,Meucci:2013gja,PhysRevC.88.025502}

The RGF results can give a satisfactory description of  
experimental $(e,e^{\prime})$ cross sections in the QE 
region \cite{Meucci:2009nm,esotici2}. In particular, the RGF provides a 
 significant asymmetry in the scaling function, in agreement with the general 
 behavior of electron scattering data that present a significant tail extended 
 to large values of the transferred energy \cite{PhysRevC.65.025502}. Moreover, 
 the RGF results can describe the shape and the magnitude of charged-current 
 QE neutrino-nucleus scattering MiniBooNE data \cite{Meucci:2011vd} and of 
 neutral-current elastic neutrino-nucleus MiniBooNE data \cite{Meucci:2011nc}.  
Numerical predictions of the RGF model for the inclusive QE electron 
scattering on oxygen and calcium isotopic chains can be found in
\cite{esotici2}.

In the present calculations 
the s.p. bound nucleon states are obtained from the relativistic
mean-field model described in Sect. \ref{sec.rmf}. 
The s.p. scattering states are eigenfunctions of the energy-dependent and  
$A-$dependent ($A$ is the mass 
number) Democratic (DEM) parametrization for the relativistic optical potential
(ROP) of \cite{Cooper:2009}, that is obtained through a fit of more 
than 200 data sets of elastic proton-nucleus scattering data on a wide
range of nuclei which is not limited to doubly closed shell nuclei. 
The different number of protons along the 
isotonic chains produces different optical potentials 
(see \cite{Cooper:2009} for more details).
For the
single-nucleon current we have adopted the relativistic free nucleon expression
denoted as CC2 \cite{DeForestJr1983232,Meucci:2003uy}.

The cross section of the inclusive QE 
	$(e,e^{\prime})$  reaction along the $N = 28, 20$, and $14$ isotonic chains at 
	$\varepsilon = 1080$ MeV and $\theta = 32^{\mathrm{o}}$ are 
	shown  in Fig.~\ref{fig:rpwia}. In a first approximation  
we have neglected FSI and calculations have been performed in the RPWIA. 
In this approach the differences between the results 
for the various isotones are entirely due to the differences in the s.p. bound
state wave functions of each isotone.  
While only the charge proton density
distribution contributes to the cross section of elastic electron scattering, 
the cross section of QE electron scattering is obtained from the 
sum of all the integrated exclusive one-nucleon knockout processes, due to the 
interaction of the probe with all the individual nucleons, protons and 
neutrons, of the nucleus and contains information on the dynamics 
of the initial nuclear ground state.  
The main role is played by protons, which give most of the  contribution.
Increasing the proton number along each chain, owing to the enhancement of the 
proton contribution,  there is a proportional increase of the QE cross 
sections. In contrast, no increase is found in the neutron
contribution , which is less significant than the proton one. 
We can see in panel (b) of Fig.~\ref{fig:rpwia} that, even if
the central density of $^{34}$Si is  reduced (see  Fig.~\ref{fig:n20-a}),
the QE cross sections  of $^{34}$Si and $^{36}$S do not show any significant
difference that could be considered as a signal of \lq\lq bubble\rq\rq\ nuclei.

In Fig.~\ref{fig:rgf} we show the QE $(e,e^{\prime})$ cross sections
 calculated in the RGF model for selected isotones of the 
 $N = 28, 20$, and $14$  chains  in the same kinematics 
 as in Fig.~\ref{fig:rpwia}. 
The general trend  of the cross sections, their magnitude, and their evolution 
with respect to the change of the proton number along each chain are generally 
similar in RPWIA and RGF. The RGF results are, however, somewhat larger. 
The FSI effects in the RGF calculations produce visible distortion effects: 
the RGF cross sections are not as symmetrical as the corresponding RPWIA ones 
and show a tail toward large values of the  energy transferred $\omega$ that 
 is related to the description of FSI  with a complex 
energy-dependent ROP.

\section{Summary and Conclusions}\label{conc}
We have presented and discussed numerical predictions for the 
cross section and the parity-violating asymmetry in elastic and quasi-elastic 
electron scattering on the $N = 28$, $20$, and $14$ isotonic chains with the aim to
investigate their evolution with increasing proton number. 

The understanding of the properties of nuclei far from the valley of stability is 
one of the major topic of interest in modern nuclear physics.   
Large efforts in this direction have been done over last years and are 
planned for the future in different laboratories worldwide. 
The use of electrons as probe provides a powerful 
tool to achieve this goal, owing to the fact that their 
interaction is well known and relatively weak with respect to the nuclear force 
and can therefore more adequately explore the details of inner nuclear structures. 
As a consequence of this weakness, the cross sections become very small and more 
difficult experiments have to be performed. The  RIB facilities  have opened the possibility 
to obtain unprecedented information into  nuclear structures not available in nature 
but which are important in astrophysics and have a relevant 
role in the nucleosynthesis.
Electron scattering experiments off
exotic nuclei have been proposed in the ELISe experiment at FAIR and in the
SCRIT project at  RIKEN. 

In this work both elastic and inclusive quasi-elastic electron scattering have
been considered. The elastic scattering can give information on the 
global properties of nuclei and, in particular, on the different behavior of 
proton and neutron density distributions. The inclusive quasi-elastic scattering 
is  affected by the dynamical properties 
 and preferably exploits the single particle aspects of the nucleus. 
In addition, when combined with the exclusive $(e,e'p)$ scattering, it is able 
to explore the evolution of the s.p. model with increasing asymmetry
between the number of neutrons and protons. 
Many interesting phenomena are predicted in this 
situation: in particular, the modification of the shell model magic numbers. 
A definite response can be obtained from the comparison with experimental 
data, which will discriminate between the different theoretical models, mainly 
referring to RMF approaches.

The calculations for the present investigation have been carried out within 
the RMF framework solving the 
relativistic Hartree-Bogoliubov equations starting from an effective 
interaction mediated by meson exchange. 
The calculated cross sections include both the hadronic 
and Coulomb final states interactions. The inclusive quasi-elastic scattering 
is calculated with the relativistic Green's function model, which conserves the 
global particle flux in all the final state channels, as it is required in an 
inclusive reaction.

The model has been compared with experimental data of elastic scattering
already available on stable isotopes  to check its reliability and subsequently
applied to calculate elastic and inclusive quasi-elastic
cross sections on the $N = 28$, $20$, and $14$ isotonic chains. 
The possible disagreement of the experimental findings from the  theoretical 
predictions will be a clear indication of the insurgence of new phenomena 
related to the proton to neutron asymmetry. 

Our results show that the  evolution of some specific observables can be 
useful to test shell effects related to the filling of s.p. orbits. 
The increase of the proton number along each chain essentially produces an 
enhancement and an extension of the proton densities. The densities of the 
proton-rich isotones are significantly extended toward larger $r$ with respect 
to those of the proton-deficient ones. Pronounced shell effects are visible in 
the nuclear interior. 
The differential cross sections calculated for elastic electron scattering
show that increasing proton number along an isotonic chain the positions of 
the diffraction minima generally shift toward smaller scattering angles, 
corresponding to lower values of the momentum transfer. The shift 
is accompanied by a simultaneous increase in the height of the corresponding 
maxima of the cross sections. A plot of the evolution of the 
position of the first minimum for each isotone in the chain shows a transition 
between proton-poor and proton-rich isotones. It is not possible to fit the 
positions of the first minima with a straight line, but it is possible to draw 
two straight lines, connecting the minima of the lighter and heavier isotones, 
with a different slope. This behavior has been found for all the 
three isotonic chains.

The neutron densities  decrease in the nuclear interior and correspondingly 
increase away from the center to preserve the normalization to the constant
number of neutrons. 
We have calculated the parity-violating asymmetry parameter as it is directly 
related to the Fourier transform of the neutron density.  
The evolution of the position of the first diffraction minimum of the 
asymmetry parameter as a function of the  neutron-to-proton asymmetry can 
provide an alternative source of information on the neutron distribution 
along each isotonic chain.

\begin{acknowledgments}

This work was partially supported by the Italian MIUR 
through the PRIN 2009 research project.

\end{acknowledgments}


\begin{thebibliography}{100}%
\makeatletter
\providecommand \@ifxundefined [1]{%
 \@ifx{#1\undefined}
}%
\providecommand \@ifnum [1]{%
 \ifnum #1\expandafter \@firstoftwo
 \else \expandafter \@secondoftwo
 \fi
}%
\providecommand \@ifx [1]{%
 \ifx #1\expandafter \@firstoftwo
 \else \expandafter \@secondoftwo
 \fi
}%
\providecommand \natexlab [1]{#1}%
\providecommand \enquote  [1]{``#1''}%
\providecommand \bibnamefont  [1]{#1}%
\providecommand \bibfnamefont [1]{#1}%
\providecommand \citenamefont [1]{#1}%
\providecommand \href@noop [0]{\@secondoftwo}%
\providecommand \href [0]{\begingroup \@sanitize@url \@href}%
\providecommand \@href[1]{\@@startlink{#1}\@@href}%
\providecommand \@@href[1]{\endgroup#1\@@endlink}%
\providecommand \@sanitize@url [0]{\catcode `\\12\catcode `\$12\catcode
  `\&12\catcode `\#12\catcode `\^12\catcode `\_12\catcode `\%12\relax}%
\providecommand \@@startlink[1]{}%
\providecommand \@@endlink[0]{}%
\providecommand \url  [0]{\begingroup\@sanitize@url \@url }%
\providecommand \@url [1]{\endgroup\@href {#1}{\urlprefix }}%
\providecommand \urlprefix  [0]{URL }%
\providecommand \Eprint [0]{\href }%
\providecommand \doibase [0]{http://dx.doi.org/}%
\providecommand \selectlanguage [0]{\@gobble}%
\providecommand \bibinfo  [0]{\@secondoftwo}%
\providecommand \bibfield  [0]{\@secondoftwo}%
\providecommand \translation [1]{[#1]}%
\providecommand \BibitemOpen [0]{}%
\providecommand \bibitemStop [0]{}%
\providecommand \bibitemNoStop [0]{.\EOS\space}%
\providecommand \EOS [0]{\spacefactor3000\relax}%
\providecommand \BibitemShut  [1]{\csname bibitem#1\endcsname}%
\let\auto@bib@innerbib\@empty
\bibitem [{\citenamefont {Hofstadter}(1956)}]{Hofstadter:1956qs}%
  \BibitemOpen
  \bibfield  {author} {\bibinfo {author} {\bibfnamefont {R.}~\bibnamefont
  {Hofstadter}},\ }\href {\doibase 10.1103/RevModPhys.28.214} {\bibfield
  {journal} {\bibinfo  {journal} {Rev. Mod. Phys.}\ }\textbf {\bibinfo {volume}
  {28}},\ \bibinfo {pages} {214} (\bibinfo {year} {1956})}\BibitemShut
  {NoStop}%
\bibitem [{\citenamefont {Donnelly}\ and\ \citenamefont
  {Walecka}(1975)}]{Donnelly:1975ze}%
  \BibitemOpen
  \bibfield  {author} {\bibinfo {author} {\bibfnamefont {T.~W.}\ \bibnamefont
  {Donnelly}}\ and\ \bibinfo {author} {\bibfnamefont {J.~D.}\ \bibnamefont
  {Walecka}},\ }\href@noop {} {\bibfield  {journal} {\bibinfo  {journal} {Ann.
  Rev. Nucl. Part. Sci.}\ }\textbf {\bibinfo {volume} {25}},\ \bibinfo {pages}
  {329} (\bibinfo {year} {1975})}\BibitemShut {NoStop}%
\bibitem [{\citenamefont {Donnelly}\ and\ \citenamefont
  {Sick}(1984)}]{Donnelly:1984rg}%
  \BibitemOpen
  \bibfield  {author} {\bibinfo {author} {\bibfnamefont {T.~W.}\ \bibnamefont
  {Donnelly}}\ and\ \bibinfo {author} {\bibfnamefont {I.}~\bibnamefont
  {Sick}},\ }\href {\doibase 10.1103/RevModPhys.56.461} {\bibfield  {journal}
  {\bibinfo  {journal} {Rev. Mod. Phys.}\ }\textbf {\bibinfo {volume} {56}},\
  \bibinfo {pages} {461} (\bibinfo {year} {1984})}\BibitemShut {NoStop}%
\bibitem [{\citenamefont {Boffi}\ \emph {et~al.}(1993)\citenamefont {Boffi},
  \citenamefont {Giusti},\ and\ \citenamefont {Pacati}}]{Boffi:1993gs}%
  \BibitemOpen
  \bibfield  {author} {\bibinfo {author} {\bibfnamefont {S.}~\bibnamefont
  {Boffi}}, \bibinfo {author} {\bibfnamefont {C.}~\bibnamefont {Giusti}}, \
  and\ \bibinfo {author} {\bibfnamefont {F.~D.}\ \bibnamefont {Pacati}},\
  }\href {\doibase 10.1016/0370-1573(93)90132-W} {\bibfield  {journal}
  {\bibinfo  {journal} {Phys. Rept.}\ }\textbf {\bibinfo {volume} {226}},\
  \bibinfo {pages} {1} (\bibinfo {year} {1993})}\BibitemShut {NoStop}%
\bibitem [{\citenamefont {Benhar}\ \emph {et~al.}(2008)\citenamefont {Benhar},
  \citenamefont {Day},\ and\ \citenamefont {Sick}}]{RevModPhys.80.189}%
  \BibitemOpen
  \bibfield  {author} {\bibinfo {author} {\bibfnamefont {O.}~\bibnamefont
  {Benhar}}, \bibinfo {author} {\bibfnamefont {D.}~\bibnamefont {Day}}, \ and\
  \bibinfo {author} {\bibfnamefont {I.}~\bibnamefont {Sick}},\ }\href {\doibase
  10.1103/RevModPhys.80.189} {\bibfield  {journal} {\bibinfo  {journal} {Rev.
  Mod. Phys.}\ }\textbf {\bibinfo {volume} {80}},\ \bibinfo {pages} {189}
  (\bibinfo {year} {2008})}\BibitemShut {NoStop}%
\bibitem [{\citenamefont {De~Vries}\ \emph {et~al.}(1987)\citenamefont
  {De~Vries}, \citenamefont {De~Jager},\ and\ \citenamefont
  {De~Vries}}]{DeJ:1987qc}%
  \BibitemOpen
  \bibfield  {author} {\bibinfo {author} {\bibfnamefont {H.}~\bibnamefont
  {De~Vries}}, \bibinfo {author} {\bibfnamefont {C.~W.}\ \bibnamefont
  {De~Jager}}, \ and\ \bibinfo {author} {\bibfnamefont {C.}~\bibnamefont
  {De~Vries}},\ }\href@noop {} {\bibfield  {journal} {\bibinfo  {journal}
  {Atom. Data Nucl. Data Tabl.}\ }\textbf {\bibinfo {volume} {36}},\ \bibinfo
  {pages} {495} (\bibinfo {year} {1987})}\BibitemShut {NoStop}%
\bibitem [{\citenamefont {Fricke}\ \emph {et~al.}(1995)\citenamefont {Fricke},
  \citenamefont {Bernhardt}, \citenamefont {Heilig}, \citenamefont {Schaller},
  \citenamefont {Schellenberg}, \citenamefont {Shera},\ and\ \citenamefont
  {De~Jager}}]{Fricke:1995zz}%
  \BibitemOpen
  \bibfield  {author} {\bibinfo {author} {\bibfnamefont {G.}~\bibnamefont
  {Fricke}}, \bibinfo {author} {\bibfnamefont {C.}~\bibnamefont {Bernhardt}},
  \bibinfo {author} {\bibfnamefont {K.}~\bibnamefont {Heilig}}, \bibinfo
  {author} {\bibfnamefont {L.~A.}\ \bibnamefont {Schaller}}, \bibinfo {author}
  {\bibfnamefont {L.}~\bibnamefont {Schellenberg}}, \bibinfo {author}
  {\bibfnamefont {E.~B.}\ \bibnamefont {Shera}}, \ and\ \bibinfo {author}
  {\bibfnamefont {C.~W.}\ \bibnamefont {De~Jager}},\ }\href@noop {} {\bibfield
  {journal} {\bibinfo  {journal} {Atom. Data Nucl. Data Tabl.}\ }\textbf
  {\bibinfo {volume} {60}},\ \bibinfo {pages} {177} (\bibinfo {year}
  {1995})}\BibitemShut {NoStop}%
\bibitem [{\citenamefont {Heisenberg}\ and\ \citenamefont
  {Blok}(1983)}]{Heisenberg:1984hb}%
  \BibitemOpen
  \bibfield  {author} {\bibinfo {author} {\bibfnamefont {J.}~\bibnamefont
  {Heisenberg}}\ and\ \bibinfo {author} {\bibfnamefont {H.~P.}\ \bibnamefont
  {Blok}},\ }\href@noop {} {\bibfield  {journal} {\bibinfo  {journal} {Ann.
  Rev. Nucl. Part. Sci.}\ }\textbf {\bibinfo {volume} {33}},\ \bibinfo {pages}
  {569} (\bibinfo {year} {1983})}\BibitemShut {NoStop}%
\bibitem [{\citenamefont {Boffi}\ \emph {et~al.}(1996)\citenamefont {Boffi},
  \citenamefont {Giusti}, \citenamefont {Pacati},\ and\ \citenamefont
  {Radici}}]{book}%
  \BibitemOpen
  \bibfield  {author} {\bibinfo {author} {\bibfnamefont {S.}~\bibnamefont
  {Boffi}}, \bibinfo {author} {\bibfnamefont {C.}~\bibnamefont {Giusti}},
  \bibinfo {author} {\bibfnamefont {F.~D.}\ \bibnamefont {Pacati}}, \ and\
  \bibinfo {author} {\bibfnamefont {M.}~\bibnamefont {Radici}},\ }\href@noop {}
  {\emph {\bibinfo {title} {Electromagnetic Response of Atomic Nuclei}}},\
  \bibinfo {series} {Oxford Studies in Nuclear Physics}, Vol.~\bibinfo {volume}
  {20}\ (\bibinfo  {publisher} {Clarendon Press},\ \bibinfo {address}
  {Oxford},\ \bibinfo {year} {1996})\BibitemShut {NoStop}%
\bibitem [{web()}]{web-benhar}%
  \BibitemOpen
  \href@noop {} {}\bibinfo {howpublished}
  {\url{http://faculty.virginia.edu/qes-archive/index.html}}\BibitemShut
  {NoStop}%
\bibitem [{\citenamefont {Frullani}\ and\ \citenamefont
  {Mougey}(1984)}]{Frullani:1984nn}%
  \BibitemOpen
  \bibfield  {author} {\bibinfo {author} {\bibfnamefont {S.}~\bibnamefont
  {Frullani}}\ and\ \bibinfo {author} {\bibfnamefont {J.}~\bibnamefont
  {Mougey}},\ }\href@noop {} {\bibfield  {journal} {\bibinfo  {journal} {Adv.
  Nucl. Phys.}\ }\textbf {\bibinfo {volume} {14}},\ \bibinfo {pages} {1}
  (\bibinfo {year} {1984})}\BibitemShut {NoStop}%
\bibitem [{\citenamefont {Bernheim}\ \emph {et~al.}(1982)\citenamefont
  {Bernheim}, \citenamefont {Bussiere}, \citenamefont {Mougey}, \citenamefont
  {Royer}, \citenamefont {Tarnowski}, \citenamefont {Frullani}, \citenamefont
  {Boffi}, \citenamefont {Giusti},\ and\ \citenamefont
  {Pacati}}]{Bernheim:1981si}%
  \BibitemOpen
  \bibfield  {author} {\bibinfo {author} {\bibfnamefont {M.}~\bibnamefont
  {Bernheim}}, \bibinfo {author} {\bibfnamefont {A.}~\bibnamefont {Bussiere}},
  \bibinfo {author} {\bibfnamefont {J.}~\bibnamefont {Mougey}}, \bibinfo
  {author} {\bibfnamefont {D.}~\bibnamefont {Royer}}, \bibinfo {author}
  {\bibfnamefont {S.}~\bibnamefont {Tarnowski}, \bibfnamefont
  {D.~Turck-Chi\`{e}ze}}, \bibinfo {author} {\bibfnamefont {S.}~\bibnamefont
  {Frullani}}, \bibinfo {author} {\bibfnamefont {S.}~\bibnamefont {Boffi}},
  \bibinfo {author} {\bibfnamefont {C.}~\bibnamefont {Giusti}}, \ and\ \bibinfo
  {author} {\bibfnamefont {F.~D.}\ \bibnamefont {Pacati}},\ }\href {\doibase
  10.1016/0375-9474(82)90020-3} {\bibfield  {journal} {\bibinfo  {journal}
  {Nucl. Phys. A}\ }\textbf {\bibinfo {volume} {375}},\ \bibinfo {pages} {381}
  (\bibinfo {year} {1982})}\BibitemShut {NoStop}%
\bibitem [{\citenamefont {Lapik\'{a}s}(1993)}]{Lapikas:1003zz}%
  \BibitemOpen
  \bibfield  {author} {\bibinfo {author} {\bibfnamefont {L.}~\bibnamefont
  {Lapik\'{a}s}},\ }\href {\doibase 10.1016/0375-9474(93)90630-G} {\bibfield
  {journal} {\bibinfo  {journal} {Nucl. Phys. A}\ }\textbf {\bibinfo {volume}
  {553}},\ \bibinfo {pages} {297} (\bibinfo {year} {1993})}\BibitemShut
  {NoStop}%
\bibitem [{\citenamefont {de~Witt~Huberts}(1990)}]{deWittHuberts:1990zy}%
  \BibitemOpen
  \bibfield  {author} {\bibinfo {author} {\bibfnamefont {P.~K.~A.}\
  \bibnamefont {de~Witt~Huberts}},\ }\href {\doibase
  10.1088/0954-3899/16/4/004} {\bibfield  {journal} {\bibinfo  {journal} {J.
  Phys. G}\ }\textbf {\bibinfo {volume} {16}},\ \bibinfo {pages} {507}
  (\bibinfo {year} {1990})}\BibitemShut {NoStop}%
\bibitem [{\citenamefont {Ud\'{\i}as}\ \emph {et~al.}(1993)\citenamefont
  {Ud\'{\i}as}, \citenamefont {Sarriguren}, \citenamefont {Moya~de Guerra},
  \citenamefont {Garrido},\ and\ \citenamefont {Caballero}}]{Udias:1993xy}%
  \BibitemOpen
  \bibfield  {author} {\bibinfo {author} {\bibfnamefont {J.~M.}\ \bibnamefont
  {Ud\'{\i}as}}, \bibinfo {author} {\bibfnamefont {P.}~\bibnamefont
  {Sarriguren}}, \bibinfo {author} {\bibfnamefont {E.}~\bibnamefont {Moya~de
  Guerra}}, \bibinfo {author} {\bibfnamefont {E.}~\bibnamefont {Garrido}}, \
  and\ \bibinfo {author} {\bibfnamefont {J.~A.}\ \bibnamefont {Caballero}},\
  }\href {\doibase 10.1103/PhysRevC.48.2731} {\bibfield  {journal} {\bibinfo
  {journal} {Phys. Rev. C}\ }\textbf {\bibinfo {volume} {48}},\ \bibinfo
  {pages} {2731} (\bibinfo {year} {1993})}\BibitemShut {NoStop}%
\bibitem [{\citenamefont {Gao}\ \emph {et~al.}(2000)\citenamefont {Gao} \emph
  {et~al.}}]{gao00}%
  \BibitemOpen
  \bibfield  {author} {\bibinfo {author} {\bibfnamefont {J.}~\bibnamefont
  {Gao}} \emph {et~al.} (\bibinfo {collaboration} {The Jefferson Lab Hall A
  Collaboration}),\ }\href {\doibase 10.1103/PhysRevLett.84.3265} {\bibfield
  {journal} {\bibinfo  {journal} {Phys. Rev. Lett.}\ }\textbf {\bibinfo
  {volume} {84}},\ \bibinfo {pages} {3265} (\bibinfo {year}
  {2000})}\BibitemShut {NoStop}%
\bibitem [{\citenamefont {Meucci}\ \emph
  {et~al.}(2001{\natexlab{a}})\citenamefont {Meucci}, \citenamefont {Giusti},\
  and\ \citenamefont {Pacati}}]{Meucci:2001qc}%
  \BibitemOpen
  \bibfield  {author} {\bibinfo {author} {\bibfnamefont {A.}~\bibnamefont
  {Meucci}}, \bibinfo {author} {\bibfnamefont {C.}~\bibnamefont {Giusti}}, \
  and\ \bibinfo {author} {\bibfnamefont {F.~D.}\ \bibnamefont {Pacati}},\
  }\href {\doibase 10.1103/PhysRevC.64.014604} {\bibfield  {journal} {\bibinfo
  {journal} {Phys. Rev. C}\ }\textbf {\bibinfo {volume} {64}},\ \bibinfo
  {pages} {014604} (\bibinfo {year} {2001}{\natexlab{a}})}\BibitemShut
  {NoStop}%
\bibitem [{\citenamefont {Giusti}\ \emph {et~al.}(2011)\citenamefont {Giusti},
  \citenamefont {Meucci}, \citenamefont {Pacati}, \citenamefont {Co'},\ and\
  \citenamefont {De~Donno}}]{Giusti:2011it}%
  \BibitemOpen
  \bibfield  {author} {\bibinfo {author} {\bibfnamefont {C.}~\bibnamefont
  {Giusti}}, \bibinfo {author} {\bibfnamefont {A.}~\bibnamefont {Meucci}},
  \bibinfo {author} {\bibfnamefont {F.~D.}\ \bibnamefont {Pacati}}, \bibinfo
  {author} {\bibfnamefont {G.}~\bibnamefont {Co'}}, \ and\ \bibinfo {author}
  {\bibfnamefont {V.}~\bibnamefont {De~Donno}},\ }\href {\doibase
  10.1103/PhysRevC.84.024615} {\bibfield  {journal} {\bibinfo  {journal} {Phys.
  Rev. C}\ }\textbf {\bibinfo {volume} {84}},\ \bibinfo {pages} {024615}
  (\bibinfo {year} {2011})}\BibitemShut {NoStop}%
\bibitem [{\citenamefont {Co'}\ \emph {et~al.}(2012)\citenamefont {Co'},
  \citenamefont {De~Donno}, \citenamefont {Finelli}, \citenamefont {Grasso},
  \citenamefont {Anguiano}, \citenamefont {Lallena}, \citenamefont {Giusti},
  \citenamefont {Meucci},\ and\ \citenamefont {Pacati}}]{esotici1}%
  \BibitemOpen
  \bibfield  {author} {\bibinfo {author} {\bibfnamefont {G.}~\bibnamefont
  {Co'}}, \bibinfo {author} {\bibfnamefont {V.}~\bibnamefont {De~Donno}},
  \bibinfo {author} {\bibfnamefont {P.}~\bibnamefont {Finelli}}, \bibinfo
  {author} {\bibfnamefont {M.}~\bibnamefont {Grasso}}, \bibinfo {author}
  {\bibfnamefont {M.}~\bibnamefont {Anguiano}}, \bibinfo {author}
  {\bibfnamefont {A.~M.}\ \bibnamefont {Lallena}}, \bibinfo {author}
  {\bibfnamefont {C.}~\bibnamefont {Giusti}}, \bibinfo {author} {\bibfnamefont
  {A.}~\bibnamefont {Meucci}}, \ and\ \bibinfo {author} {\bibfnamefont {F.~D.}\
  \bibnamefont {Pacati}},\ }\href {\doibase 10.1103/PhysRevC.85.024322}
  {\bibfield  {journal} {\bibinfo  {journal} {Phys. Rev. C}\ }\textbf {\bibinfo
  {volume} {85}},\ \bibinfo {pages} {024322} (\bibinfo {year}
  {2012})}\BibitemShut {NoStop}%
\bibitem [{\citenamefont {Tanhihata}(1995)}]{tan95}%
  \BibitemOpen
  \bibfield  {author} {\bibinfo {author} {\bibfnamefont {I.}~\bibnamefont
  {Tanhihata}},\ }\href@noop {} {\bibfield  {journal} {\bibinfo  {journal}
  {Prog. Part. Nucl. Phys.}\ }\textbf {\bibinfo {volume} {35}},\ \bibinfo
  {pages} {505} (\bibinfo {year} {1995})}\BibitemShut {NoStop}%
\bibitem [{\citenamefont {Geissel}\ \emph {et~al.}(1995)\citenamefont
  {Geissel}, \citenamefont {M{\"u}zenberg},\ and\ \citenamefont
  {Riisager}}]{gei95}%
  \BibitemOpen
  \bibfield  {author} {\bibinfo {author} {\bibfnamefont {H.}~\bibnamefont
  {Geissel}}, \bibinfo {author} {\bibfnamefont {G.}~\bibnamefont
  {M{\"u}zenberg}}, \ and\ \bibinfo {author} {\bibfnamefont {R.}~\bibnamefont
  {Riisager}},\ }\href@noop {} {\bibfield  {journal} {\bibinfo  {journal} {Ann.
  Rev. Nucl. Part. Sci.}\ }\textbf {\bibinfo {volume} {45}},\ \bibinfo {pages}
  {163} (\bibinfo {year} {1995})}\BibitemShut {NoStop}%
\bibitem [{\citenamefont {Mueller}(2001)}]{mue01}%
  \BibitemOpen
  \bibfield  {author} {\bibinfo {author} {\bibfnamefont {A.}~\bibnamefont
  {Mueller}},\ }\href@noop {} {\bibfield  {journal} {\bibinfo  {journal} {Prog.
  Part. Nucl. Phys.}\ }\textbf {\bibinfo {volume} {46}},\ \bibinfo {pages}
  {359} (\bibinfo {year} {2001})}\BibitemShut {NoStop}%
\bibitem [{\citenamefont {Suda}\ \emph {et~al.}(2001)\citenamefont {Suda},
  \citenamefont {Maruyama},\ and\ \citenamefont {Tanhihata}}]{sud01}%
  \BibitemOpen
  \bibfield  {author} {\bibinfo {author} {\bibfnamefont {T.}~\bibnamefont
  {Suda}}, \bibinfo {author} {\bibfnamefont {K.}~\bibnamefont {Maruyama}}, \
  and\ \bibinfo {author} {\bibfnamefont {I.}~\bibnamefont {Tanhihata}},\
  }\href@noop {} {\bibfield  {journal} {\bibinfo  {journal} {RIKEN Accel. Prog.
  Rep.}\ }\textbf {\bibinfo {volume} {34}},\ \bibinfo {pages} {49} (\bibinfo
  {year} {2001})}\BibitemShut {NoStop}%
\bibitem [{\citenamefont {Suda}\ \emph {et~al.}(2009)\citenamefont {Suda},
  \citenamefont {Wakasugi}, \citenamefont {Emoto}, \citenamefont {Ishii},
  \citenamefont {Ito}, \citenamefont {Kurita}, \citenamefont {Kuwajima},
  \citenamefont {Noda}, \citenamefont {Shirai}, \citenamefont {Tamae},
  \citenamefont {Tongu}, \citenamefont {Wang},\ and\ \citenamefont
  {Yano}}]{sud09}%
  \BibitemOpen
  \bibfield  {author} {\bibinfo {author} {\bibfnamefont {T.}~\bibnamefont
  {Suda}}, \bibinfo {author} {\bibfnamefont {M.}~\bibnamefont {Wakasugi}},
  \bibinfo {author} {\bibfnamefont {T.}~\bibnamefont {Emoto}}, \bibinfo
  {author} {\bibfnamefont {K.}~\bibnamefont {Ishii}}, \bibinfo {author}
  {\bibfnamefont {S.}~\bibnamefont {Ito}}, \bibinfo {author} {\bibfnamefont
  {K.}~\bibnamefont {Kurita}}, \bibinfo {author} {\bibfnamefont
  {A.}~\bibnamefont {Kuwajima}}, \bibinfo {author} {\bibfnamefont
  {A.}~\bibnamefont {Noda}}, \bibinfo {author} {\bibfnamefont {T.}~\bibnamefont
  {Shirai}}, \bibinfo {author} {\bibfnamefont {T.}~\bibnamefont {Tamae}},
  \bibinfo {author} {\bibfnamefont {H.}~\bibnamefont {Tongu}}, \bibinfo
  {author} {\bibfnamefont {S.}~\bibnamefont {Wang}}, \ and\ \bibinfo {author}
  {\bibfnamefont {Y.}~\bibnamefont {Yano}},\ }\href {\doibase
  10.1103/PhysRevLett.102.102501} {\bibfield  {journal} {\bibinfo  {journal}
  {Phys. Rev. Lett.}\ }\textbf {\bibinfo {volume} {102}},\ \bibinfo {pages}
  {102501} (\bibinfo {year} {2009})}\BibitemShut {NoStop}%
\bibitem [{\citenamefont {Katayama}\ \emph {et~al.}(2003)\citenamefont
  {Katayama}, \citenamefont {Suda},\ and\ \citenamefont {Tanhihata}}]{kat03}%
  \BibitemOpen
  \bibfield  {author} {\bibinfo {author} {\bibfnamefont {T.}~\bibnamefont
  {Katayama}}, \bibinfo {author} {\bibfnamefont {T.}~\bibnamefont {Suda}}, \
  and\ \bibinfo {author} {\bibfnamefont {I.}~\bibnamefont {Tanhihata}},\
  }\href@noop {} {\bibfield  {journal} {\bibinfo  {journal} {Physica Scripta}\
  }\textbf {\bibinfo {volume} {T104}},\ \bibinfo {pages} {129} (\bibinfo {year}
  {2003})}\BibitemShut {NoStop}%
\bibitem [{\lq\lq An $\rm{International}$ $\rm{Accelerator}$ $\rm{Facility}$
  for $\rm{Beams}$ of $\rm{Ions}$ and Antiprotons\rq\rq, GSI report
  2006()}]{gsi06}%
  \BibitemOpen
  \lq\lq An $\rm{International}$ $\rm{Accelerator}$ $\rm{Facility}$ for
  $\rm{Beams}$ of $\rm{Ions}$ and Antiprotons\rq\rq, GSI report 2006,\
  \href@noop {} {}\bibinfo {howpublished}
  {\url{http://www.gsi.de/GSI-Future/cdr/}}\BibitemShut {NoStop}%
\bibitem [{eli()}]{elise}%
  \BibitemOpen
  \href@noop {} {}\bibinfo {howpublished} {\url{
  http://www.gsi.de/forschung/fair_experiments/elise/index_e.html}}\BibitemShut
  {NoStop}%
\bibitem [{\citenamefont {Simon}(2007)}]{Simon:2007zz}%
  \BibitemOpen
  \bibfield  {author} {\bibinfo {author} {\bibfnamefont {H.}~\bibnamefont
  {Simon}},\ }\href {\doibase 10.1016/j.nuclphysa.2006.12.020} {\bibfield
  {journal} {\bibinfo  {journal} {Nucl. Phys. A}\ }\textbf {\bibinfo {volume}
  {787}},\ \bibinfo {pages} {102} (\bibinfo {year} {2007})}\BibitemShut
  {NoStop}%
\bibitem [{\citenamefont {Antonov}\ \emph {et~al.}(2011)\citenamefont {Antonov}
  \emph {et~al.}}]{Antonov:2011zza}%
  \BibitemOpen
  \bibfield  {author} {\bibinfo {author} {\bibfnamefont {A.~N.}\ \bibnamefont
  {Antonov}} \emph {et~al.},\ }\href {\doibase 10.1016/j.nima.2010.12.246}
  {\bibfield  {journal} {\bibinfo  {journal} {Nucl. Instrum. Meth. A}\ }\textbf
  {\bibinfo {volume} {637}},\ \bibinfo {pages} {60} (\bibinfo {year}
  {2011})}\BibitemShut {NoStop}%
\bibitem [{T. Suda()}]{sud10}%
  \BibitemOpen
  T. Suda,\ \href@noop {} {}\bibinfo {howpublished} {\lq\lq A construction
  proposal of an electron scattering facility for structure studies of
  short-lived nuclei\rq\rq , $\rm{Proposal}$ for $\rm{Nuclear}$ $\rm{Physics}$
  $\rm{Experiments}$ at $\rm{RIBF}$ $\rm{NP1006-SCRIT01}$} (\bibinfo {year}
  {2010})\BibitemShut {NoStop}%
\bibitem [{\citenamefont {Suda}\ \emph {et~al.}(2012)\citenamefont {Suda},
  \citenamefont {Adachi}, \citenamefont {Amagai}, \citenamefont {Enokizono},
  \citenamefont {Hara}, \citenamefont {Hori}, \citenamefont {Ichikawa},
  \citenamefont {Kurita}, \citenamefont {Miyamoto}, \citenamefont {Ogawara},
  \citenamefont {Ohnishi}, \citenamefont {Shimakura}, \citenamefont {Tamae},
  \citenamefont {Togasaki}, \citenamefont {Wakasugi}, \citenamefont {Wang},\
  and\ \citenamefont {Yanagi}}]{Suda01012012}%
  \BibitemOpen
  \bibfield  {author} {\bibinfo {author} {\bibfnamefont {T.}~\bibnamefont
  {Suda}}, \bibinfo {author} {\bibfnamefont {T.}~\bibnamefont {Adachi}},
  \bibinfo {author} {\bibfnamefont {T.}~\bibnamefont {Amagai}}, \bibinfo
  {author} {\bibfnamefont {A.}~\bibnamefont {Enokizono}}, \bibinfo {author}
  {\bibfnamefont {M.}~\bibnamefont {Hara}}, \bibinfo {author} {\bibfnamefont
  {T.}~\bibnamefont {Hori}}, \bibinfo {author} {\bibfnamefont {S.}~\bibnamefont
  {Ichikawa}}, \bibinfo {author} {\bibfnamefont {K.}~\bibnamefont {Kurita}},
  \bibinfo {author} {\bibfnamefont {T.}~\bibnamefont {Miyamoto}}, \bibinfo
  {author} {\bibfnamefont {R.}~\bibnamefont {Ogawara}}, \bibinfo {author}
  {\bibfnamefont {T.}~\bibnamefont {Ohnishi}}, \bibinfo {author} {\bibfnamefont
  {Y.}~\bibnamefont {Shimakura}}, \bibinfo {author} {\bibfnamefont
  {T.}~\bibnamefont {Tamae}}, \bibinfo {author} {\bibfnamefont
  {M.}~\bibnamefont {Togasaki}}, \bibinfo {author} {\bibfnamefont
  {M.}~\bibnamefont {Wakasugi}}, \bibinfo {author} {\bibfnamefont
  {S.}~\bibnamefont {Wang}}, \ and\ \bibinfo {author} {\bibfnamefont
  {K.}~\bibnamefont {Yanagi}},\ }\href {\doibase 10.1093/ptep/pts043}
  {\bibfield  {journal} {\bibinfo  {journal} {Progress of Theoretical and
  Experimental Physics}\ }\textbf {\bibinfo {volume} {2012}},\ \bibinfo {pages}
  {03C008} (\bibinfo {year} {2012})}\BibitemShut {NoStop}%
\bibitem [{\citenamefont {Garrido}\ and\ \citenamefont {Moya~de
  Guerra}(1999)}]{Garrido:1999zy}%
  \BibitemOpen
  \bibfield  {author} {\bibinfo {author} {\bibfnamefont {E.}~\bibnamefont
  {Garrido}}\ and\ \bibinfo {author} {\bibfnamefont {E.}~\bibnamefont {Moya~de
  Guerra}},\ }\href {\doibase 10.1016/S0375-9474(99)00116-5} {\bibfield
  {journal} {\bibinfo  {journal} {Nucl. Phys. A}\ }\textbf {\bibinfo {volume}
  {650}},\ \bibinfo {pages} {387} (\bibinfo {year} {1999})}\BibitemShut
  {NoStop}%
\bibitem [{\citenamefont {Garrido}\ and\ \citenamefont {Moya~de
  Guerra}(2000)}]{Garrido:2000ht}%
  \BibitemOpen
  \bibfield  {author} {\bibinfo {author} {\bibfnamefont {E.}~\bibnamefont
  {Garrido}}\ and\ \bibinfo {author} {\bibfnamefont {E.}~\bibnamefont {Moya~de
  Guerra}},\ }\href {\doibase 10.1016/S0370-2693(00)00852-2} {\bibfield
  {journal} {\bibinfo  {journal} {Phys. Lett. B}\ }\textbf {\bibinfo {volume}
  {488}},\ \bibinfo {pages} {68} (\bibinfo {year} {2000})}\BibitemShut
  {NoStop}%
\bibitem [{\citenamefont {Ershov}\ \emph {et~al.}(2005)\citenamefont {Ershov},
  \citenamefont {Danilin},\ and\ \citenamefont {Vaagen}}]{Ershov:2005kq}%
  \BibitemOpen
  \bibfield  {author} {\bibinfo {author} {\bibfnamefont {S.~N.}\ \bibnamefont
  {Ershov}}, \bibinfo {author} {\bibfnamefont {B.~V.}\ \bibnamefont {Danilin}},
  \ and\ \bibinfo {author} {\bibfnamefont {J.~S.}\ \bibnamefont {Vaagen}},\
  }\href {\doibase 10.1103/PhysRevC.72.044606} {\bibfield  {journal} {\bibinfo
  {journal} {Phys. Rev. C}\ }\textbf {\bibinfo {volume} {72}},\ \bibinfo
  {pages} {044606} (\bibinfo {year} {2005})}\BibitemShut {NoStop}%
\bibitem [{\citenamefont {Wang}\ and\ \citenamefont
  {Ren}(2004)}]{PhysRevC.70.034303}%
  \BibitemOpen
  \bibfield  {author} {\bibinfo {author} {\bibfnamefont {Z.}~\bibnamefont
  {Wang}}\ and\ \bibinfo {author} {\bibfnamefont {Z.}~\bibnamefont {Ren}},\
  }\href {\doibase 10.1103/PhysRevC.70.034303} {\bibfield  {journal} {\bibinfo
  {journal} {Phys. Rev. C}\ }\textbf {\bibinfo {volume} {70}},\ \bibinfo
  {pages} {034303} (\bibinfo {year} {2004})}\BibitemShut {NoStop}%
\bibitem [{\citenamefont {Antonov}\ \emph {et~al.}(2004)\citenamefont
  {Antonov}, \citenamefont {Gaidarov}, \citenamefont {Kadrev}, \citenamefont
  {Hodgson},\ and\ \citenamefont {Moya~de Guerra}}]{Antonov04}%
  \BibitemOpen
  \bibfield  {author} {\bibinfo {author} {\bibfnamefont {A.~N.}\ \bibnamefont
  {Antonov}}, \bibinfo {author} {\bibfnamefont {M.~K.}\ \bibnamefont
  {Gaidarov}}, \bibinfo {author} {\bibfnamefont {D.~N.}\ \bibnamefont
  {Kadrev}}, \bibinfo {author} {\bibfnamefont {P.~E.}\ \bibnamefont {Hodgson}},
  \ and\ \bibinfo {author} {\bibfnamefont {E.}~\bibnamefont {Moya~de Guerra}},\
  }\href {\doibase 10.1142/S0218301304002430} {\bibfield  {journal} {\bibinfo
  {journal} {Int. J. Mod. Phys. E}\ }\textbf {\bibinfo {volume} {13}},\
  \bibinfo {pages} {759} (\bibinfo {year} {2004})}\BibitemShut {NoStop}%
\bibitem [{\citenamefont {Antonov}\ \emph {et~al.}(2005)\citenamefont
  {Antonov}, \citenamefont {Kadrev}, \citenamefont {Gaidarov}, \citenamefont
  {Moya~de Guerra}, \citenamefont {Sarriguren}, \citenamefont {Ud\'{\i}as},
  \citenamefont {Lukyanov}, \citenamefont {Zemlyanaya},\ and\ \citenamefont
  {Krumova}}]{PhysRevC.72.044307}%
  \BibitemOpen
  \bibfield  {author} {\bibinfo {author} {\bibfnamefont {A.~N.}\ \bibnamefont
  {Antonov}}, \bibinfo {author} {\bibfnamefont {D.~N.}\ \bibnamefont {Kadrev}},
  \bibinfo {author} {\bibfnamefont {M.~K.}\ \bibnamefont {Gaidarov}}, \bibinfo
  {author} {\bibfnamefont {E.}~\bibnamefont {Moya~de Guerra}}, \bibinfo
  {author} {\bibfnamefont {P.}~\bibnamefont {Sarriguren}}, \bibinfo {author}
  {\bibfnamefont {J.~M.}\ \bibnamefont {Ud\'{\i}as}}, \bibinfo {author}
  {\bibfnamefont {V.~K.}\ \bibnamefont {Lukyanov}}, \bibinfo {author}
  {\bibfnamefont {E.~V.}\ \bibnamefont {Zemlyanaya}}, \ and\ \bibinfo {author}
  {\bibfnamefont {G.~Z.}\ \bibnamefont {Krumova}},\ }\href {\doibase
  10.1103/PhysRevC.72.044307} {\bibfield  {journal} {\bibinfo  {journal} {Phys.
  Rev. C}\ }\textbf {\bibinfo {volume} {72}},\ \bibinfo {pages} {044307}
  (\bibinfo {year} {2005})}\BibitemShut {NoStop}%
\bibitem [{\citenamefont {Bertulani}(2007)}]{PhysRevC.75.024606}%
  \BibitemOpen
  \bibfield  {author} {\bibinfo {author} {\bibfnamefont {C.~A.}\ \bibnamefont
  {Bertulani}},\ }\href {\doibase 10.1103/PhysRevC.75.024606} {\bibfield
  {journal} {\bibinfo  {journal} {Phys. Rev. C}\ }\textbf {\bibinfo {volume}
  {75}},\ \bibinfo {pages} {024606} (\bibinfo {year} {2007})}\BibitemShut
  {NoStop}%
\bibitem [{\citenamefont {Khan}\ \emph {et~al.}(2008)\citenamefont {Khan},
  \citenamefont {Grasso}, \citenamefont {Margueron},\ and\ \citenamefont
  {Van~Giai}}]{Khan:2007ji}%
  \BibitemOpen
  \bibfield  {author} {\bibinfo {author} {\bibfnamefont {E.}~\bibnamefont
  {Khan}}, \bibinfo {author} {\bibfnamefont {M.}~\bibnamefont {Grasso}},
  \bibinfo {author} {\bibfnamefont {J.}~\bibnamefont {Margueron}}, \ and\
  \bibinfo {author} {\bibfnamefont {N.}~\bibnamefont {Van~Giai}},\ }\href
  {\doibase 10.1016/j.nuclphysa.2007.11.012} {\bibfield  {journal} {\bibinfo
  {journal} {Nucl. Phys. A}\ }\textbf {\bibinfo {volume} {800}},\ \bibinfo
  {pages} {37} (\bibinfo {year} {2008})}\BibitemShut {NoStop}%
\bibitem [{\citenamefont {Grasso}\ \emph {et~al.}(2009)\citenamefont {Grasso},
  \citenamefont {Gaudefroy}, \citenamefont {Khan}, \citenamefont
  {Nik$\mathrm{\check{s}}$i\'c}, \citenamefont {Piekarewicz}, \citenamefont
  {Sorlin}, \citenamefont {Van~Giai},\ and\ \citenamefont
  {Vretenar}}]{PhysRevC.79.034318}%
  \BibitemOpen
  \bibfield  {author} {\bibinfo {author} {\bibfnamefont {M.}~\bibnamefont
  {Grasso}}, \bibinfo {author} {\bibfnamefont {L.}~\bibnamefont {Gaudefroy}},
  \bibinfo {author} {\bibfnamefont {E.}~\bibnamefont {Khan}}, \bibinfo {author}
  {\bibfnamefont {T.}~\bibnamefont {Nik$\mathrm{\check{s}}$i\'c}}, \bibinfo
  {author} {\bibfnamefont {J.}~\bibnamefont {Piekarewicz}}, \bibinfo {author}
  {\bibfnamefont {O.}~\bibnamefont {Sorlin}}, \bibinfo {author} {\bibfnamefont
  {N.}~\bibnamefont {Van~Giai}}, \ and\ \bibinfo {author} {\bibfnamefont
  {D.}~\bibnamefont {Vretenar}},\ }\href {\doibase 10.1103/PhysRevC.79.034318}
  {\bibfield  {journal} {\bibinfo  {journal} {Phys. Rev. C}\ }\textbf {\bibinfo
  {volume} {79}},\ \bibinfo {pages} {034318} (\bibinfo {year}
  {2009})}\BibitemShut {NoStop}%
\bibitem [{\citenamefont {Chu}\ \emph {et~al.}(2009)\citenamefont {Chu},
  \citenamefont {Ren}, \citenamefont {Dong},\ and\ \citenamefont
  {Wang}}]{PhysRevC.79.044313}%
  \BibitemOpen
  \bibfield  {author} {\bibinfo {author} {\bibfnamefont {Y.}~\bibnamefont
  {Chu}}, \bibinfo {author} {\bibfnamefont {Z.}~\bibnamefont {Ren}}, \bibinfo
  {author} {\bibfnamefont {T.}~\bibnamefont {Dong}}, \ and\ \bibinfo {author}
  {\bibfnamefont {Z.~W.}\ \bibnamefont {Wang}},\ }\href {\doibase
  10.1103/PhysRevC.79.044313} {\bibfield  {journal} {\bibinfo  {journal} {Phys.
  Rev. C}\ }\textbf {\bibinfo {volume} {79}},\ \bibinfo {pages} {044313}
  (\bibinfo {year} {2009})}\BibitemShut {NoStop}%
\bibitem [{\citenamefont {Roca-Maza}\ \emph {et~al.}(2008)\citenamefont
  {Roca-Maza}, \citenamefont {Centelles}, \citenamefont {Salvat},\ and\
  \citenamefont {Vi\~nas}}]{RocaMaza:2008cg}%
  \BibitemOpen
  \bibfield  {author} {\bibinfo {author} {\bibfnamefont {X.}~\bibnamefont
  {Roca-Maza}}, \bibinfo {author} {\bibfnamefont {M.}~\bibnamefont
  {Centelles}}, \bibinfo {author} {\bibfnamefont {F.}~\bibnamefont {Salvat}}, \
  and\ \bibinfo {author} {\bibfnamefont {X.}~\bibnamefont {Vi\~nas}},\ }\href
  {\doibase 10.1103/PhysRevC.78.044332} {\bibfield  {journal} {\bibinfo
  {journal} {Phys. Rev. C}\ }\textbf {\bibinfo {volume} {78}},\ \bibinfo
  {pages} {044332} (\bibinfo {year} {2008})}\BibitemShut {NoStop}%
\bibitem [{\citenamefont {Roca-Maza}\ \emph {et~al.}(2013)\citenamefont
  {Roca-Maza}, \citenamefont {Centelles}, \citenamefont {Salvat},\ and\
  \citenamefont {Vi\~nas}}]{RocaMaza:2012hv}%
  \BibitemOpen
  \bibfield  {author} {\bibinfo {author} {\bibfnamefont {X.}~\bibnamefont
  {Roca-Maza}}, \bibinfo {author} {\bibfnamefont {M.}~\bibnamefont
  {Centelles}}, \bibinfo {author} {\bibfnamefont {F.}~\bibnamefont {Salvat}}, \
  and\ \bibinfo {author} {\bibfnamefont {X.}~\bibnamefont {Vi\~nas}},\ }\href
  {\doibase 10.1103/PhysRevC.87.014304} {\bibfield  {journal} {\bibinfo
  {journal} {Phys. Rev. C}\ }\textbf {\bibinfo {volume} {87}},\ \bibinfo
  {pages} {014304} (\bibinfo {year} {2013})}\BibitemShut {NoStop}%
\bibitem [{\citenamefont {Dong}\ \emph {et~al.}(2009)\citenamefont {Dong},
  \citenamefont {Chu}, \citenamefont {Ren},\ and\ \citenamefont
  {Wang}}]{PhysRevC.79.014317}%
  \BibitemOpen
  \bibfield  {author} {\bibinfo {author} {\bibfnamefont {T.}~\bibnamefont
  {Dong}}, \bibinfo {author} {\bibfnamefont {Y.}~\bibnamefont {Chu}}, \bibinfo
  {author} {\bibfnamefont {Z.}~\bibnamefont {Ren}}, \ and\ \bibinfo {author}
  {\bibfnamefont {Z.}~\bibnamefont {Wang}},\ }\href {\doibase
  10.1103/PhysRevC.79.014317} {\bibfield  {journal} {\bibinfo  {journal} {Phys.
  Rev. C}\ }\textbf {\bibinfo {volume} {79}},\ \bibinfo {pages} {014317}
  (\bibinfo {year} {2009})}\BibitemShut {NoStop}%
\bibitem [{\citenamefont {Dong}\ \emph {et~al.}(2008)\citenamefont {Dong},
  \citenamefont {Ren},\ and\ \citenamefont {Wang}}]{PhysRevC.77.064302}%
  \BibitemOpen
  \bibfield  {author} {\bibinfo {author} {\bibfnamefont {T.}~\bibnamefont
  {Dong}}, \bibinfo {author} {\bibfnamefont {Z.}~\bibnamefont {Ren}}, \ and\
  \bibinfo {author} {\bibfnamefont {Z.}~\bibnamefont {Wang}},\ }\href {\doibase
  10.1103/PhysRevC.77.064302} {\bibfield  {journal} {\bibinfo  {journal} {Phys.
  Rev. C}\ }\textbf {\bibinfo {volume} {77}},\ \bibinfo {pages} {064302}
  (\bibinfo {year} {2008})}\BibitemShut {NoStop}%
\bibitem [{\citenamefont {Liu}\ \emph {et~al.}(2012)\citenamefont {Liu},
  \citenamefont {Ren},\ and\ \citenamefont {Dong}}]{Liu:2012zj}%
  \BibitemOpen
  \bibfield  {author} {\bibinfo {author} {\bibfnamefont {J.}~\bibnamefont
  {Liu}}, \bibinfo {author} {\bibfnamefont {Z.}~\bibnamefont {Ren}}, \ and\
  \bibinfo {author} {\bibfnamefont {T.}~\bibnamefont {Dong}},\ }\href {\doibase
  10.1016/j.nuclphysa.2012.06.002} {\bibfield  {journal} {\bibinfo  {journal}
  {Nucl. Phys. A}\ }\textbf {\bibinfo {volume} {888}},\ \bibinfo {pages} {45}
  (\bibinfo {year} {2012})}\BibitemShut {NoStop}%
\bibitem [{\citenamefont {Dong}\ \emph {et~al.}(2012)\citenamefont {Dong},
  \citenamefont {Chu},\ and\ \citenamefont {Ren}}]{Dong:2012ed}%
  \BibitemOpen
  \bibfield  {author} {\bibinfo {author} {\bibfnamefont {T.}~\bibnamefont
  {Dong}}, \bibinfo {author} {\bibfnamefont {Y.}~\bibnamefont {Chu}}, \ and\
  \bibinfo {author} {\bibfnamefont {Z.}~\bibnamefont {Ren}},\ }\href {\doibase
  10.1088/1742-6596/381/1/012135} {\bibfield  {journal} {\bibinfo  {journal}
  {J. Phys. Conf. Ser.}\ }\textbf {\bibinfo {volume} {381}},\ \bibinfo {pages}
  {012135} (\bibinfo {year} {2012})}\BibitemShut {NoStop}%
\bibitem [{\citenamefont {Meucci}\ \emph
  {et~al.}(2013{\natexlab{a}})\citenamefont {Meucci}, \citenamefont {Vorabbi},
  \citenamefont {Giusti}, \citenamefont {Pacati},\ and\ \citenamefont
  {Finelli}}]{esotici2}%
  \BibitemOpen
  \bibfield  {author} {\bibinfo {author} {\bibfnamefont {A.}~\bibnamefont
  {Meucci}}, \bibinfo {author} {\bibfnamefont {M.}~\bibnamefont {Vorabbi}},
  \bibinfo {author} {\bibfnamefont {C.}~\bibnamefont {Giusti}}, \bibinfo
  {author} {\bibfnamefont {F.~D.}\ \bibnamefont {Pacati}}, \ and\ \bibinfo
  {author} {\bibfnamefont {P.}~\bibnamefont {Finelli}},\ }\href {\doibase
  10.1103/PhysRevC.87.054620} {\bibfield  {journal} {\bibinfo  {journal} {Phys.
  Rev. C}\ }\textbf {\bibinfo {volume} {87}},\ \bibinfo {pages} {054620}
  (\bibinfo {year} {2013}{\natexlab{a}})}\BibitemShut {NoStop}%
\bibitem [{\citenamefont {Donnelly}\ \emph {et~al.}(1989)\citenamefont
  {Donnelly}, \citenamefont {Dubach},\ and\ \citenamefont
  {Sick}}]{Donnelly1989589}%
  \BibitemOpen
  \bibfield  {author} {\bibinfo {author} {\bibfnamefont {T.~W.}\ \bibnamefont
  {Donnelly}}, \bibinfo {author} {\bibfnamefont {J.}~\bibnamefont {Dubach}}, \
  and\ \bibinfo {author} {\bibfnamefont {I.}~\bibnamefont {Sick}},\ }\href
  {\doibase 10.1016/0375-9474(89)90432-6} {\bibfield  {journal} {\bibinfo
  {journal} {Nuclear Physics A}\ }\textbf {\bibinfo {volume} {503}},\ \bibinfo
  {pages} {589 } (\bibinfo {year} {1989})}\BibitemShut {NoStop}%
\bibitem [{\citenamefont {Donnelly}\ and\ \citenamefont
  {Peccei}(1979)}]{Donnelly19791}%
  \BibitemOpen
  \bibfield  {author} {\bibinfo {author} {\bibfnamefont {T.~W.}\ \bibnamefont
  {Donnelly}}\ and\ \bibinfo {author} {\bibfnamefont {R.~D.}\ \bibnamefont
  {Peccei}},\ }\href {\doibase 10.1016/0370-1573(79)90010-3} {\bibfield
  {journal} {\bibinfo  {journal} {Physics Reports}\ }\textbf {\bibinfo {volume}
  {50}},\ \bibinfo {pages} {1 } (\bibinfo {year} {1979})}\BibitemShut {NoStop}%
\bibitem [{\citenamefont {Horowitz}(1998)}]{PhysRevC.57.3430}%
  \BibitemOpen
  \bibfield  {author} {\bibinfo {author} {\bibfnamefont {C.~J.}\ \bibnamefont
  {Horowitz}},\ }\href {\doibase 10.1103/PhysRevC.57.3430} {\bibfield
  {journal} {\bibinfo  {journal} {Phys. Rev. C}\ }\textbf {\bibinfo {volume}
  {57}},\ \bibinfo {pages} {3430} (\bibinfo {year} {1998})}\BibitemShut
  {NoStop}%
\bibitem [{\citenamefont {Abrahamyan}\ \emph {et~al.}(2012)\citenamefont
  {Abrahamyan} \emph {et~al.}}]{Abrahamyan:2012gp}%
  \BibitemOpen
  \bibfield  {author} {\bibinfo {author} {\bibfnamefont {S.}~\bibnamefont
  {Abrahamyan}} \emph {et~al.} (\bibinfo {collaboration} {PREX
  Collaboration}),\ }\href {\doibase 10.1103/PhysRevLett.108.112502} {\bibfield
   {journal} {\bibinfo  {journal} {Phys. Rev. Lett.}\ }\textbf {\bibinfo
  {volume} {108}},\ \bibinfo {pages} {112502} (\bibinfo {year}
  {2012})}\BibitemShut {NoStop}%
\bibitem [{\citenamefont {Horowitz}\ \emph {et~al.}(2012)\citenamefont
  {Horowitz}, \citenamefont {Ahmed}, \citenamefont {Jen}, \citenamefont
  {Rakhman}, \citenamefont {Souder}, \citenamefont {Dalton}, \citenamefont
  {Liyanage}, \citenamefont {Paschke}, \citenamefont {Saenboonruang},
  \citenamefont {Silwal}, \citenamefont {Franklin}, \citenamefont {Friend},
  \citenamefont {Quinn}, \citenamefont {Kumar}, \citenamefont {McNulty},
  \citenamefont {Mercado}, \citenamefont {Riordan}, \citenamefont {Wexler},
  \citenamefont {Michaels},\ and\ \citenamefont
  {Urciuoli}}]{PhysRevC.85.032501}%
  \BibitemOpen
  \bibfield  {author} {\bibinfo {author} {\bibfnamefont {C.~J.}\ \bibnamefont
  {Horowitz}}, \bibinfo {author} {\bibfnamefont {Z.}~\bibnamefont {Ahmed}},
  \bibinfo {author} {\bibfnamefont {C.~M.}\ \bibnamefont {Jen}}, \bibinfo
  {author} {\bibfnamefont {A.}~\bibnamefont {Rakhman}}, \bibinfo {author}
  {\bibfnamefont {P.~A.}\ \bibnamefont {Souder}}, \bibinfo {author}
  {\bibfnamefont {M.~M.}\ \bibnamefont {Dalton}}, \bibinfo {author}
  {\bibfnamefont {N.}~\bibnamefont {Liyanage}}, \bibinfo {author}
  {\bibfnamefont {K.~D.}\ \bibnamefont {Paschke}}, \bibinfo {author}
  {\bibfnamefont {K.}~\bibnamefont {Saenboonruang}}, \bibinfo {author}
  {\bibfnamefont {R.}~\bibnamefont {Silwal}}, \bibinfo {author} {\bibfnamefont
  {G.~B.}\ \bibnamefont {Franklin}}, \bibinfo {author} {\bibfnamefont
  {M.}~\bibnamefont {Friend}}, \bibinfo {author} {\bibfnamefont
  {B.}~\bibnamefont {Quinn}}, \bibinfo {author} {\bibfnamefont {K.~S.}\
  \bibnamefont {Kumar}}, \bibinfo {author} {\bibfnamefont {D.}~\bibnamefont
  {McNulty}}, \bibinfo {author} {\bibfnamefont {L.}~\bibnamefont {Mercado}},
  \bibinfo {author} {\bibfnamefont {S.}~\bibnamefont {Riordan}}, \bibinfo
  {author} {\bibfnamefont {J.}~\bibnamefont {Wexler}}, \bibinfo {author}
  {\bibfnamefont {R.~W.}\ \bibnamefont {Michaels}}, \ and\ \bibinfo {author}
  {\bibfnamefont {G.~M.}\ \bibnamefont {Urciuoli}},\ }\href {\doibase
  10.1103/PhysRevC.85.032501} {\bibfield  {journal} {\bibinfo  {journal} {Phys.
  Rev. C}\ }\textbf {\bibinfo {volume} {85}},\ \bibinfo {pages} {032501}
  (\bibinfo {year} {2012})}\BibitemShut {NoStop}%
\bibitem [{PREX-II, Proposal to Jefferson Lab PAC 38()}]{prex2}%
  \BibitemOpen
  PREX-II, Proposal to Jefferson Lab PAC 38,\ \href@noop {} {}\bibinfo
  {howpublished}
  {\url{http://hallaweb.jlab.org/parity/prex/prexII.pdf}}\BibitemShut {NoStop}%
\bibitem [{\citenamefont {Tarbert}\ \emph {et~al.}(2013)\citenamefont {Tarbert}
  \emph {et~al.}}]{mainz-neutronskin}%
  \BibitemOpen
  \bibfield  {author} {\bibinfo {author} {\bibfnamefont {C.~M.}\ \bibnamefont
  {Tarbert}} \emph {et~al.},\ }\href@noop {} {\  (\bibinfo {year} {2013})},\
  \Eprint {http://arxiv.org/abs/1311.0168} {arXiv:1311.0168 [nucl-ex]}
  \BibitemShut {NoStop}%
\bibitem [{\lq\lq$\rm{CREX}$: parity-violating measurements of the weak charge
  distribution of $^{48}\rm{Ca}$ to 0.03 fm accuracy\rq\rq, Proposal to
  Jefferson Lab PAC 39()}]{crex}%
  \BibitemOpen
  \lq\lq$\rm{CREX}$: parity-violating measurements of the weak charge
  distribution of $^{48}\rm{Ca}$ to 0.03 fm accuracy\rq\rq, Proposal to
  Jefferson Lab PAC 39,\ \href@noop {} {}\bibinfo {howpublished}
  {\url{http://hallaweb.jlab.org/parity/prex}}\BibitemShut {NoStop}%
\bibitem [{\citenamefont {Sorlin}\ and\ \citenamefont
  {Porquet}(2013)}]{sorlin}%
  \BibitemOpen
  \bibfield  {author} {\bibinfo {author} {\bibfnamefont {O.}~\bibnamefont
  {Sorlin}}\ and\ \bibinfo {author} {\bibfnamefont {M.-G.}\ \bibnamefont
  {Porquet}},\ }\href {\doibase 10.10088/0031-8949/2013/T152/014003} {\bibfield
   {journal} {\bibinfo  {journal} {Phys. Scr.}\ }\textbf {\bibinfo {volume}
  {T152}},\ \bibinfo {pages} {014003} (\bibinfo {year} {2013})}\BibitemShut
  {NoStop}%
\bibitem [{\citenamefont {Serot}\ and\ \citenamefont
  {Walecka}(1986)}]{Serot:1984ey}%
  \BibitemOpen
  \bibfield  {author} {\bibinfo {author} {\bibfnamefont {B.~D.}\ \bibnamefont
  {Serot}}\ and\ \bibinfo {author} {\bibfnamefont {J.~D.}\ \bibnamefont
  {Walecka}},\ }\href@noop {} {\bibfield  {journal} {\bibinfo  {journal} {Adv.
  Nucl. Phys.}\ }\textbf {\bibinfo {volume} {16}},\ \bibinfo {pages} {1}
  (\bibinfo {year} {1986})}\BibitemShut {NoStop}%
\bibitem [{\citenamefont {Reinhard}(1989)}]{Rein:1989}%
  \BibitemOpen
  \bibfield  {author} {\bibinfo {author} {\bibfnamefont {P.~G.}\ \bibnamefont
  {Reinhard}},\ }\href@noop {} {\bibfield  {journal} {\bibinfo  {journal} {Rep.
  Prog. Phys.}\ }\textbf {\bibinfo {volume} {52}},\ \bibinfo {pages} {439}
  (\bibinfo {year} {1989})}\BibitemShut {NoStop}%
\bibitem [{\citenamefont {Ring}(1996)}]{Rign:1996}%
  \BibitemOpen
  \bibfield  {author} {\bibinfo {author} {\bibfnamefont {P.}~\bibnamefont
  {Ring}},\ }\href@noop {} {\bibfield  {journal} {\bibinfo  {journal} {Prog.
  Part. Nucl. Phys.}\ }\textbf {\bibinfo {volume} {37}},\ \bibinfo {pages}
  {193} (\bibinfo {year} {1996})}\BibitemShut {NoStop}%
\bibitem [{\citenamefont {Serot}\ and\ \citenamefont
  {Walecka}(1997)}]{Serot:1997}%
  \BibitemOpen
  \bibfield  {author} {\bibinfo {author} {\bibfnamefont {B.~D.}\ \bibnamefont
  {Serot}}\ and\ \bibinfo {author} {\bibfnamefont {J.~D.}\ \bibnamefont
  {Walecka}},\ }\href@noop {} {\bibfield  {journal} {\bibinfo  {journal} {Int.
  J. Mod. Phys. E}\ }\textbf {\bibinfo {volume} {6}},\ \bibinfo {pages} {515}
  (\bibinfo {year} {1997})}\BibitemShut {NoStop}%
\bibitem [{\citenamefont {Vretenar}\ \emph {et~al.}(2005)\citenamefont
  {Vretenar}, \citenamefont {Afanasjev}, \citenamefont {Lalazissis},\ and\
  \citenamefont {Ring}}]{Vretenar:2005zz}%
  \BibitemOpen
  \bibfield  {author} {\bibinfo {author} {\bibfnamefont {D.}~\bibnamefont
  {Vretenar}}, \bibinfo {author} {\bibfnamefont {A.}~\bibnamefont {Afanasjev}},
  \bibinfo {author} {\bibfnamefont {G.}~\bibnamefont {Lalazissis}}, \ and\
  \bibinfo {author} {\bibfnamefont {P.}~\bibnamefont {Ring}},\ }\href {\doibase
  10.1016/j.physrep.2004.10.001} {\bibfield  {journal} {\bibinfo  {journal}
  {Phys. Rept.}\ }\textbf {\bibinfo {volume} {409}},\ \bibinfo {pages} {101}
  (\bibinfo {year} {2005})}\BibitemShut {NoStop}%
\bibitem [{\citenamefont {Meucci}\ \emph {et~al.}(2003)\citenamefont {Meucci},
  \citenamefont {Capuzzi}, \citenamefont {Giusti},\ and\ \citenamefont
  {Pacati}}]{Meucci:2003uy}%
  \BibitemOpen
  \bibfield  {author} {\bibinfo {author} {\bibfnamefont {A.}~\bibnamefont
  {Meucci}}, \bibinfo {author} {\bibfnamefont {F.}~\bibnamefont {Capuzzi}},
  \bibinfo {author} {\bibfnamefont {C.}~\bibnamefont {Giusti}}, \ and\ \bibinfo
  {author} {\bibfnamefont {F.~D.}\ \bibnamefont {Pacati}},\ }\href {\doibase
  10.1103/PhysRevC.67.054601} {\bibfield  {journal} {\bibinfo  {journal} {Phys.
  Rev. C}\ }\textbf {\bibinfo {volume} {67}},\ \bibinfo {pages} {054601}
  (\bibinfo {year} {2003})}\BibitemShut {NoStop}%
\bibitem [{\citenamefont {Meucci}\ \emph {et~al.}(2004)\citenamefont {Meucci},
  \citenamefont {Giusti},\ and\ \citenamefont {Pacati}}]{Meucci:2003cv}%
  \BibitemOpen
  \bibfield  {author} {\bibinfo {author} {\bibfnamefont {A.}~\bibnamefont
  {Meucci}}, \bibinfo {author} {\bibfnamefont {C.}~\bibnamefont {Giusti}}, \
  and\ \bibinfo {author} {\bibfnamefont {F.~D.}\ \bibnamefont {Pacati}},\
  }\href {\doibase 10.1016/j.nuclphysa.2004.04.108} {\bibfield  {journal}
  {\bibinfo  {journal} {Nuclear Physics A}\ }\textbf {\bibinfo {volume}
  {739}},\ \bibinfo {pages} {277} (\bibinfo {year} {2004})}\BibitemShut
  {NoStop}%
\bibitem [{\citenamefont {Meucci}\ \emph {et~al.}(2005)\citenamefont {Meucci},
  \citenamefont {Giusti},\ and\ \citenamefont {Pacati}}]{Meucci:2005pk}%
  \BibitemOpen
  \bibfield  {author} {\bibinfo {author} {\bibfnamefont {A.}~\bibnamefont
  {Meucci}}, \bibinfo {author} {\bibfnamefont {C.}~\bibnamefont {Giusti}}, \
  and\ \bibinfo {author} {\bibfnamefont {F.~D.}\ \bibnamefont {Pacati}},\
  }\href {\doibase 10.1016/j.nuclphysa.2005.04.007} {\bibfield  {journal}
  {\bibinfo  {journal} {Nuclear Physics A}\ }\textbf {\bibinfo {volume}
  {756}},\ \bibinfo {pages} {359} (\bibinfo {year} {2005})}\BibitemShut
  {NoStop}%
\bibitem [{\citenamefont {Meucci}\ \emph {et~al.}(2009)\citenamefont {Meucci},
  \citenamefont {Caballero}, \citenamefont {Giusti}, \citenamefont {Pacati},\
  and\ \citenamefont {Ud\'{\i}as}}]{Meucci:2009nm}%
  \BibitemOpen
  \bibfield  {author} {\bibinfo {author} {\bibfnamefont {A.}~\bibnamefont
  {Meucci}}, \bibinfo {author} {\bibfnamefont {J.~A.}\ \bibnamefont
  {Caballero}}, \bibinfo {author} {\bibfnamefont {C.}~\bibnamefont {Giusti}},
  \bibinfo {author} {\bibfnamefont {F.~D.}\ \bibnamefont {Pacati}}, \ and\
  \bibinfo {author} {\bibfnamefont {J.~M.}\ \bibnamefont {Ud\'{\i}as}},\ }\href
  {\doibase 10.1103/PhysRevC.80.024605} {\bibfield  {journal} {\bibinfo
  {journal} {Phys. Rev. C}\ }\textbf {\bibinfo {volume} {80}},\ \bibinfo
  {pages} {024605} (\bibinfo {year} {2009})}\BibitemShut {NoStop}%
\bibitem [{\citenamefont {Meucci}\ \emph
  {et~al.}(2011{\natexlab{a}})\citenamefont {Meucci}, \citenamefont
  {Caballero}, \citenamefont {Giusti},\ and\ \citenamefont
  {Ud\'{\i}as}}]{Meucci:2011pi}%
  \BibitemOpen
  \bibfield  {author} {\bibinfo {author} {\bibfnamefont {A.}~\bibnamefont
  {Meucci}}, \bibinfo {author} {\bibfnamefont {J.~A.}\ \bibnamefont
  {Caballero}}, \bibinfo {author} {\bibfnamefont {C.}~\bibnamefont {Giusti}}, \
  and\ \bibinfo {author} {\bibfnamefont {J.~M.}\ \bibnamefont {Ud\'{\i}as}},\
  }\href {\doibase 10.1103/PhysRevC.83.064614} {\bibfield  {journal} {\bibinfo
  {journal} {Phys. Rev. C}\ }\textbf {\bibinfo {volume} {83}},\ \bibinfo
  {pages} {064614} (\bibinfo {year} {2011}{\natexlab{a}})}\BibitemShut
  {NoStop}%
\bibitem [{\citenamefont {Meucci}\ and\ \citenamefont
  {Giusti}(2012)}]{Meucci:ant}%
  \BibitemOpen
  \bibfield  {author} {\bibinfo {author} {\bibfnamefont {A.}~\bibnamefont
  {Meucci}}\ and\ \bibinfo {author} {\bibfnamefont {C.}~\bibnamefont
  {Giusti}},\ }\href {\doibase 10.1103/PhysRevD.85.093002} {\bibfield
  {journal} {\bibinfo  {journal} {Phys. Rev. D}\ }\textbf {\bibinfo {volume}
  {85}},\ \bibinfo {pages} {093002} (\bibinfo {year} {2012})}\BibitemShut
  {NoStop}%
\bibitem [{\citenamefont {Meucci}\ \emph
  {et~al.}(2011{\natexlab{b}})\citenamefont {Meucci}, \citenamefont {Giusti},\
  and\ \citenamefont {Pacati}}]{Meucci:2011nc}%
  \BibitemOpen
  \bibfield  {author} {\bibinfo {author} {\bibfnamefont {A.}~\bibnamefont
  {Meucci}}, \bibinfo {author} {\bibfnamefont {C.}~\bibnamefont {Giusti}}, \
  and\ \bibinfo {author} {\bibfnamefont {F.~D.}\ \bibnamefont {Pacati}},\
  }\href {\doibase 10.1103/PhysRevD.84.113003} {\bibfield  {journal} {\bibinfo
  {journal} {Phys. Rev. D}\ }\textbf {\bibinfo {volume} {84}},\ \bibinfo
  {pages} {113003} (\bibinfo {year} {2011}{\natexlab{b}})}\BibitemShut
  {NoStop}%
\bibitem [{\citenamefont {Meucci}\ \emph
  {et~al.}(2011{\natexlab{c}})\citenamefont {Meucci}, \citenamefont {Barbaro},
  \citenamefont {Caballero}, \citenamefont {Giusti},\ and\ \citenamefont
  {Ud\'{\i}as}}]{Meucci:2011vd}%
  \BibitemOpen
  \bibfield  {author} {\bibinfo {author} {\bibfnamefont {A.}~\bibnamefont
  {Meucci}}, \bibinfo {author} {\bibfnamefont {M.~B.}\ \bibnamefont {Barbaro}},
  \bibinfo {author} {\bibfnamefont {J.~A.}\ \bibnamefont {Caballero}}, \bibinfo
  {author} {\bibfnamefont {C.}~\bibnamefont {Giusti}}, \ and\ \bibinfo {author}
  {\bibfnamefont {J.~M.}\ \bibnamefont {Ud\'{\i}as}},\ }\href {\doibase
  10.1103/PhysRevLett.107.172501} {\bibfield  {journal} {\bibinfo  {journal}
  {Phys. Rev. Lett.}\ }\textbf {\bibinfo {volume} {107}},\ \bibinfo {pages}
  {172501} (\bibinfo {year} {2011}{\natexlab{c}})}\BibitemShut {NoStop}%
\bibitem [{\citenamefont {Finelli}\ \emph {et~al.}(2006)\citenamefont
  {Finelli}, \citenamefont {Kaiser}, \citenamefont {Vretenar},\ and\
  \citenamefont {Weise}}]{Finelli:2005ni}%
  \BibitemOpen
  \bibfield  {author} {\bibinfo {author} {\bibfnamefont {P.}~\bibnamefont
  {Finelli}}, \bibinfo {author} {\bibfnamefont {N.}~\bibnamefont {Kaiser}},
  \bibinfo {author} {\bibfnamefont {D.}~\bibnamefont {Vretenar}}, \ and\
  \bibinfo {author} {\bibfnamefont {W.}~\bibnamefont {Weise}},\ }\href
  {\doibase 10.1016/j.nuclphysa.2006.02.007} {\bibfield  {journal} {\bibinfo
  {journal} {Nucl. Phys. A}\ }\textbf {\bibinfo {volume} {770}},\ \bibinfo
  {pages} {1} (\bibinfo {year} {2006})}\BibitemShut {NoStop}%
\bibitem [{\citenamefont {Bogner}\ \emph {et~al.}(2010)\citenamefont {Bogner},
  \citenamefont {Furnstahl},\ and\ \citenamefont {Schwenk}}]{Bogner:2009bt}%
  \BibitemOpen
  \bibfield  {author} {\bibinfo {author} {\bibfnamefont {S.}~\bibnamefont
  {Bogner}}, \bibinfo {author} {\bibfnamefont {R.}~\bibnamefont {Furnstahl}}, \
  and\ \bibinfo {author} {\bibfnamefont {A.}~\bibnamefont {Schwenk}},\ }\href
  {\doibase 10.1016/j.ppnp.2010.03.001} {\bibfield  {journal} {\bibinfo
  {journal} {Prog. Part. Nucl. Phys.}\ }\textbf {\bibinfo {volume} {65}},\
  \bibinfo {pages} {94} (\bibinfo {year} {2010})}\BibitemShut {NoStop}%
\bibitem [{\citenamefont {Nik$\mathrm{\check{s}}$i\'c}\ \emph
  {et~al.}(2002)\citenamefont {Nik$\mathrm{\check{s}}$i\'c}, \citenamefont
  {Vretenar}, \citenamefont {Finelli},\ and\ \citenamefont
  {Ring}}]{PhysRevC.66.024306}%
  \BibitemOpen
  \bibfield  {author} {\bibinfo {author} {\bibfnamefont {T.}~\bibnamefont
  {Nik$\mathrm{\check{s}}$i\'c}}, \bibinfo {author} {\bibfnamefont
  {D.}~\bibnamefont {Vretenar}}, \bibinfo {author} {\bibfnamefont
  {P.}~\bibnamefont {Finelli}}, \ and\ \bibinfo {author} {\bibfnamefont
  {P.}~\bibnamefont {Ring}},\ }\href {\doibase 10.1103/PhysRevC.66.024306}
  {\bibfield  {journal} {\bibinfo  {journal} {Phys. Rev. C}\ }\textbf {\bibinfo
  {volume} {66}},\ \bibinfo {pages} {024306} (\bibinfo {year}
  {2002})}\BibitemShut {NoStop}%
\bibitem [{\citenamefont {Decharge}\ and\ \citenamefont
  {Gogny}(1980)}]{Decharge:1979fa}%
  \BibitemOpen
  \bibfield  {author} {\bibinfo {author} {\bibfnamefont {J.}~\bibnamefont
  {Decharge}}\ and\ \bibinfo {author} {\bibfnamefont {D.}~\bibnamefont
  {Gogny}},\ }\href {\doibase 10.1103/PhysRevC.21.1568} {\bibfield  {journal}
  {\bibinfo  {journal} {Phys. Rev. C}\ }\textbf {\bibinfo {volume} {21}},\
  \bibinfo {pages} {1568} (\bibinfo {year} {1980})}\BibitemShut {NoStop}%
\bibitem [{\citenamefont {Berger}\ \emph {et~al.}(1984)\citenamefont {Berger},
  \citenamefont {Girod},\ and\ \citenamefont {Gogny}}]{Berger:1984zz}%
  \BibitemOpen
  \bibfield  {author} {\bibinfo {author} {\bibfnamefont {J.}~\bibnamefont
  {Berger}}, \bibinfo {author} {\bibfnamefont {M.}~\bibnamefont {Girod}}, \
  and\ \bibinfo {author} {\bibfnamefont {D.}~\bibnamefont {Gogny}},\ }\href
  {\doibase 10.1016/0375-9474(84)90240-9} {\bibfield  {journal} {\bibinfo
  {journal} {Nucl. Phys. A}\ }\textbf {\bibinfo {volume} {428}},\ \bibinfo
  {pages} {23} (\bibinfo {year} {1984})}\BibitemShut {NoStop}%
\bibitem [{\citenamefont {Serra}\ \emph {et~al.}(2001)\citenamefont {Serra},
  \citenamefont {Rummel},\ and\ \citenamefont {Ring}}]{Serra:2001pv}%
  \BibitemOpen
  \bibfield  {author} {\bibinfo {author} {\bibfnamefont {M.}~\bibnamefont
  {Serra}}, \bibinfo {author} {\bibfnamefont {A.}~\bibnamefont {Rummel}}, \
  and\ \bibinfo {author} {\bibfnamefont {P.}~\bibnamefont {Ring}},\ }\href
  {\doibase 10.1103/PhysRevC.65.014304} {\bibfield  {journal} {\bibinfo
  {journal} {Phys. Rev. C}\ }\textbf {\bibinfo {volume} {65}},\ \bibinfo
  {pages} {014304} (\bibinfo {year} {2001})}\BibitemShut {NoStop}%
\bibitem [{\citenamefont {Frosch}\ \emph {et~al.}(1968)\citenamefont {Frosch},
  \citenamefont {Hofstadter}, \citenamefont {McCarthy}, \citenamefont
  {N\"oldeke}, \citenamefont {van Oostrum}, \citenamefont {Yearian},
  \citenamefont {Clark}, \citenamefont {Herman},\ and\ \citenamefont
  {Ravenhall}}]{PhysRev.174.1380}%
  \BibitemOpen
  \bibfield  {author} {\bibinfo {author} {\bibfnamefont {R.~F.}\ \bibnamefont
  {Frosch}}, \bibinfo {author} {\bibfnamefont {R.}~\bibnamefont {Hofstadter}},
  \bibinfo {author} {\bibfnamefont {J.~S.}\ \bibnamefont {McCarthy}}, \bibinfo
  {author} {\bibfnamefont {G.~K.}\ \bibnamefont {N\"oldeke}}, \bibinfo {author}
  {\bibfnamefont {K.~J.}\ \bibnamefont {van Oostrum}}, \bibinfo {author}
  {\bibfnamefont {M.~R.}\ \bibnamefont {Yearian}}, \bibinfo {author}
  {\bibfnamefont {B.~C.}\ \bibnamefont {Clark}}, \bibinfo {author}
  {\bibfnamefont {R.}~\bibnamefont {Herman}}, \ and\ \bibinfo {author}
  {\bibfnamefont {D.~G.}\ \bibnamefont {Ravenhall}},\ }\href {\doibase
  10.1103/PhysRev.174.1380} {\bibfield  {journal} {\bibinfo  {journal} {Phys.
  Rev.}\ }\textbf {\bibinfo {volume} {174}},\ \bibinfo {pages} {1380} (\bibinfo
  {year} {1968})}\BibitemShut {NoStop}%
\bibitem [{\citenamefont {Lightbody}\ \emph {et~al.}(1983)\citenamefont
  {Lightbody}, \citenamefont {Bellicard}, \citenamefont {Cavedon},
  \citenamefont {Frois}, \citenamefont {Goutte}, \citenamefont {Huet},
  \citenamefont {Leconte}, \citenamefont {Nakada}, \citenamefont {Ho},
  \citenamefont {Platchkov}, \citenamefont {Turck-Chieze}, \citenamefont
  {de~Jager}, \citenamefont {Lapik\'as},\ and\ \citenamefont
  {de~Witt~Huberts}}]{PhysRevC.27.113}%
  \BibitemOpen
  \bibfield  {author} {\bibinfo {author} {\bibfnamefont {J.~W.}\ \bibnamefont
  {Lightbody}}, \bibinfo {author} {\bibfnamefont {J.~B.}\ \bibnamefont
  {Bellicard}}, \bibinfo {author} {\bibfnamefont {J.~M.}\ \bibnamefont
  {Cavedon}}, \bibinfo {author} {\bibfnamefont {B.}~\bibnamefont {Frois}},
  \bibinfo {author} {\bibfnamefont {D.}~\bibnamefont {Goutte}}, \bibinfo
  {author} {\bibfnamefont {M.}~\bibnamefont {Huet}}, \bibinfo {author}
  {\bibfnamefont {P.}~\bibnamefont {Leconte}}, \bibinfo {author} {\bibfnamefont
  {A.}~\bibnamefont {Nakada}}, \bibinfo {author} {\bibfnamefont {P.~X.}\
  \bibnamefont {Ho}}, \bibinfo {author} {\bibfnamefont {S.~K.}\ \bibnamefont
  {Platchkov}}, \bibinfo {author} {\bibfnamefont {S.}~\bibnamefont
  {Turck-Chieze}}, \bibinfo {author} {\bibfnamefont {C.~W.}\ \bibnamefont
  {de~Jager}}, \bibinfo {author} {\bibfnamefont {J.~J.}\ \bibnamefont
  {Lapik\'as}}, \ and\ \bibinfo {author} {\bibfnamefont {P.~K.~A.}\
  \bibnamefont {de~Witt~Huberts}},\ }\href {\doibase 10.1103/PhysRevC.27.113}
  {\bibfield  {journal} {\bibinfo  {journal} {Phys. Rev. C}\ }\textbf {\bibinfo
  {volume} {27}},\ \bibinfo {pages} {113} (\bibinfo {year} {1983})}\BibitemShut
  {NoStop}%
\bibitem [{\citenamefont {Vretenar}\ \emph {et~al.}(2000)\citenamefont
  {Vretenar}, \citenamefont {Finelli}, \citenamefont {Ventura}, \citenamefont
  {Lalazissis},\ and\ \citenamefont {Ring}}]{PhysRevC.61.064307}%
  \BibitemOpen
  \bibfield  {author} {\bibinfo {author} {\bibfnamefont {D.}~\bibnamefont
  {Vretenar}}, \bibinfo {author} {\bibfnamefont {P.}~\bibnamefont {Finelli}},
  \bibinfo {author} {\bibfnamefont {A.}~\bibnamefont {Ventura}}, \bibinfo
  {author} {\bibfnamefont {G.~A.}\ \bibnamefont {Lalazissis}}, \ and\ \bibinfo
  {author} {\bibfnamefont {P.}~\bibnamefont {Ring}},\ }\href {\doibase
  10.1103/PhysRevC.61.064307} {\bibfield  {journal} {\bibinfo  {journal} {Phys.
  Rev. C}\ }\textbf {\bibinfo {volume} {61}},\ \bibinfo {pages} {064307}
  (\bibinfo {year} {2000})}\BibitemShut {NoStop}%
\bibitem [{\citenamefont {Horowitz}\ \emph {et~al.}(2001)\citenamefont
  {Horowitz}, \citenamefont {Pollock}, \citenamefont {Souder},\ and\
  \citenamefont {Michaels}}]{PhysRevC.63.025501}%
  \BibitemOpen
  \bibfield  {author} {\bibinfo {author} {\bibfnamefont {C.~J.}\ \bibnamefont
  {Horowitz}}, \bibinfo {author} {\bibfnamefont {S.~J.}\ \bibnamefont
  {Pollock}}, \bibinfo {author} {\bibfnamefont {P.~A.}\ \bibnamefont {Souder}},
  \ and\ \bibinfo {author} {\bibfnamefont {R.}~\bibnamefont {Michaels}},\
  }\href {\doibase 10.1103/PhysRevC.63.025501} {\bibfield  {journal} {\bibinfo
  {journal} {Phys. Rev. C}\ }\textbf {\bibinfo {volume} {63}},\ \bibinfo
  {pages} {025501} (\bibinfo {year} {2001})}\BibitemShut {NoStop}%
\bibitem [{\citenamefont {Moreno}\ \emph {et~al.}(2011)\citenamefont {Moreno},
  \citenamefont {Sarriguren}, \citenamefont {Moya~de Guerra}, \citenamefont
  {Ud\'{\i}as}, \citenamefont {Donnelly},\ and\ \citenamefont
  {Sick}}]{1742-6596-312-9-092044}%
  \BibitemOpen
  \bibfield  {author} {\bibinfo {author} {\bibfnamefont {O.}~\bibnamefont
  {Moreno}}, \bibinfo {author} {\bibfnamefont {P.}~\bibnamefont {Sarriguren}},
  \bibinfo {author} {\bibfnamefont {E.}~\bibnamefont {Moya~de Guerra}},
  \bibinfo {author} {\bibfnamefont {J.~M.}\ \bibnamefont {Ud\'{\i}as}},
  \bibinfo {author} {\bibfnamefont {T.~W.}\ \bibnamefont {Donnelly}}, \ and\
  \bibinfo {author} {\bibfnamefont {I.}~\bibnamefont {Sick}},\ }\href
  {http://stacks.iop.org/1742-6596/312/i=9/a=092044} {\bibfield  {journal}
  {\bibinfo  {journal} {Journal of Physics: Conference Series}\ }\textbf
  {\bibinfo {volume} {312}},\ \bibinfo {pages} {092044} (\bibinfo {year}
  {2011})}\BibitemShut {NoStop}%
\bibitem [{\citenamefont {Bastin}\ \emph {et~al.}(2007)\citenamefont {Bastin},
  \citenamefont {Gr\'evy}, \citenamefont {Sohler}, \citenamefont {Sorlin},
  \citenamefont {Dombr\'adi}, \citenamefont {Achouri}, \citenamefont
  {Ang\'elique}, \citenamefont {Azaiez}, \citenamefont {Baiborodin},
  \citenamefont {Borcea}, \citenamefont {Bourgeois}, \citenamefont {Buta},
  \citenamefont {B\"urger}, \citenamefont {Chapman}, \citenamefont {Dalouzy},
  \citenamefont {Dlouhy}, \citenamefont {Drouard}, \citenamefont {Elekes},
  \citenamefont {Franchoo}, \citenamefont {Iacob}, \citenamefont {Laurent},
  \citenamefont {Lazar}, \citenamefont {Liang}, \citenamefont {Li\'enard},
  \citenamefont {Mrazek}, \citenamefont {Nalpas}, \citenamefont {Negoita},
  \citenamefont {Orr}, \citenamefont {Penionzhkevich}, \citenamefont
  {Podoly\'ak}, \citenamefont {Pougheon}, \citenamefont {Roussel-Chomaz},
  \citenamefont {Saint-Laurent}, \citenamefont {Stanoiu}, \citenamefont
  {Stefan}, \citenamefont {Nowacki},\ and\ \citenamefont
  {Poves}}]{PhysRevLett.99.022503}%
  \BibitemOpen
  \bibfield  {author} {\bibinfo {author} {\bibfnamefont {B.}~\bibnamefont
  {Bastin}}, \bibinfo {author} {\bibfnamefont {S.}~\bibnamefont {Gr\'evy}},
  \bibinfo {author} {\bibfnamefont {D.}~\bibnamefont {Sohler}}, \bibinfo
  {author} {\bibfnamefont {O.}~\bibnamefont {Sorlin}}, \bibinfo {author}
  {\bibfnamefont {Z.}~\bibnamefont {Dombr\'adi}}, \bibinfo {author}
  {\bibfnamefont {N.~L.}\ \bibnamefont {Achouri}}, \bibinfo {author}
  {\bibfnamefont {J.~C.}\ \bibnamefont {Ang\'elique}}, \bibinfo {author}
  {\bibfnamefont {F.}~\bibnamefont {Azaiez}}, \bibinfo {author} {\bibfnamefont
  {D.}~\bibnamefont {Baiborodin}}, \bibinfo {author} {\bibfnamefont
  {R.}~\bibnamefont {Borcea}}, \bibinfo {author} {\bibfnamefont
  {C.}~\bibnamefont {Bourgeois}}, \bibinfo {author} {\bibfnamefont
  {A.}~\bibnamefont {Buta}}, \bibinfo {author} {\bibfnamefont {A.}~\bibnamefont
  {B\"urger}}, \bibinfo {author} {\bibfnamefont {R.}~\bibnamefont {Chapman}},
  \bibinfo {author} {\bibfnamefont {J.~C.}\ \bibnamefont {Dalouzy}}, \bibinfo
  {author} {\bibfnamefont {Z.}~\bibnamefont {Dlouhy}}, \bibinfo {author}
  {\bibfnamefont {A.}~\bibnamefont {Drouard}}, \bibinfo {author} {\bibfnamefont
  {Z.}~\bibnamefont {Elekes}}, \bibinfo {author} {\bibfnamefont
  {S.}~\bibnamefont {Franchoo}}, \bibinfo {author} {\bibfnamefont
  {S.}~\bibnamefont {Iacob}}, \bibinfo {author} {\bibfnamefont
  {B.}~\bibnamefont {Laurent}}, \bibinfo {author} {\bibfnamefont
  {M.}~\bibnamefont {Lazar}}, \bibinfo {author} {\bibfnamefont
  {X.}~\bibnamefont {Liang}}, \bibinfo {author} {\bibfnamefont
  {E.}~\bibnamefont {Li\'enard}}, \bibinfo {author} {\bibfnamefont
  {J.}~\bibnamefont {Mrazek}}, \bibinfo {author} {\bibfnamefont
  {L.}~\bibnamefont {Nalpas}}, \bibinfo {author} {\bibfnamefont
  {F.}~\bibnamefont {Negoita}}, \bibinfo {author} {\bibfnamefont {N.~A.}\
  \bibnamefont {Orr}}, \bibinfo {author} {\bibfnamefont {Y.}~\bibnamefont
  {Penionzhkevich}}, \bibinfo {author} {\bibfnamefont {Z.}~\bibnamefont
  {Podoly\'ak}}, \bibinfo {author} {\bibfnamefont {F.}~\bibnamefont
  {Pougheon}}, \bibinfo {author} {\bibfnamefont {P.}~\bibnamefont
  {Roussel-Chomaz}}, \bibinfo {author} {\bibfnamefont {M.~G.}\ \bibnamefont
  {Saint-Laurent}}, \bibinfo {author} {\bibfnamefont {M.}~\bibnamefont
  {Stanoiu}}, \bibinfo {author} {\bibfnamefont {I.}~\bibnamefont {Stefan}},
  \bibinfo {author} {\bibfnamefont {F.}~\bibnamefont {Nowacki}}, \ and\
  \bibinfo {author} {\bibfnamefont {A.}~\bibnamefont {Poves}},\ }\href
  {\doibase 10.1103/PhysRevLett.99.022503} {\bibfield  {journal} {\bibinfo
  {journal} {Phys. Rev. Lett.}\ }\textbf {\bibinfo {volume} {99}},\ \bibinfo
  {pages} {022503} (\bibinfo {year} {2007})}\BibitemShut {NoStop}%
\bibitem [{\citenamefont {Sorlin}\ and\ \citenamefont
  {Porquet}(2008)}]{Sorlin2008602}%
  \BibitemOpen
  \bibfield  {author} {\bibinfo {author} {\bibfnamefont {O.}~\bibnamefont
  {Sorlin}}\ and\ \bibinfo {author} {\bibfnamefont {M.-G.}\ \bibnamefont
  {Porquet}},\ }\href {\doibase http://dx.doi.org/10.1016/j.ppnp.2008.05.001}
  {\bibfield  {journal} {\bibinfo  {journal} {Progress in Particle and Nuclear
  Physics}\ }\textbf {\bibinfo {volume} {61}},\ \bibinfo {pages} {602 }
  (\bibinfo {year} {2008})}\BibitemShut {NoStop}%
\bibitem [{\citenamefont {Todd-Rutel}\ \emph {et~al.}(2004)\citenamefont
  {Todd-Rutel}, \citenamefont {Piekarewicz},\ and\ \citenamefont
  {Cottle}}]{todd}%
  \BibitemOpen
  \bibfield  {author} {\bibinfo {author} {\bibfnamefont {B.~G.}\ \bibnamefont
  {Todd-Rutel}}, \bibinfo {author} {\bibfnamefont {J.}~\bibnamefont
  {Piekarewicz}}, \ and\ \bibinfo {author} {\bibfnamefont {P.~D.}\ \bibnamefont
  {Cottle}},\ }\href {\doibase 10.1103/PhysRevC.69.021301} {\bibfield
  {journal} {\bibinfo  {journal} {Phys. Rev. C}\ }\textbf {\bibinfo {volume}
  {69}},\ \bibinfo {pages} {021301} (\bibinfo {year} {2004})}\BibitemShut
  {NoStop}%
\bibitem [{\citenamefont {Chu}\ \emph {et~al.}(2010)\citenamefont {Chu},
  \citenamefont {Ren}, \citenamefont {Wang},\ and\ \citenamefont {Dong}}]{chu}%
  \BibitemOpen
  \bibfield  {author} {\bibinfo {author} {\bibfnamefont {Y.}~\bibnamefont
  {Chu}}, \bibinfo {author} {\bibfnamefont {Z.}~\bibnamefont {Ren}}, \bibinfo
  {author} {\bibfnamefont {Z.}~\bibnamefont {Wang}}, \ and\ \bibinfo {author}
  {\bibfnamefont {T.}~\bibnamefont {Dong}},\ }\href {\doibase
  10.1103/PhysRevC.82.024320} {\bibfield  {journal} {\bibinfo  {journal} {Phys.
  Rev. C}\ }\textbf {\bibinfo {volume} {82}},\ \bibinfo {pages} {024320}
  (\bibinfo {year} {2010})}\BibitemShut {NoStop}%
\bibitem [{\citenamefont {Wang}\ \emph {et~al.}(2013)\citenamefont {Wang},
  \citenamefont {Ren},\ and\ \citenamefont {Dong}}]{Wang:2013rwg}%
  \BibitemOpen
  \bibfield  {author} {\bibinfo {author} {\bibfnamefont {Z.}~\bibnamefont
  {Wang}}, \bibinfo {author} {\bibfnamefont {Z.}~\bibnamefont {Ren}}, \ and\
  \bibinfo {author} {\bibfnamefont {T.}~\bibnamefont {Dong}},\ }\href@noop {}
  {\  (\bibinfo {year} {2013})},\ \Eprint {http://arxiv.org/abs/1305.3027}
  {arXiv:1305.3027 [nucl-th]} \BibitemShut {NoStop}%
\bibitem [{\citenamefont {Wang}\ \emph {et~al.}(2011)\citenamefont {Wang},
  \citenamefont {Gu}, \citenamefont {Zhang},\ and\ \citenamefont
  {Dong}}]{wang}%
  \BibitemOpen
  \bibfield  {author} {\bibinfo {author} {\bibfnamefont {Y.}~\bibnamefont
  {Wang}}, \bibinfo {author} {\bibfnamefont {J.}~\bibnamefont {Gu}}, \bibinfo
  {author} {\bibfnamefont {X.}~\bibnamefont {Zhang}}, \ and\ \bibinfo {author}
  {\bibfnamefont {J.}~\bibnamefont {Dong}},\ }\href@noop {} {\bibfield
  {journal} {\bibinfo  {journal} {Chin. Phys. Lett.}\ }\textbf {\bibinfo
  {volume} {28}},\ \bibinfo {pages} {102101} (\bibinfo {year}
  {2011})}\BibitemShut {NoStop}%
\bibitem [{\citenamefont {Yao}\ \emph {et~al.}(2013)\citenamefont {Yao},
  \citenamefont {Mei},\ and\ \citenamefont {Li}}]{yao}%
  \BibitemOpen
  \bibfield  {author} {\bibinfo {author} {\bibfnamefont {J.}~\bibnamefont
  {Yao}}, \bibinfo {author} {\bibfnamefont {H.}~\bibnamefont {Mei}}, \ and\
  \bibinfo {author} {\bibfnamefont {Z.}~\bibnamefont {Li}},\ }\href@noop {}
  {\bibfield  {journal} {\bibinfo  {journal} {Phys. Lett. B}\ }\textbf
  {\bibinfo {volume} {723}},\ \bibinfo {pages} {459} (\bibinfo {year}
  {2013})}\BibitemShut {NoStop}%
\bibitem [{\citenamefont {Meucci}\ \emph
  {et~al.}(2001{\natexlab{b}})\citenamefont {Meucci}, \citenamefont {Giusti},\
  and\ \citenamefont {Pacati}}]{Meucci:2001ja}%
  \BibitemOpen
  \bibfield  {author} {\bibinfo {author} {\bibfnamefont {A.}~\bibnamefont
  {Meucci}}, \bibinfo {author} {\bibfnamefont {C.}~\bibnamefont {Giusti}}, \
  and\ \bibinfo {author} {\bibfnamefont {F.~D.}\ \bibnamefont {Pacati}},\
  }\href {\doibase 10.1103/PhysRevC.64.064615} {\bibfield  {journal} {\bibinfo
  {journal} {Phys. Rev. C}\ }\textbf {\bibinfo {volume} {64}},\ \bibinfo
  {pages} {064615} (\bibinfo {year} {2001}{\natexlab{b}})}\BibitemShut
  {NoStop}%
\bibitem [{\citenamefont {Meucci}(2002)}]{Meucci:2001ty}%
  \BibitemOpen
  \bibfield  {author} {\bibinfo {author} {\bibfnamefont {A.}~\bibnamefont
  {Meucci}},\ }\href {\doibase 10.1103/PhysRevC.65.044601} {\bibfield
  {journal} {\bibinfo  {journal} {Phys. Rev. C}\ }\textbf {\bibinfo {volume}
  {65}},\ \bibinfo {pages} {044601} (\bibinfo {year} {2002})}\BibitemShut
  {NoStop}%
\bibitem [{\citenamefont {Radici}\ \emph {et~al.}(2003)\citenamefont {Radici},
  \citenamefont {Meucci},\ and\ \citenamefont {Dickhoff}}]{Radici:2003zz}%
  \BibitemOpen
  \bibfield  {author} {\bibinfo {author} {\bibfnamefont {M.}~\bibnamefont
  {Radici}}, \bibinfo {author} {\bibfnamefont {A.}~\bibnamefont {Meucci}}, \
  and\ \bibinfo {author} {\bibfnamefont {W.~H.}\ \bibnamefont {Dickhoff}},\
  }\href {\doibase 10.1140/epja/i2002-10137-2} {\bibfield  {journal} {\bibinfo
  {journal} {Eur. Phys. J. A}\ }\textbf {\bibinfo {volume} {17}},\ \bibinfo
  {pages} {65} (\bibinfo {year} {2003})}\BibitemShut {NoStop}%
\bibitem [{\citenamefont {Capuzzi}\ \emph {et~al.}(1991)\citenamefont
  {Capuzzi}, \citenamefont {Giusti},\ and\ \citenamefont
  {Pacati}}]{Capuzzi:1991qd}%
  \BibitemOpen
  \bibfield  {author} {\bibinfo {author} {\bibfnamefont {F.}~\bibnamefont
  {Capuzzi}}, \bibinfo {author} {\bibfnamefont {C.}~\bibnamefont {Giusti}}, \
  and\ \bibinfo {author} {\bibfnamefont {F.~D.}\ \bibnamefont {Pacati}},\
  }\href {\doibase 10.1016/0375-9474(91)90269-C} {\bibfield  {journal}
  {\bibinfo  {journal} {Nuclear Physics A}\ }\textbf {\bibinfo {volume}
  {524}},\ \bibinfo {pages} {681 } (\bibinfo {year} {1991})}\BibitemShut
  {NoStop}%
\bibitem [{\citenamefont {Capuzzi}\ \emph {et~al.}(2005)\citenamefont
  {Capuzzi}, \citenamefont {Giusti}, \citenamefont {Pacati},\ and\
  \citenamefont {Kadrev}}]{Capuzzi:2004au}%
  \BibitemOpen
  \bibfield  {author} {\bibinfo {author} {\bibfnamefont {F.}~\bibnamefont
  {Capuzzi}}, \bibinfo {author} {\bibfnamefont {C.}~\bibnamefont {Giusti}},
  \bibinfo {author} {\bibfnamefont {F.~D.}\ \bibnamefont {Pacati}}, \ and\
  \bibinfo {author} {\bibfnamefont {D.~N.}\ \bibnamefont {Kadrev}},\ }\href
  {\doibase 10.1016/j.aop.2004.12.005} {\bibfield  {journal} {\bibinfo
  {journal} {Annals of Physics (N.Y.)}\ }\textbf {\bibinfo {volume} {317}},\
  \bibinfo {pages} {492 } (\bibinfo {year} {2005})}\BibitemShut {NoStop}%
\bibitem [{\citenamefont {Giusti}\ and\ \citenamefont
  {Meucci}(2011)}]{Giusti:cortona11}%
  \BibitemOpen
  \bibfield  {author} {\bibinfo {author} {\bibfnamefont {C.}~\bibnamefont
  {Giusti}}\ and\ \bibinfo {author} {\bibfnamefont {A.}~\bibnamefont
  {Meucci}},\ }\href@noop {} {\bibfield  {journal} {\bibinfo  {journal}
  {Journal of Physics: Conference Series}\ }\textbf {\bibinfo {volume} {336}},\
  \bibinfo {pages} {012025} (\bibinfo {year} {2011})}\BibitemShut {NoStop}%
\bibitem [{\citenamefont {Meucci}\ \emph
  {et~al.}(2013{\natexlab{b}})\citenamefont {Meucci}, \citenamefont {Giusti},\
  and\ \citenamefont {Vorabbi}}]{Meucci:2013gja}%
  \BibitemOpen
  \bibfield  {author} {\bibinfo {author} {\bibfnamefont {A.}~\bibnamefont
  {Meucci}}, \bibinfo {author} {\bibfnamefont {C.}~\bibnamefont {Giusti}}, \
  and\ \bibinfo {author} {\bibfnamefont {M.}~\bibnamefont {Vorabbi}},\ }\href
  {\doibase 10.1103/PhysRevD.88.013006} {\bibfield  {journal} {\bibinfo
  {journal} {Phys. Rev. D}\ }\textbf {\bibinfo {volume} {88}},\ \bibinfo
  {pages} {013006} (\bibinfo {year} {2013}{\natexlab{b}})},\ \Eprint
  {http://arxiv.org/abs/1305.5466} {arXiv:1305.5466 [nucl-th]} \BibitemShut
  {NoStop}%
\bibitem [{\citenamefont {Gonz\'alez-Jim\'enez}\ \emph
  {et~al.}(2013)\citenamefont {Gonz\'alez-Jim\'enez}, \citenamefont
  {Caballero}, \citenamefont {Meucci}, \citenamefont {Giusti}, \citenamefont
  {Barbaro}, \citenamefont {Ivanov},\ and\ \citenamefont
  {Ud\'{\i}as}}]{PhysRevC.88.025502}%
  \BibitemOpen
  \bibfield  {author} {\bibinfo {author} {\bibfnamefont {R.}~\bibnamefont
  {Gonz\'alez-Jim\'enez}}, \bibinfo {author} {\bibfnamefont {J.~A.}\
  \bibnamefont {Caballero}}, \bibinfo {author} {\bibfnamefont {A.}~\bibnamefont
  {Meucci}}, \bibinfo {author} {\bibfnamefont {C.}~\bibnamefont {Giusti}},
  \bibinfo {author} {\bibfnamefont {M.~B.}\ \bibnamefont {Barbaro}}, \bibinfo
  {author} {\bibfnamefont {M.~V.}\ \bibnamefont {Ivanov}}, \ and\ \bibinfo
  {author} {\bibfnamefont {J.~M.}\ \bibnamefont {Ud\'{\i}as}},\ }\href
  {\doibase 10.1103/PhysRevC.88.025502} {\bibfield  {journal} {\bibinfo
  {journal} {Phys. Rev. C}\ }\textbf {\bibinfo {volume} {88}},\ \bibinfo
  {pages} {025502} (\bibinfo {year} {2013})}\BibitemShut {NoStop}%
\bibitem [{\citenamefont {Horikawa}\ \emph {et~al.}(1980)\citenamefont
  {Horikawa}, \citenamefont {Lenz},\ and\ \citenamefont {Mukhopadhyay}}]{hori}%
  \BibitemOpen
  \bibfield  {author} {\bibinfo {author} {\bibfnamefont {Y.}~\bibnamefont
  {Horikawa}}, \bibinfo {author} {\bibfnamefont {F.}~\bibnamefont {Lenz}}, \
  and\ \bibinfo {author} {\bibfnamefont {N.~C.}\ \bibnamefont {Mukhopadhyay}},\
  }\href@noop {} {\bibfield  {journal} {\bibinfo  {journal} {Phys. Rev. C}\
  }\textbf {\bibinfo {volume} {22}},\ \bibinfo {pages} {1680} (\bibinfo {year}
  {1980})}\BibitemShut {NoStop}%
\bibitem [{\citenamefont {Maieron}\ \emph {et~al.}(2002)\citenamefont
  {Maieron}, \citenamefont {Donnelly},\ and\ \citenamefont
  {Sick}}]{PhysRevC.65.025502}%
  \BibitemOpen
  \bibfield  {author} {\bibinfo {author} {\bibfnamefont {C.}~\bibnamefont
  {Maieron}}, \bibinfo {author} {\bibfnamefont {T.~W.}\ \bibnamefont
  {Donnelly}}, \ and\ \bibinfo {author} {\bibfnamefont {I.}~\bibnamefont
  {Sick}},\ }\href {\doibase 10.1103/PhysRevC.65.025502} {\bibfield  {journal}
  {\bibinfo  {journal} {Phys. Rev. C}\ }\textbf {\bibinfo {volume} {65}},\
  \bibinfo {pages} {025502} (\bibinfo {year} {2002})}\BibitemShut {NoStop}%
\bibitem [{\citenamefont {Cooper}\ \emph {et~al.}(2009)\citenamefont {Cooper},
  \citenamefont {Hama},\ and\ \citenamefont {Clark}}]{Cooper:2009}%
  \BibitemOpen
  \bibfield  {author} {\bibinfo {author} {\bibfnamefont {E.~D.}\ \bibnamefont
  {Cooper}}, \bibinfo {author} {\bibfnamefont {S.}~\bibnamefont {Hama}}, \ and\
  \bibinfo {author} {\bibfnamefont {B.~C.}\ \bibnamefont {Clark}},\ }\href
  {\doibase 10.1103/PhysRevC.80.034605} {\bibfield  {journal} {\bibinfo
  {journal} {Phys. Rev. C}\ }\textbf {\bibinfo {volume} {80}},\ \bibinfo
  {pages} {034605} (\bibinfo {year} {2009})}\BibitemShut {NoStop}%
\bibitem [{\citenamefont {de~Forest~Jr.}(1983)}]{DeForestJr1983232}%
  \BibitemOpen
  \bibfield  {author} {\bibinfo {author} {\bibfnamefont {T.}~\bibnamefont
  {de~Forest~Jr.}},\ }\href {\doibase 10.1016/0375-9474(83)90124-0} {\bibfield
  {journal} {\bibinfo  {journal} {Nuclear Physics A}\ }\textbf {\bibinfo
  {volume} {392}},\ \bibinfo {pages} {232 } (\bibinfo {year}
  {1983})}\BibitemShut {NoStop}%
\end{thebibliography}
%
\end{document}